\documentclass[a4paper,fleqn,usenatbib]{mnras}
\usepackage{graphicx}
\usepackage{amsmath}
\usepackage{hyperref}
\usepackage{float}
\usepackage{siunitx}
\usepackage{subcaption}
\usepackage{braket}
\usepackage{color}
\usepackage{epsfig}
\usepackage{enumitem}

\usepackage{newtxtext,newtxmath}

\usepackage[T1]{fontenc}
\usepackage{ae,aecompl}

\def\kmsmpc{\,{\rm km}\,{\rm s}^{-1}{\rm Mpc}^{-1}}

 at 13 truept
\def\citejap#1{\citeauthor{#1}\ \citeyear{#1}}
\def\citefix#1{\citeauthor{#1}\ (\citeyear{#1})}
\def\fixv{\rlap{\phantom{$\strut^b$}}}

\def\m@th{\mathsurround=0pt }
\def\eqalign#1{\null\,\vcenter{\openup1\jot \m@th
 \ialign{\strut\hfil$\displaystyle{##}$&$\displaystyle{{}##}$\hfil
 \crcr#1\crcr}}\,}

\begin{document}

\twocolumn

\title[The Halo Mass Function]{An empirical measurement of the Halo Mass Function from the combination of GAMA\,DR4, SDSS\,DR12, and REFLEX\,II data}
\author[Driver et al] 
{Simon\,P.\,Driver$^{1}$,
Aaron\,S.\,G.\,Robotham$^{1}$,
Danail\,Obreschkow$^{1}$, 
John\,A.\,Peacock$^{2}$, 
Ivan\,K.\,Baldry$^{3}$, \newauthor
Sabine\,Bellstedt$^1$,
Joss Bland-Hawthorn$^4$,
Sarah\,Brough$^{5}$,  
Michelle\,Cluver$^{6}$, 
Benne\,W.\,Holwerda$^{7}$, \newauthor  
Andrew Hopkins$^{8}$, 
Claudia Lagos$^1$, 
Jochen Liske$^{9}$,
Jon Loveday$^{10}$, 
Steven\,Phillipps$^{11}$, 
Edward\,N.\,Taylor$^5$ \\
$^1$ International Centre for Radio Astronomy Research (ICRAR), University of Western Australia, Crawley, WA 6009, Australia \\
$^2$ Institute for Astronomy, University of Edinburgh, Royal Observatory, Edinburgh EH9 3HJ, UK \\
$^3$ Astrophysics Research Institute, Liverpool John Moores University, IC2, Liverpool Science Park, 146 Brownlow Hill, Liverpool, L3 5RF \\
$^4$ Sydney Astrophotonic Instrumentation Labs, School of Physics, the University of Sydney, Sydney NSW 2006, Australia \\ 
$^5$ School of Physics, University of New South Wales, NSW 2052, Australia \\
$^6$ Centre for Astrophysics and Supercomputing, Swinburne University of Technology, Hawthorn, VIC 3122, Australia \\
$^7$ Department of Physics and Astronomy, 102 Natural Science Building, University of Louisville, Louisville KY 40292, USA \\
$^8$ Australian Astronomical Optics, Macquarie University, 105 Delhi Rd, North Ryde, NSW 2113, Australia \\
$^{9}$ Hamburger Sternwarte, Universit\"at Hamburg, Gojenbergsweg 112, 21029 Hamburg, Germany \\ 
$^{10}$ Astronomy Centre, University of Sussex, Falmer, Brighton BN1 9QH, UK, \\
$^{11}$ Astrophysics Group, School of Physics, Tyndall Avenue, University of Bristol, Bristol, BS8 1TL, UK }

\pubyear{2022} \volume{000}


\maketitle
\label{firstpage}

\vspace{-2.0cm}

\begin{abstract}
We construct the halo mass function (HMF) from the GAMA galaxy group catalogue over the mass range $10^{12.7}$\,M$_{\odot}$ to $10^{15.5}$\,M$_{\odot}$, and find good agreement with the expectation from $\Lambda$CDM. In comparison to previous studies, this result extends the mass range over which the HMF has now been measured over by an order of magnitude. We combine the GAMA\,DR4 HMF with similar data from the SDSS\,DR12 and REFLEX\,II surveys, and fit a four-parameter Murray-Robotham-Power (MRP) function, valid at $\tilde{z} \approx 0.1$, yielding:
a density normalisation of: $\log_{10}(\phi_*$\,Mpc$^{3})= -3.96^{+0.55}_{-0.82}$, 
a high mass turn-over of: $\log_{10}(M_*$\,M$_{\odot}^{-1})=14.13^{+0.43}_{-0.40}$, 
a low mass power law slope of: $\alpha =-1.68^{+0.21}_{-0.24}$, 
and a high mass softening parameter of: $\beta=0.63^{+0.25}_{-0.11}$.
If we fold in the constraint on $\Omega_M$ from Planck\,2018 Cosmology, we are able to reduce these uncertainties further, but this relies on the assumption that the power-law trend can be extrapolated from $10^{12.7}$M$_{\odot}$ to zero mass.
Throughout, we highlight the effort needed to improve on our HMF measurement: improved halo mass estimates that {\it do not} rely on calibration to simulations; reduced halo mass uncertainties needed to mitigate the strong Eddington Bias that arises from the steepness of the HMF low mass slope; and deeper wider area spectroscopic surveys. 
To our halo mass limit of $10^{12.7}$\,M$_{\odot}$, we are directly resolving (`seeing') $41 \pm 5$ per cent of the total mass density, i.e. $\Omega_{M,>12.7}=0.128 \pm 0.016$, opening the door for the direct construction of 3D dark matter mass maps at Mpc resolution.
\end{abstract}


\section{Introduction}
One of the definitive predictions from the $\Lambda$ Cold Dark Matter paradigm ($\Lambda$CDM), is the form (shape and amplitude), of the underlying dark matter Halo Mass Function (HMF, see early works by \citeauthor{frenk1988} \citeyear{frenk1988} and \citeauthor{brainerd1992} \citeyear{brainerd1992} for example), and its evolution over time (see \citeauthor{reed2003} \citeyear{reed2003}; \citeauthor{lukic2007} \citeyear{lukic2007}; \citeauthor{watson2013} \citeyear{watson2013}). The HMF describes the number density of dark matter haloes, either per mass interval or per log mass interval, with both forms in common usage. At redshift zero the HMF can be described as a near mass-divergent power-law distribution, with a high mass cutoff correlated with the mass assembly time since the Big Bang. 

The HMF form can be derived heuristically from Press-Schechter theory for the gravitational collapse of over-dense condensates (\citeauthor{press1974} \citeyear{press1974}). 
The theory of the HMF was subsequently placed on a sounder basis via the
random-trajectories approach, which allowed an understanding
of the fate of material in underdense regions and a solution of the
`cloud-in-cloud’ problem \citep{peacock1990,bond1991}.

The initial HMF, established at the time of decoupling, then evolves through the process of hierarchical assembly of smaller dark matter haloes. This results in an HMF with both a time- and scale-invariant power-law at masses below some time-dependent cut-off mass (see for example \citeauthor{jenkins2001} \citeyear{jenkins2001}; \citeauthor{reed2007} \citeyear{reed2007}). The HMF also emerges naturally from numerical simulations of the evolving dark matter distribution, albeit with 10-20 per cent variation within a particular simulation \citep{ondaro-meallea2021}, and a 10 to 20 per cent variation between simulations \citep{murray2013}.

\begin{figure*}
	\centering     
	\includegraphics[width=\textwidth]{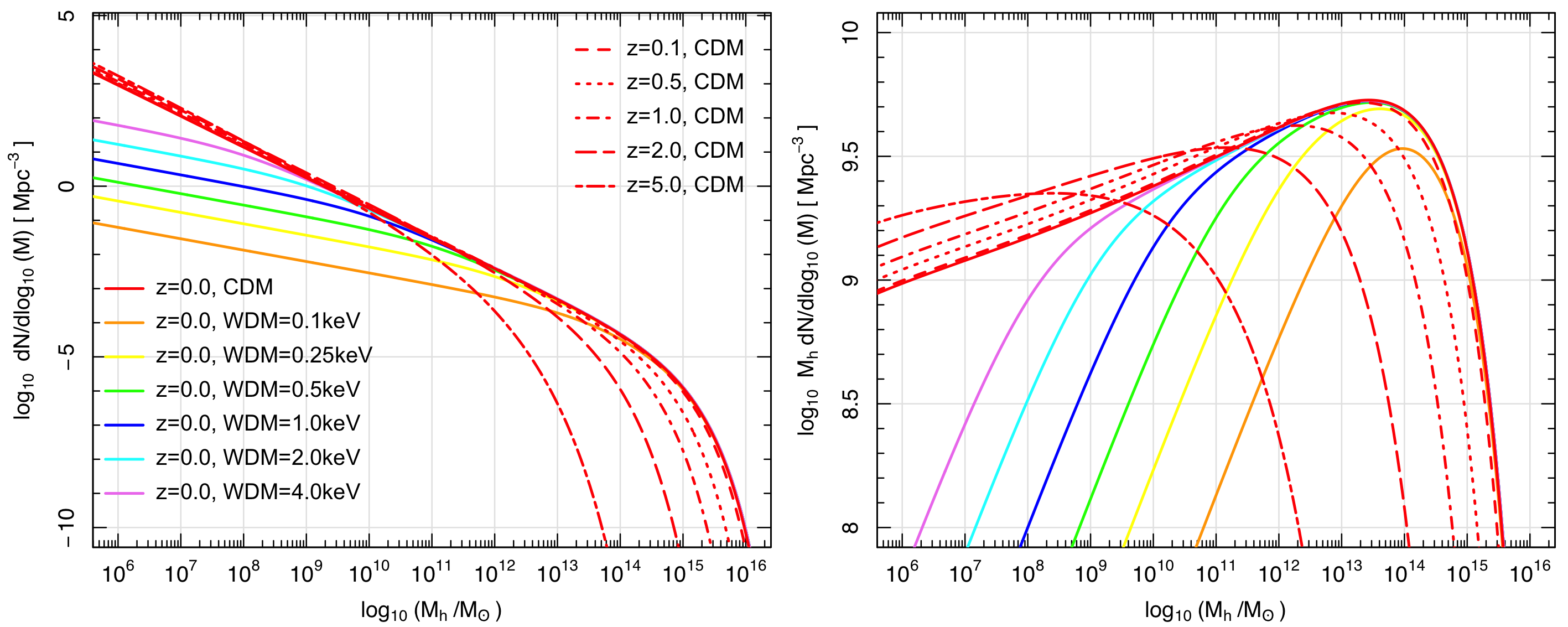}
	\caption{(left) Analytic predictions of the HMF over a very broad halo mass range (lines), its predicted evolution with redshift (red lines), and its dependence on the dark matter particle mass (coloured lines). (right) The same predictions as the left panel, but now showing the contribution of each decade of halo mass to the total matter density, and highlighting the mass range where most mass is predicted to reside. All the curves shown are taken from \citefix{murray2013}; see also \citefix{murray2021} for a recent revision and online tool.}
	\label{fig:cdmpredictions}
\end{figure*}

Recently, \cite{murray2018} demonstrated that the HMF can be described by a four parameter function (hereafter the MRP function, see Eqn.\,\ref{eqn:mrp}) to within 5 per cent accuracy at any mass interval. The MRP function closely relates to the Schechter function \citep{schechter1976}, which is commonly used for fitting galaxy luminosity and/or stellar mass functions (e.g. Driver et al. submitted) and which can be derived from Press-Schechter formalism. The one difference between the MRP and Schechter functions is the addition of a parameter ($\beta$) to soften or sharpen the exponential cut-off at high mass.

This exponential cut-off at the high mass end of the HMF is expected to evolve significantly (\citejap{reed2003}; see red lines in Fig.\,\ref{fig:cdmpredictions}), as progressively larger haloes form over time from the hierarchical merging of the dark matter haloes. This results in the emergence of massive dark matter haloes ($\geq 10^{14.5}$\,M$_{\odot}$) hosting rich clusters of galaxies around a redshift of unity \citep{allen2011}. Hence the existence and evolving density of high-redshift high mass clusters is a key and stringent test of $\Lambda$CDM \citep{allen2011,elgordo2}. 

The dark matter particle mass can also influence the low mass end of the HMF, as illustrated in Fig.\,\ref{fig:cdmpredictions}, with any low mass cut-off directly linked to the dark matter particle mass for some early-decoupling weakly interacting massive particle (WIMPs: e.g. \citeauthor{murray2013} \citeyear{murray2013}). For $\Lambda$CDM this mass cut-off is around Earth mass, but in warm or hot (neutrino) dark matter it can be as high as $10^{13}$\,M$_{\odot}$ for dark matter particles of around 0.1\,keV (see Fig.\,\ref{fig:cdmpredictions}). This is due to the propensity of Hot or Warm Dark Matter to free stream during the era in which the particles are relativistic. If the dark matter consists of a very heavy particle ($>$ many keV), or even primordial black holes ($0.5$M$_{\odot}$ -- $100$M$_{\odot}$) then no detectable low mass cutoff would be expected.

The HMF is hence a critical probe of $\Lambda$CDM, with at least three distinct testable facets: \begin{itemize}[noitemsep,topsep=0pt]
\setlength{\itemindent}{1em}
    \item the high mass cut-off and its evolution with redshift; 
    \item the power-law slope and amplitude of the HMF; 
    \item and the existence and location of any low mass cut-off or perceived flattening in the low mass slope. 
\end{itemize}
A slightly more subtle test is the first moment of the HMF, the integration over all masses (see the right-hand panel of Fig.\,\ref{fig:cdmpredictions}), which should be consistent with the total matter density. For completely cold dark matter, all dark matter particles should be accounted for when summing over haloes down to infinitesimal mass; but in the case of WDM or HDM the effects of free streaming mean that there will be a subset of mass that is not associated with haloes, so that the integral of the local HMF may lie below the density inferred at recombination by Cosmic Microwave Background (CMB) studies. Fig.\,\ref{fig:cdmpredictions} illustrates these two tests and hence the importance of measuring the HMF. Fig.\,\ref{fig:cdmpredictions} (right panel) highlights how the majority of matter is predicted to reside in intermediate- to low-mass haloes (i.e. $\sim 60$ per cent in the range $10^{14.5}$M$_{\odot} - 10^{11}$M$_{\odot}$), hence demonstrating the importance of establishing group catalogues to low halo masses.

On the observational side the measurement of the HMF appears tractable and relatively straightforward \citep[see][]{bahcall1993}, but does hinge critically on group/cluster identification, and especially on robust halo mass estimates from group/cluster sizes and velocity dispersions. In general, the observational and analysis pathway is a simple task: construct a group catalogue, estimate masses, and convert to a volume limited space density. 
Comparison of the observed and predicted HMF can be made in multiple redshift slices, and over as broad a mass range as the observations permit. Typically this direct approach has been pursued in two ways, (1) via X-ray detection of galaxy clusters, e.g. \cite{bohringer2017}, and (2) via group finding  within large spectroscopic surveys, e.g. \cite{eke2004}. Both \cite{bohringer2017} and \cite{eke2006} show good agreement ($\sim 10-20$ per cent), with the HMF predicted by $\Lambda$CDM over the halo mass range of $10^{13.75}$\,M$_{\odot}-10^{15.25}$\,M$_{\odot}$ but also have significant subtleties, not least of which is a reliance on numerical simulations of the matter (and plasma for the X-ray pathway) to calibrate the mass estimates.

In the former, one has to rely on the correct calibration of the $L_X-M$ relation \citep{stanek2006,hoekstra2011}, or rather the better understood $L_X-T$ relation, and any variation with redshift \citep{leauthaud2010}. One also has to worry about virialisation, variance in the plasma properties (temperature especially in what may well be a mixed multi-phase medium), and the general bias towards selecting denser and higher-$L_X$ clusters, given detection is inevitably subject to some X-ray flux and flux contrast limit. The final amplitude of the predicted HMF is also dependent on the underlying clustering strength, with lower $\sigma_8$ values giving rise to higher than expected numbers of rich clusters. 
Moreover, as the X-ray flux of intermediate mass groups is minimal, either because the plasma is much cooler or not present, this method may only be viable at moderate to high group/cluster masses (i.e. $M \geq 10^{13}$\,M$_{\odot}$), hence not capable of probing to lower halo masses (i.e. $\sim 30$ per cent of the total predicted dark matter content). Progress in this area may be possible through the stacking of X-ray data at the locations of known galaxy groups: see the promising results from using SDSS selected galaxy samples to stack ROSAT data by \cite{anderson2015} \citep[see also more general discussion on this topic in][]{driver2021}, or via stacking at millimeter wavelengths via the Sunyaev-Zeldovich effect \citep{singari2020}.

For the large wide-area spectroscopic surveys, e.g. 2dFGRS \citep{colless2001}, SDSS \citep{sdss}, GAMA \citep{driver2009,driver2011}, DEVILS \citep{davies2018}, WAVES \citep{driver2016,driver2019}, DESI \citep{desi} etc., one has to worry about spectroscopic completeness that biases against lower-mass groups with fewer members, biases in the group finding algorithm (missed groups and/or false positives), and the inherent uncertainties in converting the measured redshifts into velocity dispersions and robust halo masses. This latter aspect becomes especially hard when the multiplicity, i.e. number of group members, is only a few. 

These problems, inherent in both the X-ray and spectroscopic survey approach, are tractable but the results are susceptible to strong biases, some of which arise from the large random uncertainties. Paramount amongst these is the Eddington bias that emerges from the uncertainty in the halo masses, and whose impact is exacerbated by the steepness of the HMF (see Fig.\,\ref{fig:cdmpredictions}). Hence a significant systematic mass shift can arise as the haloes are scattered to higher or lower masses during the measurement process.

One further aspect worth considering is the practice of calibrating the group-finding algorithms to numerical simulations, i.e. linking-lengths, and the mass scaling factor ($A$). This calibration process fundamentally links the empirical results to a specific semi-analytic model built upon an underlying dark matter simulation (see \citeauthor{eke2004} \citeyear{eke2004}; \citeauthor{robotham2011} \citeyear{robotham2011}). We note that the recent analysis of \cite{tempel2014} attempts to avoid this issue by calibrating the tangential linking length to the mean observed galaxy separation within redshift slices. Even more concerning are techniques that involve abundance matching, in which halo masses are assigned based on some rank order, and where the recovered HMF, by construction, is required to match that of the adopted simulation (see the review by \citejap{weschler2018}).

A more statistical approach to probing the HMF comes from weak lensing studies \citep[see e.g.][]{viola2015,dong2019,rana2021}, in which an adopted HMF combined with assumptions of the DM profiles, can be used to predict the lensing signature and compared to that measured. This method too has issues, mainly the need for a calibration, or adoption, of a specific HMF form and a universal DM profile shape that is poorly constrained especially at intermediate halo masses. It also has the disadvantage of not providing mass on a halo-by-halo basses, but the significant advantage of not needing to provide mass on a halo-by-halo basis, mitigating the concern over the Eddington Bias.

Finally, a viable alternative to measuring and comparing the HMF is to measure and compare the line-of-sight velocity dispersions with the 3D velocity dispersions from simulations. This approach bypasses the additional uncertainties around halo mass estimation, \citep[see e.g.][]{caldwell2016}, but is not without its own issues given the often limited number of velocity dispersion measurements for each halo. However, perhaps a stronger reason for pursuing the halo mass measurement pathway is that fundamentally we are interested in the determination of masses on a halo-by-halo basis, partly to study the variations of galaxy properties (and evolutionary pathways) as a function of halo mass, but also to investigate how $\Omega_M$, as constrained by cosmological measurements, is broken down into discrete self-gravitating clumps.

Current and future spectroscopic surveys have, in varying degrees, built their science cases around the measurement of both halo masses and the HMF (GAMA; \citeauthor{driver2011} \citeyear{driver2011}; WAVES, \citeauthor{driver2016} \citeyear{driver2016}, \citeauthor{driver2019} \citeyear{driver2019}; DEVILS, \citeauthor{davies2018} \citeyear{davies2018}), and while in practice the pathway looks straightforward the reality has proved more elusive. 

Here we attempt to construct an HMF from the GAMA survey and in this work look to follow an {\it empirical} pathway, while articulating the difficulties in doing so. Where possible we explore the dependency of our fitted HMF MRP parameters on some of the issues raised above, and compare our HMF to the few existing published measurements, which include: 2PIGG \citep{eke2006}; SDSS \citep{tempel2014}; and REFLEX\,II \citep{bohringer2017}. We later combine these data to provide a joint constraint on the MRP HMF parameters, and compare to the prediction from $\Lambda$CDM. Throughout we attempt to highlight key issues that need to be addressed to produce precision HMF measurements as new surveys that warrant such robustness come online (e.g. DESI, WAVES). Future papers will address biases in more detail, and seek to improve further on our halo mass estimates and more robust errors, as well as using bespoke simulations to better understand the systematics that emerge through the group-finding process and its calibration. 

In Section\,\ref{sec:GAMAGroupCat} we describe the GAMA group catalogue ($G^3C$), the mass measurements, and the mass errors. In Section\,\ref{sec:GAMAHMF} we describe our methodology for constructing the GAMA HMF, and show our attempts to fit the MRP function. In Section\,\ref{sec:ExternalEstimates} we combine the GAMA data with that from the SDSS, 2PIGG, and REFLEX\,II datasets, to provide a final joint HMF constraint, and discuss some of the broader and ultimately more speculative implications.

We use the Planck 2018 cosmology throughout, namely $\Omega_{M}=0.3147 \pm 0.0074$, $\Omega_{\Lambda}=1-\Omega_{M}$, and $H_0=67.37 \pm 0.54\kmsmpc$  \citep[][Table\,1, Col.\,6 -- Combined]{planck2018}.
For comparison with studies that use other cosmologies (especially the common units with explicit powers of $h$), our numbers should be scaled as follows:

~

\noindent
Halo masses :

~

$M \propto H_0^{-1}$, i.e. our units are M$_{\odot} h^{-1}_{\rm P18}$

~

\noindent
Space densities : 

~

$\phi \propto H_0^3$, i.e. our units are Mpc$^{-3}h^{3}_{\rm P18}$

~

\noindent
where $h_{\rm P18}=H_0/(67.37$\,km\,s$^{-1}$\,Mpc$^{-1}$).

\section{The GAMA Galaxy Group Catalogue ($G^3C$)}
\label{sec:GAMAGroupCat}

\begin{figure*}
	\centering     
	\includegraphics[width=\textwidth]{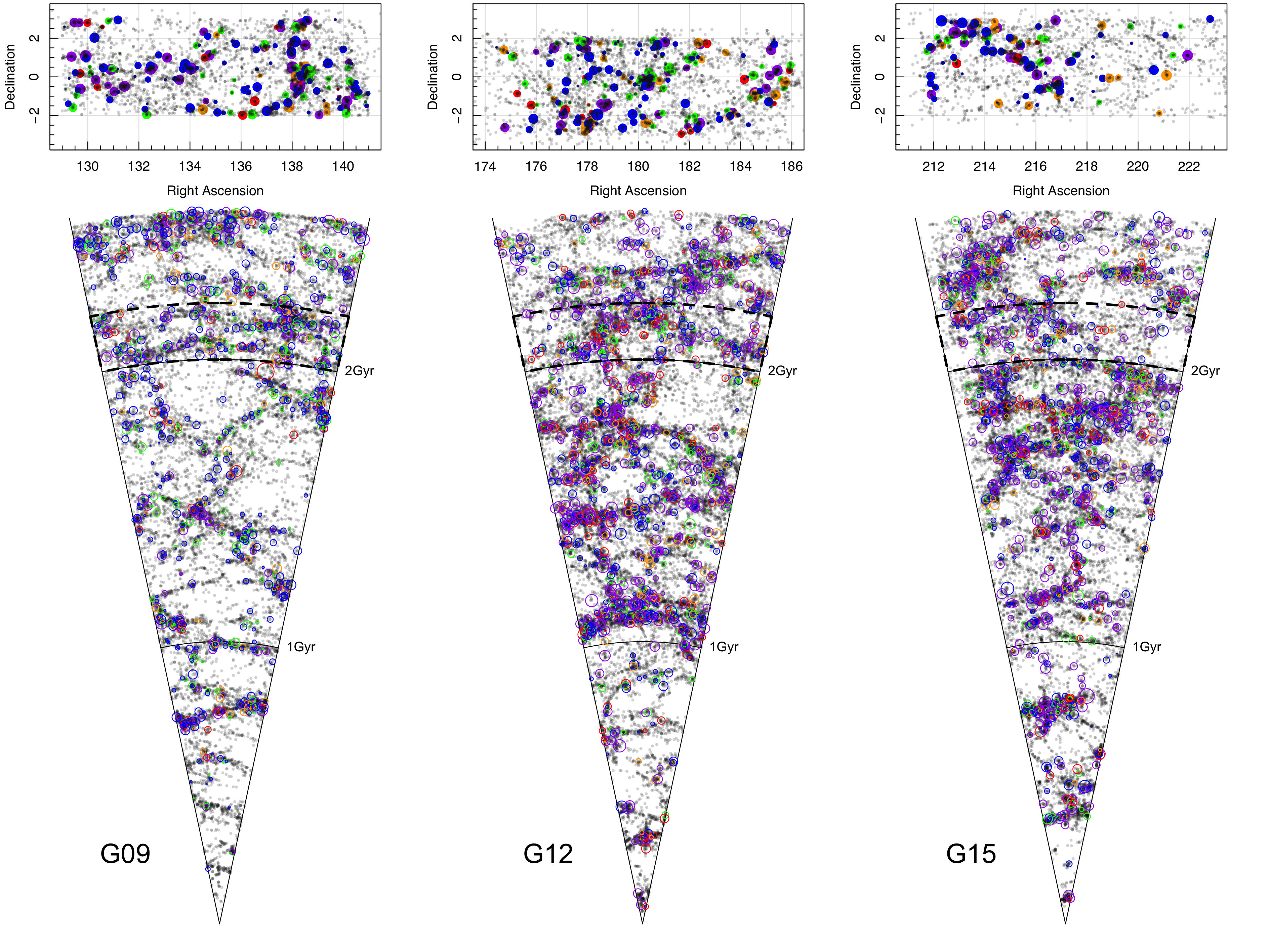}
	\caption{Each panel shows a cone plot of the GAMA group (coloured circles) and galaxy (grey dots) distributions to a maximum redshift of 0.3, indicating look-back time (lower cones), and in (upper panels) Right Ascension and Declination for a narrow redshift slice indicated by the dashed rectangles in the lower panels. The group circles are coloured according to multiplicity with `blue', `green', `orange', `red' and `purple' denoting multiplicities ($N_{\rm FoF}$) of 3, 4, 5, 6, $>6$ respectively. Circle sizes are scaled according to $\log_{10}(M_{\rm FoF})$.
}
	\label{fig:G3C}
\end{figure*}

\begin{figure*}
	\centering     
	\includegraphics[width=\textwidth]{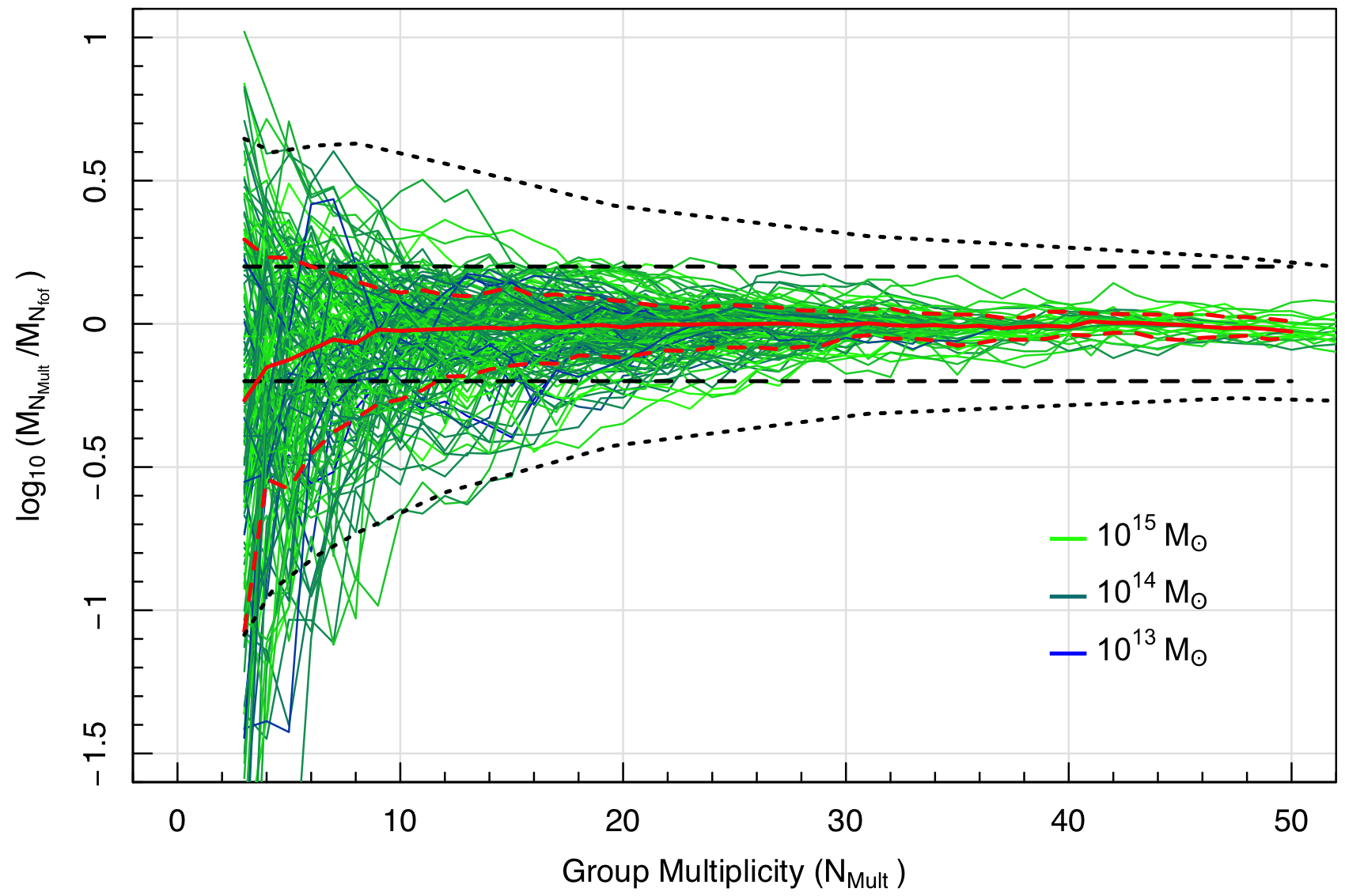}
	\caption{The determination of our mass error as a function of multiplicity. For all GAMA groups with more than 20 members we calculate the change in the mass as the multiplicity decreases (blue-green coloured lines). The trail colour indicates the original halo mass (see key). In red are the 16, 50 and 84 per cent quantiles. The dotted black lines show the initial error estimate from \citefix{robotham2011} while the dashed black lines show the mean weak lensing error estimate from \citefix{viola2015}.
	}
	\label{fig:masserr}
\end{figure*}

The Galaxy And Mass Assembly Survey (GAMA: see \citeauthor{driver2011} \citeyear{driver2011}; \citeauthor{liske2015} \citeyear{liske2015}; Driver et al., submitted), has provided over 300\,000 spectroscopic redshifts from the Anglo Australian Telescope's AAOmega facility, combined with pre-existing literature measurements. The GAMA survey covers 250 square degrees of sky in five distinct regions (G02, G09, G12, G15 \& G23), and the three primary equatorial regions (G09+G12+G15) cover 179.92 square degrees. Within these equatorial regions GAMA achieved over 98 per cent redshift completeness (see \citeauthor{liske2015} \citeyear{liske2015}), to the flux limit of $r_{\rm SDSS} = 19.8$ mag. 

Using a friends-of-friends algorithm, calibrated to the GALFORM semi-analytic model \citep{bower2006} built onto the Millennium dark matter simulation \citep{springel2005}, \cite{robotham2011} constructed the GAMA Galaxy Group catalogue (hereafter $G^3C$). The reliance of the catalogue on the simulations will be discussed at various stages below, but the final $G^3C$ catalogue consists of 26\,194 galaxy groups and pairs, of which 9\,718 groups have a multiplicity of 3 or greater, and 3\,061 with a multiplicity of 5 or more. The catalogue identifies more massive groups to $z>0.4$, but for the present analysis we impose an upper limit of $z=0.25$ in order to reduce complications from evolution of the HMF and also redshift-dependent systematics in mass estimation. This cut yields 5241 groups with multiplicity $\ge3$ and 1732 groups with multiplicity $\ge5$. The mean redshift of the latter subset is 0.153. 

Fig.\,\ref{fig:G3C} shows the spatial distribution of the group catalogue for multiplicity $\geq 5$ systems. Each of the three equatorial GAMA regions is shown independently with the top panels showing the RA and Dec distribution for a narrow redshift slice, while the lower panels show the light cones in lookback time and where the dashed line indicates the selected redshift slice. Symbol sizes are linked to $\log_{10}M_{\rm FoF}$, and the underlying grey data points show the full GAMA spectroscopic catalogue. Symbol colours denote multiplicity (see caption). The large scale structure, which traces the underlying dark matter distribution, is clearly evident, as is the known general under-density of the G09 region. Fly-through animations of these cones are available from the GAMA DR4 website\footnote{\url{http://www.gama-survey.org/dr4}}.

\subsection{$G^3C$ velocity dispersion and total mass estimates}
Total masses for the $G^3C$ sample were estimated from the velocity dispersion ($\sigma$), which in turn were derived from the group member CMB frame redshifts using the {\sc gapper} estimator (see \citeauthor{beers1990} \citeyear{beers1990}). This estimator includes consideration of: the magnitude distribution of the objects contributing to the velocity dispersion; the redshift of the cluster/group equalling that of the brightest member; and the mean uncertainty in the redshifts from our intrinsic redshift measurement error of $\pm 50$\,km\,s$^{-1}$ \citep{baldry2014,liske2015}. For full details as to how the $G^3C$ velocity dispersions are derived see equations 16 and 17 from \cite{robotham2011}. 

The total dynamical masses ($M_{\rm FoF}$), are then determined, assuming fully virialised haloes, via:
\begin{equation}
M_{\rm FoF} = \frac{A}{G}\sigma^2 R_{50}, \label{eqn:mass}
\end{equation}
where $A$ is some normalising factor (typically $\sim 10$), $G$ is the Gravitational Constant, $\sigma$ is the line-of-sight velocity dispersion, and $R_{50}$ is the projected radius containing 50 per cent of the group members (see section\,4.2 of \citejap{robotham2011} for a discussion concerning the optimal radius). In a virial equation such as Eqn.\,\ref{eqn:mass}, one would normally write $R_{50}$ in proper units -- but we choose to use comoving units, so that all lengths in this paper are comoving (although velocities are in physical units). The same choice was made by \cite{robotham2011}, and it had no impact on the results of that paper because the scaling factor was calibrated as a function of redshift and richness using simulations: thus any change in $A(z)$ from moving between proper and comoving lengths is automatically accounted for. The values of $A$ for the present updated catalogue were determined similarly, by running the same group finder and mass estimation algorithm on 9 equivalent-volume mock catalogues  (see \citejap{merson2013}) with identical properties to GAMA (i.e. volumes, magnitude limits, redshift errors, and incompleteness). 
This aspect is discussed further in Section\,\ref{sec:distances}.

The values for $A$ were found to range from $10.7$ to $19.2$ depending on group multiplicity, redshift and the adopted magnitude limits. Hence the mass value, {\sc MassAfunc}, in the $G^3C$ catalogue ({\sc G3CFoFGroupv10}) uses an optimised $A$ value for each group, based on its redshift and multiplicity, whereas the {\sc MassA} masses simply use the formula shown in Eqn.\,\ref{eqn:mass} with comoving $R_{50}$ values and a constant value of $A=10$. 

While the {\sc MassAfunc} values are designed to be as correct as possible in a $\Lambda$CDM universe, there may be some concern about excessive reliance on inexact simulations. We therefore also consider the simpler and less fine-tuned alternative of adopting a single typical value of $A$. The mean ratio between {\sc MassAfunc} and {\sc MassA} for groups within our nominal limits ($z<0.25$ and  $M>10^{13}$\,M$_{\odot}$) is $M_{A}/M_{\rm Afunc}=0.72$, and hence we will adopt $A=13.9$ as the best choice for the case of fixed $A$. We make the resulting mass estimates our primary choice for the HMF analysis, although we also report results based on the full {\sc MassAfunc} estimates.

The robustness of the $G^3C$ halo masses have been independently confirmed through comparison with weak lensing constraints from the ESO VLT Survey Telescope's Kilo Degree Survey: see \cite{viola2015}. Their equation 38 shows good consistency between KiDS and GAMA (Eqn.\,\ref{eqn:mass}), in the mass normalisation ($1.00 \pm 0.15$) around $10^{14}$\,M$_{\odot}$, with an average scatter (i.e. mass error) of $\sigma_{\log_{10}M} \sim 0.20$. However the weak lensing does find evidence for non-isothermal behaviour of the mass profiles and this is discussed further in Section\,\ref{sec:isothermal}.

\cite{chauhan2021} also tested the \cite{robotham2011} mass estimates against an independent semi-analytic model, SHARK \citep{lagos2018}, showing that the inferred $M_{\rm FoF}$ were in good agreement with the intrinsic halo masses of the model for groups containing $\geq 5$ members. Hence we will adopt $N_{\rm Fof} \geq 5$ for our baseline HMF measurement but also show results for a range of $N_{\rm FoF}$ cuts.

\subsection{$G^3$ mass error estimates}
Individual mass errors for each halo are not provided within the $G^3C$ catalogue, although the weak lensing results do provide an indication of the average mass error. Here we derive an approximate mass error as a function of group multiplicity, by identifying those groups with more then 20 members, and for which the mass estimates should be stable and reliable. This constitutes 154 groups and for each of these groups we gradually remove group members, by removing the system with the faintest flux, and recomputing the group's $\sigma$ and $R_{50}$ values, and via Eqn.\,\ref{eqn:mass} (with $A=13.9$) the halo mass.

Fig.\,\ref{fig:masserr} shows as blue-green lines, a trail for each of the 154 groups (with $N_{\rm FoF} >20$). These trace out the recovered-to-original mass ratio, i.e. $\log_{10}(M_N/M_{\rm max})$, as the multiplicity decreases. The solid red line shows the median of these trails, while the dashed red lines show the limits that enclose 84 (upper line) or 16 (lower line) per cent of the trails. These lines suggest a very small error at high multiplicity, which grows significantly as the multiplicity is reduced. The trails are colour-coded by mass, as labelled, and no obvious trend with mass is seen.

The dashed black horizontal lines represent the global mass uncertainty seen by \cite{viola2015} and are consistent with the mean/median mulitplicity of our sample of $N=7.6/5.0$. The
dotted black lines are a transcription of the error estimate shown in Fig.\,8 from \cite{robotham2011} and also defined by their Eqn.\,20; these imply significantly greater uncertainty. The difference between our estimate and those provided in \cite{robotham2011} is unclear, but we note that our results are consistent with a recent analysis by Meyer (2021; UWA Master's thesis), who explored the robustness of $G^3C$ masses based on extensive numerical simulations. We therefore adopt the mass uncertainty based on the sum in quadrature of our multiplicity implied error and an error floor of $\sigma_{\log_{10}M}=0.10$ (motivated by the work of Meyer et al.) We also implement, for our {\sc MassA} values only, the correction implied by the median (solid red line from Fig.\,~\ref{fig:masserr}), which suggests some bias towards lower masses at very low multiplicities (i.e. $\Delta \log_{10}M \sim 0.1$ at $N{\rm FoF} \sim 6$).

Fig.\,\ref{fig:masserr} raises an interesting question, as given the significance of the mass error, its uncertainty, and the susceptibility to a severe Eddington bias, it is not clear whether a more reliable HMF is built by including low multiplicity systems to maximise the sample size, or by selecting a smaller high-multiplicity high-fidelity sample. To address this we will explore a range of multiplicity cuts when generating the GAMA HMF in Section\,\ref{sec:GAMAHMF} and adopt $N_{\rm FoF} \geq 5$ as our primary cut given the significant increase in uncertainty at $N_{\rm FoF} < 5$.

\subsection{Some digressions}
Before moving on to derive the GAMA HMF, it is worth highlighting a number of issues in the construction of the group catalogue to maximise transparency and motivate future work.

\subsubsection{The mass estimation formula \label{sec:isothermal}}
The functional form shown in Eqn.\,\ref{eqn:mass} resembles that of a singular isothermal sphere, where the mass within a radius $R$ is $M=2\sigma^2 R/G$ (see e.g. \citejap{BT}).
A divergent mass is generally circumvented by using a truncated isothermal sphere model (see e.g. \citejap{brainerd1996}, \citejap{brimioulle2013}), which relies on introducing some limiting radius. This is typically taken as the $r_{200}$ value, which represents the radius within which the average density is $200 \times$ the critical density. In this case, we expect the total mass to scale $\propto R^3$, so that $M\propto \sigma^3$. Any radius can be used in combination with an appropriate [$A$,$R$], pairing; see Eqn.\,\ref{eqn:mass}. But generally some uncertain extrapolation is required from the mass within the half-light radius (which can generally be measured robustly: see \citejap{strigari2007}), to the total mass.

Ultimately, the value of $A$ and the choice of $R$ are critical, and can act to shift the entire HMF to higher or lower masses. The presence of sub-structure can also bias masses high if not accounted for \citep[see e.g.][]{tempel2017,old2018,tucker2020}.

Two lensing studies have looked to verify the $G^3C$ masses \citep{han2015,viola2015}, and find consistency in the mean mass at around $10^{14}$\,M$_{\odot}$, but a shallower trend with $\sigma$ of $M \propto \sigma^2$ (so two thirds of the expected slope). This shallower $\sigma$ dependence would predict no groups/clusters above a mass of $10^{14.7}$\,M$_{\odot}$, which is contrary to observational constraints (e.g. the Coma cluster, \citeauthor{sohn2017} \citeyear{sohn2017}; or El Gordo \citeauthor{elgordo} \citeyear{elgordo}, \citeauthor{elgordo2} \citeyear{elgordo}). It is hard to know how much of a concern this is for the present work, since the slope of the $M-\sigma$ relation is hard to measure accurately given the few extremely massive objects in the GAMA samples used for lensing studies. We can take some reassurance from the fact that the weak lensing masses scale very nearly linearly with total $r$-band group luminosity, and this alternative mass proxy correlates very well with our dynamical estimates.

Nevertheless, it would be desirable to improve the absolute calibration of group mass estimates, preferably in a way that is independent of simulations. A radical proposal by \cite{caldwell2016} is that we should move away from mass comparisons entirely, and compare velocity dispersion distributions instead. This idea clearly has some merit as it dispenses with the need for a scaling parameter, i.e. $A$. This would however require careful consideration of line-of-sight velocities versus 3D velocities, which is non trivial \citep[see][]{elahi2018}.

\subsubsection{The value of the mass calibration scaling value, $A$}
By adopting a constant value of $A=13.9$ for our {\sc MassA} values, we lessen our detailed reliance on simulations for mass calibration. Ultimately this parameter can be probed empirically, via logic that goes back to \cite{zwicky1933} and his attempt to reconcile the measured velocity dispersion of Coma ($\approx 1000$\,km\,s$^{-1}$) with that predicted by its visible mass alone of 80\,km\,s$^{-1}$. Zwicky's derivation of $A$ was based on the virial theorem, coupled with the assumption of an isothermal mass distribution, and where he found $A=1.667$ \citep{zwicky1933}, and that appeared to under-predict the mass content by a factor of 400.

Since \cite{zwicky1933}, our measurements of Coma have improved substantially, including direct measurements of the total dynamical mass. A robust total mass of  $M_{200}=\smash{1.88^{+0.65}_{-0.56}} \times 10^{15}$\,$h^{-1}$\,M$_{\odot}$ within an $r_{200}$ radius of $1.99^{+0.21}_{-0.22}$\,$h^{-1}$\,Mpc has been derived from gravitational lensing \citep{kubo2007}. This mass and size are on the whole consistent with a range of earlier estimates stretching from $0.8$ to $1.9 \times 10^{15}$\,$h^{-1}$\,M$_{\odot}$. As Coma's line-of-sight velocity dispersion is $947 \pm 31$\,km\,s$^{-1}$ \citep{sohn2017}, and the projected half-light radius is R$_{\rm 50}=1.4$\,$h^{-1}$Mpc \citep{doi1995}, this implies a value of $A \approx 6 \pm 3$ which is less than half of our adopted value of $A=13.9$. More precision and clarity over $A$ is clearly required, and hopefully as detailed studies of more clusters emerge one will eventually be able to obtain a fully simulation independent estimate of $A$ and its dependency on halo mass and other variables.

\subsubsection{Comoving distance or proper (angular diameter) distance \label{sec:distances}}

As discussed earlier, we have chosen to base our mass estimate on the comoving sizes of clusters, rather than using the angular diameter distance to derive physical radii. This follows the convention established by \cite{eke2004} and \cite{robotham2011}, but over a wide range of redshifts it might be considered problematic. If the virial relation applied rigorously in proper coordinates, then use of comoving half-light radii would require a scaling $A(z)\propto 1/(1+z)$. This is not a concern for the current work, for a number of reasons.
Firstly, the redshifts probed are relatively local ($z<0.25$), so that there is no real ability to probe evolution, and we are interested only in $A$ at a single effective redshift. Furthermore, the redshift-dependent group selection effects complicate any simple virial relation, meaning that it is better to work empirically rather than enforce a simplistic virial relation.

However, this issue will be important in future deeper studies that probe to significant redshifts. We can anticipate an increase in the frequency of non- or partially- virialised haloes as our surveys advance to higher redshifts, and also as we probe to lower masses where the systems may exhibit more structure (i.e. be less dynamically relaxed). 

Note that by fitting for $A$($z$) one effectively folds in this dependency into $A$, however by opting for a fixed $A$ we have removed this dependency, perhaps reopening the issue as to which distance one should use. As our $R_{50}$ values are inevitably well within the $r_{200}$ radii, and very much within the bound region, the case for using the angular diameter distance starts to looks stronger. Later we will report results from both but continue to adopt the convention defined by \cite{eke2004} and \cite{robotham2011}.


\subsubsection{Linking lengths and overdensities}
Critical to the operation of the group finding friends-of-friends algorithm are the linking-lengths in the spatial and redshift directions \citep[see for example:][]{duarte2014}. These define whether a galaxy is or is not a member of a group. The linking lengths are derived by testing against a mock catalogue, and modifying the lengths until one recovers the known haloes. This is typically done in a bijective manner,  \citep{robotham2011}, i.e. a false positive carries the same penalty as a missing group or cluster, and one looks to find the optimal lengths that minimise both. 

At high mass, linking-length uncertainties are a relatively weak concern, as the interloper density is relatively low. Increasing the linking lengths has a fairly minimal impact at high masses and high multiplicities, although can lead to the merging of nearby groups into super-clusters. 

At low mass the addition of a single bogus member, or loss of a real member, can introduce significant mass errors. Further investigation is needed to explore this dependence in detail, and it is left for future work. 

One possible advancement will be to fold in the expected group profile shape into the friends-of-friends search algorithm, essentially moving towards halo-finding as used in numerical simulations (for a summary of halo finders, see \citejap{knebe2013} and for an initial attempt in this direction see \citeauthor{tempel2018} \citeyear{tempel2018}). In due course it will be important to build complete end-to-end Monte-Carlo simulations that model the reliance that includes the linking length uncertainty.

In all of this, we should bear in mind that there is an ambiguity in defining the 'true' masses. We have calibrated to $M_{200}$, which is the mass within the
$r_{200}$ radius.  These values ultimately depend, by their definition, on the critical density. An alternative convention is to define $r_{200}$ as the radius within which the mean density is 200 times the {\it background\/} value (only 63 times the critical density for our fiducial cosmology). Defining $M_{200}$ via the background density would shift the HMF to higher masses. This shift can be estimated using NFW halo profiles \citep{NFW1997}: for a typical NFW concentration of 5, the mass defined at 200 times the background is 1.4 times the mass at 200 times critical, and this correction factor is rather insensitive to mass. Since $N$-body halo catalogues are computed using algorithms similar to FoF, which scale with the mean particle density, it could be argued that defining $M_{200}$ with respect to the background would be a more consistent approach. But we continue to use the definition with respect to the critical density, for consistency with the literature on gravitational lensing.

\section{The GAMA HMF}
\label{sec:GAMAHMF}
The $G^3C$ catalogue contains mass estimates down to $10^{10}$\,M$_{\odot}$ and extends out to $z \approx 0.6$, however masses below $10^{12}$\,M$_{\odot}$ become increasingly unreliable, due to low multiplicity, irregularity, and the propensity to be impacted more significantly by an interloper. We hence confine ourselves to $z<0.25$ where the number of multiplicity $\geq3$ groups is $N_g = 5246$. As discussed later, we will construct multiple HMFs with various $N_{\rm FoF}$ selections, and show figures in the main text based on the results from $N_{\rm FoF} \geq 5$  but with all results reported in the tables, and all derived HMFs shown in the Appendices. Ultimately $N_{\rm FoF} \geq 5$ represents a trade-off between sample size and fidelity in our measurements, while the redshift limit ensures that any evolution of the HMF will have a negligible impact on our measurements.

\begin{figure*}
	\centering     
	\includegraphics[width=\textwidth]{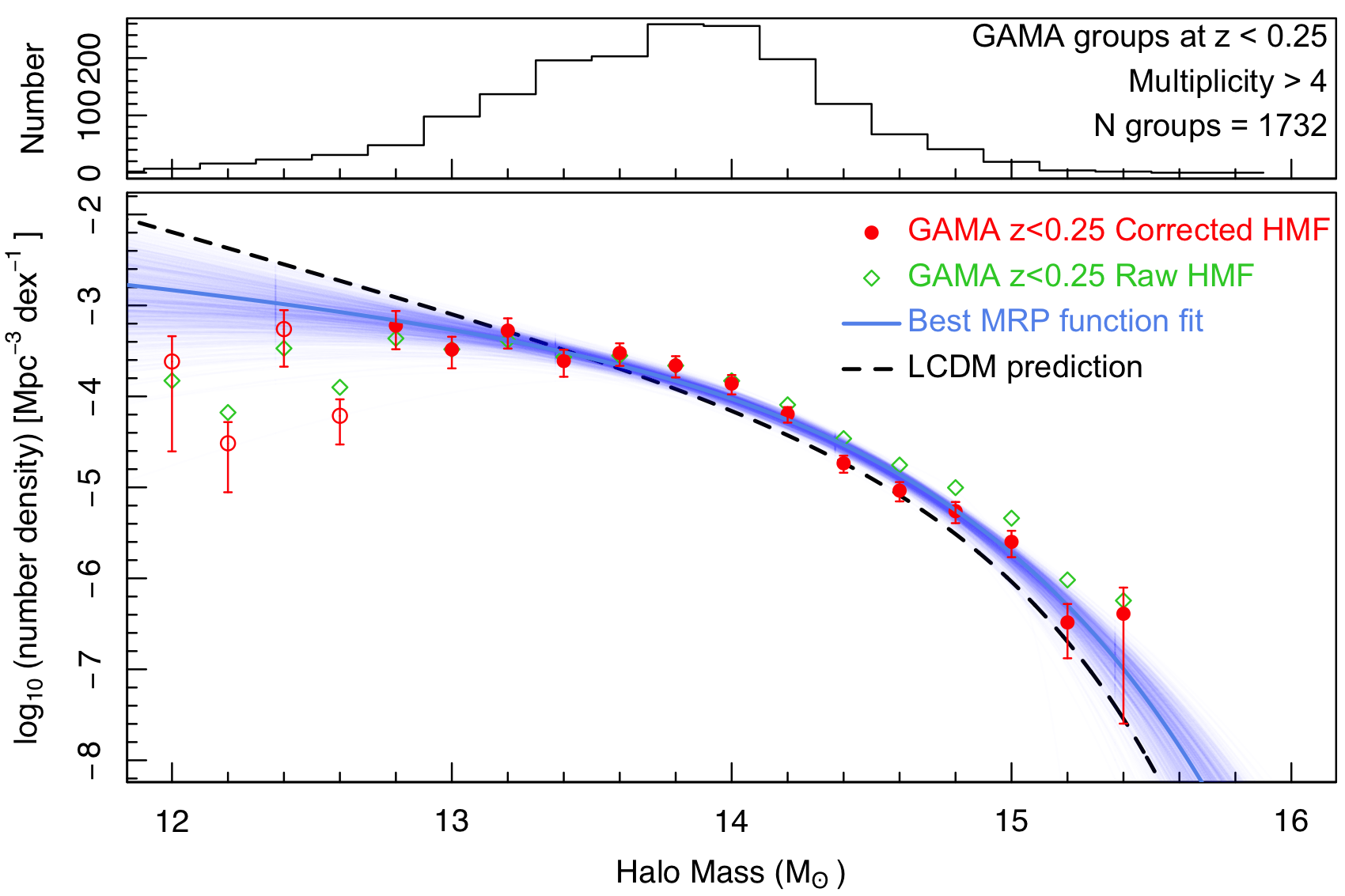}
	\caption{(upper) The original histogram of the number of GAMA groups at $z<0.25$. (main panel) The recovered HMF before (green arrows) and after (red points) the Eddington bias correction. The black dashed line shows the expected MRP function from \citefix{murray2021}, while the blue line shows the best MRP function fit and the fainter blue lines are from the Monte-Carlo realisations. Solid symbols denote where the data are complete and hence to which the MRP function is fitted, and the open symbol the onset of incompleteness.
	The dashed line shows the $\Lambda$CDM prediction for our fiducial cosmology, evaluated at the effective redshift $z_{\rm eff}=0.1$. 
	}
	\label{fig:gamahmf}
\end{figure*}

\begin{figure*}
	\centering     
	\includegraphics[width=\textwidth]{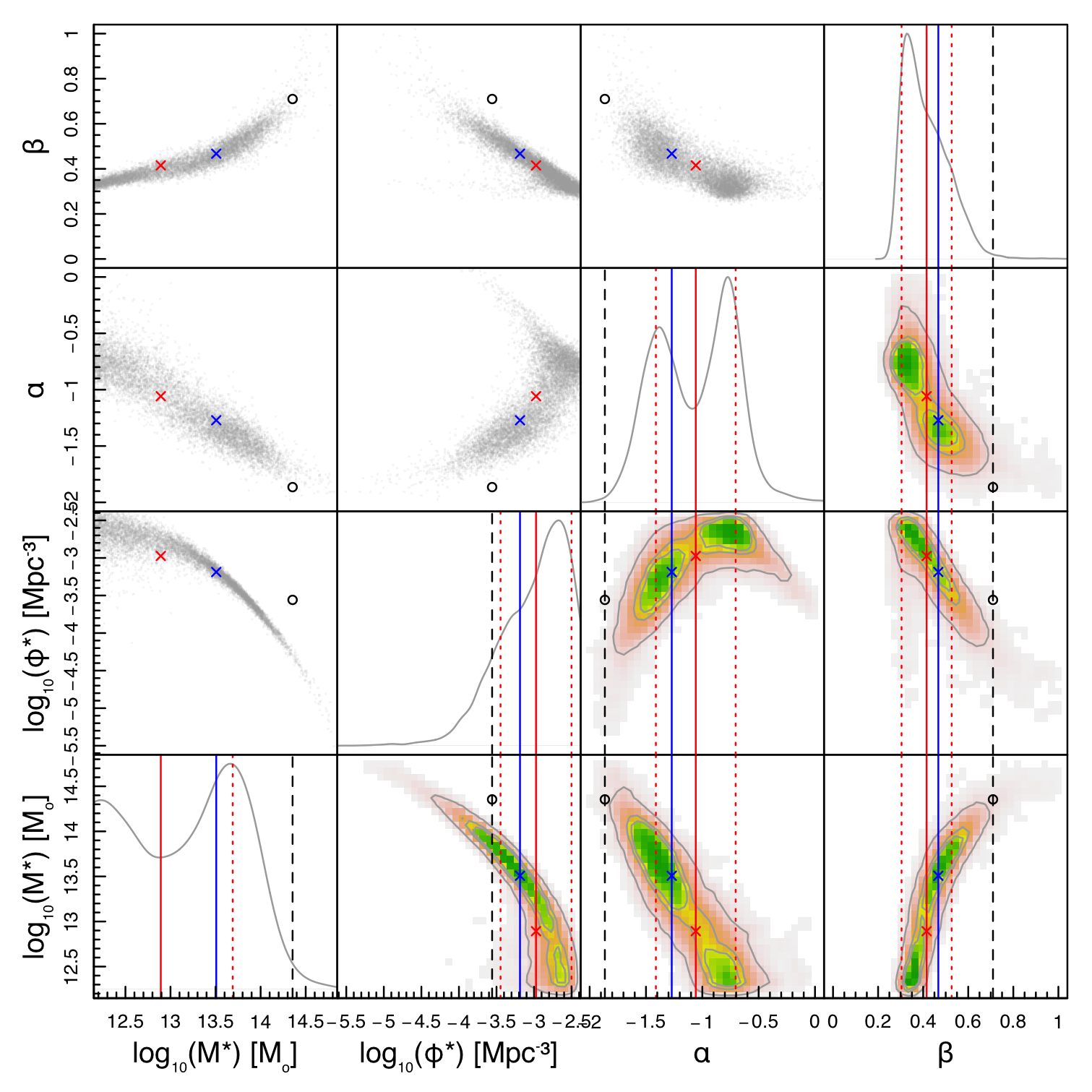}
	\caption{The co-variances of the fitted MRP parameters to the data shown in Fig.\,\ref{fig:gamahmf}. The blue crosses and blue solid lines show the location of the best fit within each panel. The red cross and lines shows the mean of the fitted values, and the red dotted lines the $\pm1\sigma$ error ranges based on the standard deviation of the distribution. Note that due to the correlation and complex shape of the error distribution the simple mean of the fitted data (red crosses) do not always represent viable fits. Also shown are the parameters recommended by MRP adjusted to $\tilde{z}=0.5$ (black circles). The lower right quadrants show the 1, 2 and 3 $\sigma$ error contours (grey contours encircling green, yellow and orange shaded regions).
	\label{fig:gamacov}}
\end{figure*}

For clarity, we first provide a summary of the method we implement, and where Fig.\,\ref{fig:gamahmf} shows the resulting HMF:
\begin{description}
\item[{(1)}] Construct the raw HMF (halo density per log mass interval), using a $1/V_{\rm max}$ estimator, where the $V_{\rm max}$ of the group is based on the limiting $z$ of the $n^{\rm{th}}$ group member
\item[{(2)}] Run a Monte-Carlo of the following ($1001\times$):
\vspace{-0.2cm}
\begin{description}
\item[\hspace{0.15cm} {(i)}] Peturb each group's mass by its mass uncertainty assuming a lognormal error distribution.
\item[\hspace{0.15cm} {(ii)}] Re-derive the HMF using the same $1/V_{\rm max}$ estimates.
\end{description}
\vspace{-0.2cm}
\item[{(3})] Determine the mean of the revised density estimated in each mass interval, and the 16/84 per cent quantiles from the Monte-Carlo simulations.
\item[{(4)}] Determine the multiplicative offset between the original HMF and the median Monte-Carlo HMF for each mass interval.
\item[{(5)}] Reduce the original HMF values by this multiplier to produce the final HMF (thereby correcting for the Eddington bias inherent in the original measurement).
\item[{(6)}] Assign the quantile error in each mass interval from the Monte Carlo simulation as the Eddington bias correction error.
\item[{(7)}] Determine the Poisson and Cosmic Variance errors for the original distribution assuming root-$n$ statiscs and the CV formula from \cite{driver2011}.
\item[{(8)}] Combine the Eddington bias uncertainty and the Poisson error in quadrature to obtain the total error on the final HMF.
\end{description}

In Section\,\ref{sec:gamahmf} we elaborate on this summary, to provide additional details and highlight some of the subtleties, in Section\,\ref{sec:gamafit} we describe our MRP function, our fitting process and fitting errors, and in the remaining sections some further digressions.

\subsection{Determination of the GAMA HMF \label{sec:gamahmf}}
We construct the GAMA HMF via a standard $1/V_{\rm max}$ method, modified to operate on groups, i.e.
\begin{equation}
\phi(\log_{10}[M/M_{\odot}])=\sum\limits_{i=0}^{N_{\rm group}}\left(\frac{w_i}{V_{\rm max}^{i}\Delta\log_{10}M}\right) \label{eqn:vmax},
\end{equation}
where $V_{\rm max}^{i}$ is the maximum comoving volume over which the $i^{\rm{th}}$ group can be detected, $w_i$ is the multiplicative Eddington bias correction for that mass interval (discussed later), $\Delta \log_{10}M$ is the bin width, and $N_{\rm group}$ the total number of groups. The $V_{\rm max}$ values represent the distance at which the $n^{\rm{th}}$ brightest group member is no longer detectable, and where $n$ is the selected multiplicity limit for acceptance as a group. These $V_{\rm max}$ (or rather $z_{\rm max}$) values for each individual galaxy are reported in the {\sc StellarMassesv19} DMU as {\sc z19pt8}. 

In practice, for each group we identify the members of the group, using {\sc G3CGals} within the GroupFinding DMU, and rank them by apparent $r$-band flux. We match this catalogue to the {\sc StellarMassesv19} catalogue, and adopt the redshift limit of the $n^{\rm{th}}$ member. For any group whose redshift exceeds its $n^{\rm{th}}$ galaxy redshift limit, we reset the limit to its current redshift and for any group whose $z_{\rm max}$ exceeds our redshift boundary we reset it to $z_{\rm max}=0.25$. We convert $z_{\rm max}$ to $V_{\rm max}$ values using Planck 2018 Cosmology. 

The raw HMF, $\phi(\log[M/\rm{M}_{\odot}])$ (see Section\,1), is then determined from the sum of the $1/V_{\rm max}$ values within each mass interval (see Eqn.\,\ref{eqn:vmax}) and initially with $w_i=1$, resulting in the green points shown in Fig.\,\ref{fig:gamahmf} (main panel).

Because of the steepness of the HMF, and the significant mass errors, the Eddington bias is expected to be significant. We forward evolve the measured HMF distribution via a set of Monte-Carlo simulations, to estimate the size of this bias ($w_i$). To linear order, $w_i=\phi(M)_1/\phi(M)_{<MC>}$, where $\phi(M)_1$ is our initial measurement, with $w_i=1$, and $\phi(M)_{<MC>}$ is the average of our Monte-Carlo measurements, after perturbing the mass of each group independently. This mass perturbation for each group is given by a lognormal distribution with mean zero and a standard deviation given by the group's $\smash{\sigma_{\log_{10}M}}$ error. Hence we are using a forward propagation method to estimate the severity of the Eddington bias. We then use this empirically derived Eddington bias per mass bin to debias the observed HMF and recover the intrinsic HMF.

\begin{table*}
\caption{The GAMA HMF values derived in this work adopting a multiplicity lower limit of $N_{\rm FoF}=5$ and as plotted in Fig.\,\ref{fig:gamahmf}. Values above the horizontal bar should be considered credible and those below not. Hence Column\,4 rows\,1:16 represent our final GAMA HMF. The errors are given as linear fractions. \label{tab:gamahmf}}
\begin{tabular}{cccccccc} \hline
$\log_{10}(M/M_{\odot})$\fixv & N & $\log_{10}\phi$ & $\log_{10}\phi_{\rm corr}$ & $\sigma_{\rm Poisson}$ & $\sigma_{\rm Monte-Carlo}$ & $\sigma_{\rm CosVar}$ & $\sigma_{\rm Combined}$\\ 
(bin centre) & (linear) & Mpc$^{-3}$ & Mpc$^{-3}$ & (Frac.) & (Frac.) & (Frac.) & (Frac.)\\ \hline  \hline
$15.4$\fixv&$2$&$-6.943$&$-6.389$&$0.71$&$0.62$&$0.07$&$0.94$ \\ 
$15.2$&$4$&$-6.717$&$-6.485$&$0.50$&$0.32$&$0.07$&$0.60$ \\ 
$15.0$&$19$&$-6.038$&$-5.599$&$0.23$&$0.23$&$0.07$&$0.32$ \\ 
$14.8$&$41$&$-5.702$&$-5.261$&$0.16$&$0.21$&$0.07$&$0.26$ \\ 
$14.6$&$67$&$-5.454$&$-5.035$&$0.12$&$0.21$&$0.07$&$0.24$ \\ 
$14.4$&$120$&$-5.163$&$-4.735$&$0.09$&$0.19$&$0.07$&$0.21$ \\ 
$14.2$&$198$&$-4.791$&$-4.195$&$0.07$&$0.18$&$0.07$&$0.19$ \\ 
$14.0$&$256$&$-4.528$&$-3.859$&$0.06$&$0.23$&$0.07$&$0.24$ \\ 
$13.8$&$259$&$-4.360$&$-3.659$&$0.06$&$0.26$&$0.07$&$0.26$ \\ 
$13.6$&$203$&$-4.250$&$-3.523$&$0.07$&$0.27$&$0.07$&$0.28$ \\ 
$13.4$&$196$&$-4.266$&$-3.611$&$0.07$&$0.32$&$0.07$&$0.33$ \\ 
$13.2$&$137$&$-4.082$&$-3.277$&$0.09$&$0.35$&$0.07$&$0.36$ \\ 
$13.0$&$98$&$-4.181$&$-3.484$&$0.10$&$0.37$&$0.07$&$0.38$ \\ 
$12.8$&$48$&$-4.062$&$-3.222$&$0.14$&$0.43$&$0.07$&$0.45$ \\ \hline
$12.6$\fixv&$31$&$-4.599$&$-4.213$&$0.18$&$0.48$&$0.07$&$0.52$ \\ 
$12.4$&$23$&$-4.171$&$-3.260$&$0.21$&$0.58$&$0.07$&$0.61$ \\ 
$12.2$&$16$&$-4.876$&$-4.515$&$0.25$&$0.67$&$0.07$&$0.71$ \\ 
$12.0$&$7$&$-4.526$&$-3.616$&$0.38$&$0.81$&$0.07$&$0.90$ \\ 
$11.8$&$2$&$-6.148$&$-6.667$&$0.71$&$0.90$&$0.07$&$1.00$ \\ 
$11.6$&$2$&$-5.146$&$-4.512$&$0.71$&$0.92$&$0.07$&$1.00$ \\ 
$11.4$&$1$&$-6.290$&$-6.628$&$1.00$&$0.95$&$0.07$&$1.00$ \\ \hline
\end{tabular}
\end{table*}

An advantage of this method is that we can use the mass error for each group, rather than an average mass error. The downside is that the Eddington bias correction is only lowest order in the mass errors, and it may be inaccurate for large biases. This is mitigated by the uniform slope of the HMF, as the Eddington bias should be mass-invariant for any power law distribution, except around the `knee' where the Eddington bias becomes more extreme or incompleteness impedes. The Monte-Carlo is repeated 1001 times to minimise statistical noise. We also note here an interesting subtlety: as detection is based on the $n^{\rm{th}}$ members' flux, yet mass is dependent on the orbital velocity dispersions of the detected members, the Eddington bias does not impact the sample selection but only the masses of the selected systems. Hence we do not need to apply any mass-based Eddington bias to systems below our detection limit. As detection is based on fluxes that are robust to relatively high precision, we do not believe that an additional detection Eddington bias will be significant.

Finally, we estimate the Poisson, Monte-Carlo and cosmic variance errors for each log mass interval. The former is derived from the square root of the number of groups and the Monte-Carlo error from the 16 and 84 per cent quantiles in the $\phi(M)$ values within each mass bin (due to the non-Normal behaviour of the HMF distributions). For the cosmic variance errors, we use Eqn~2 from \cite{driver2011}, to determine a single cosmic variance value based on the total volume surveyed (i.e. $\pm 7$ per cent).

Fig.\,\ref{fig:gamahmf} (main panel, red circles) shows the GAMA HMF derived using the method described above for a multiplicity limit of $n \geq 5$ (see figures in Appendix A for a range of multiplicity cuts). In the upper panel, we show the direct histogram that peaks at $\sim 10^{13.9}$\,M$_{\odot}$ and information on the selection and final number of groups. In the lower panel we show the reconstructed HMF using the $1/V_{\rm max}$ weighting (green arrows), and the final Eddington bias corrected HMF (red data points). The errors include the Poisson and Monte-Carlo errors added in quadrature. The cosmic variance errors are not indicated but are generally less than the Monte-Carlo error (see Table\,\ref{tab:gamahmf} Col 6 v Col 7). 

The depth of our sample means that our estimate of the HMF applies at an effective redshift that is substantially different from zero: a simple estimate of this redshift can be derived from the mean redshift of our group catalogue, which as stated earlier is $z_{\rm eff}=0.153$. In practice, however, we will assume that our measurement applies at the slightly smaller round figure of $z_{\rm eff}=0.1$, and we show theoretical models corresponding to that value. This choice is driven by the desire to compare with other determinations of the HMF from samples that are slightly more local than GAMA (see Section\,\ref{sec:ExternalEstimates}). We did consider adjusting the different measurements to allow for slightly different effective redshifts in each case, but the required corrections would be small and within the noise. The evolution of the HMF can be approximated by a shift to smaller masses at higher $z$ and a mass-conserving increase in comoving number density by the same factor. For our fiducial cosmology, the shift between $z=0$ and $z=0.1$ is 0.08 dex, and the shift between $z=0.1$ and $z=0.153$ is 0.04 dex.

\subsection{Fitting the GAMA HMF with the MRP function \label{sec:gamafit}}
In Fig.\,\ref{fig:gamahmf} the dotted black line shows the recommended MRP function as defined in \cite{murray2021} but adjusted to our median redshift of $\tilde{z} \approx 0.1$. This adjustment of $-0.075$\,dex in $(M/M_{\odot})$ and +0.075\,dex in $\phi(\log_{10}(M/M_{\odot}))$ is based on the predicted evolution of the HMF from $z=0.1$ to $z=0.0$ (cf. Fig.\,\ref{fig:cdmpredictions}). The definition of the MRP function is repeated here for clarity, and in logarithmic mass intervals, as:
\begin{equation*}
\phi(\log_{10}(M/M_{\odot}))  \equiv \frac{dn}{d(\log_{10}(M/\rm{M}_{\odot}))}
\end{equation*}
\begin{equation}
=\ln(10)\phi_*\beta\,\left(\frac{M}{M_*}\right)^{\alpha+1}\exp\left[-\left(\frac{M}{M_*}\right)^\beta\right] \label{eqn:mrp},
\end{equation}
where $\phi_*$ is the space-density at the characteristic mass point that acts as the vertical normalisation, $M_*$ is the characteristic halo mass that acts as the horizontal normalisation, $\alpha$ is the low mass slope parameter, and $\beta$ the high mass exponential softening parameter \citep[see also][]{trevisan2017}.
This can also be expressed in linear mass intervals as:
\begin{equation}
\phi(M)\equiv \frac{dn}{dM}=\frac{\phi_*\beta}{M_*}\left(\frac{M}{M_*}\right)^{\alpha}\exp\left[-\left(\frac{M}{M_*}\right)^{\beta}\right] 
\end{equation}
Following from this the total mass density, i.e. the integral of the MRP to zero mass, can be given in terms of the complete $\Gamma$ function, by:
\begin{equation}
\rho_M = M_*\phi_* \Gamma( (2+\alpha)/\beta ) \label{eqn:mrp2}
\end{equation}
Note that from Eqn.\,\ref{eqn:mrp2} values of $\alpha > -2$ are required for a finite value for $\rho$. 

We fit the MRP function to our data by minimising the $\chi^2$ (see Eqn.\,\ref{eqn:fit}), with the addition of an appropriate factor to account for the empty high mass bins (in which groups are visible across the full survey volume should they exist), i.e.
\begin{equation}
\sum^{i=n}_{i=1}\left(\frac{\log_{10}(\phi_i)-\log_{10}(\bar{\phi_i})}{\frac{\sqrt{\Delta \phi_{i,{\rm Ran}}^2+\Delta \phi_{i,{\rm MC}}^2}}{(\ln(10)\phi_i)}}\right)^2+2\sum^{i=n+10}_{i=n+1}\bar{\phi_i}V_{\rm limit}\Delta\log_{10} M\label{eqn:fit}
\end{equation}
here $\phi_i$ represents our binned measurements (Table\,\ref{tab:gamahmf}, Col.\,4), $\bar{\phi_i}$ the expectation from Eqn.\,\ref{eqn:mrp}, and $\sigma_i$ the data uncertainty of the $i^{\rm{th}}$ bin (Table\,\ref{tab:gamahmf}, Col.\,8). For the penalty factor we calculate $\bar{\phi_i}$ for the first ten unoccupied bins (as one increases in halo mass from the highest detected group), where $V_{\rm limit}$ is the maximum volume surveyed, and $\Delta \log_{10} M$ the bin width. The penalty factor derives from the exact Poisson probability of finding an empty bin when the expected number of system is non-zero, and it ensures the fit does not over-predict at the high mass end where no groups were detected.

The data and errors, are shown in Fig.\,\ref{fig:gamahmf} and reported in Table\,\ref{tab:gamahmf}. Here the errors are derived in a posterior fashion from either $\sqrt(n)$ statistics ($\sigma_{\rm Poisson}$) where $n$ is the number of groups in the bin, or from the Monte-Carlo refitting ($\sigma_{\rm Monte-Carlo}$). The errors shown in Fig.\,\ref{fig:gamahmf} are the $\sigma_{\rm Combined}$ errors (see Table\,\ref{tab:gamahmf}, Col.\,8) which are the sum in quadrature of $\sigma_{\rm Poisson}$ and $\sigma_{\rm Monte-Carlo}$ only. The best-fit MRP values to these data, and for our other selections discussed latter, are shown in Table\,\ref{tab:mrpfits}. Fig.\,\ref{fig:gamacov} highlights the co-variances of the fitted parameters (as indicated).

\begin{table*}
\renewcommand{\arraystretch}{1.6}
\caption{MRP function fits to various HMF datasets (as indicated) and where, for example, `GAMA5' refers to the GAMA-only sample with a multiplicity limit of 5. Note the fits are only valid over the range of the data, i.e. $10^{12.7} $M$_{\odot} < M_h < 10^{15.5}$M$_{\odot}$ except for the final row, `Omega', where the requirement to converge to the Planck 2018 $\Omega_M$ value is included. Note that the fits are for an effective redshift of $\tilde{z} \approx 0.1$  and to convert to $z = 0$ one should add 0.075\,dex to $\log_{10}(M_*/M_{\odot})$ and subtract 0.075\,dex from $\log_{10}(\phi*)$}.\label{tab:mrpfits}
\begin{tabular}{c|c|c|c|c} \hline
Sample\fixv & $\log_{10}(M_*/M_{\odot})$ & $\log_{10}(\phi_*)$ & $\alpha$ & $\beta$ \\ 
       &                            &  Mpc$^{-3}$dex$^{-1}$ & & \\ \hline \hline
GAMA3\fixv & $11.65_{-0.45}^{+1.37}$& $-1.87_{-0.78}^{+0.03}$& $-0.78_{-0.44}^{+0.24}$& $0.31_{-0.02}^{+0.12}$ \\ 
GAMA4 & $10.44_{-1.24}^{+3.22}$& $-2.49_{-0.83}^{+0.17}$& $-0.47_{-1.03}^{+0.25}$& $0.24_{-0.05}^{+0.22}$ \\ 
GAMA5 & $13.51_{-1.51}^{+0.26}$& $-3.19_{-0.28}^{+0.67}$& $-1.27_{-0.18}^{+0.57}$& $0.47_{-0.15}^{+0.05}$ \\ 
GAMA6 & $14.42_{-0.84}^{+0.28}$& $-4.32_{-0.54}^{+1.03}$& $-1.75_{-0.10}^{+0.35}$& $0.57_{-0.17}^{+0.18}$ \\ 
GAMA7 & $14.51_{-0.59}^{+0.42}$& $-4.45_{-0.83}^{+0.72}$& $-1.68_{-0.16}^{+0.29}$& $0.60_{-0.14}^{+0.40}$ \\ 
GAMA8 & $14.88_{-0.81}^{+0.07}$& $-5.11_{-0.21}^{+1.20}$& $-1.77_{-0.04}^{+0.41}$& $0.88_{-0.38}^{+0.22}$ \\ 
OrigErr5 & $13.03_{-0.42}^{+0.99}$& $-2.77_{-1.16}^{+0.04}$& $-1.12_{-0.66}^{+0.29}$& $0.45_{-0.04}^{+0.19}$ \\ 
AngDiam5 & $13.52_{-0.34}^{+0.47}$& $-3.22_{-0.61}^{+0.30}$& $-1.60_{-0.19}^{+0.21}$& $0.40_{-0.04}^{+0.08}$ \\ 
MassA5 & $13.14_{-0.54}^{+0.71}$& $-2.98_{-0.80}^{+0.26}$& $-1.45_{-0.27}^{+0.32}$& $0.37_{-0.04}^{+0.10}$ \\ 
MassAfunc5 & $13.68_{-1.62}^{+0.21}$& $-3.46_{-0.30}^{+0.84}$& $-1.35_{-0.21}^{+0.74}$& $0.59_{-0.24}^{+0.04}$ \\ 
MassWL5 & $12.50_{-0.52}^{+0.25}$& $-3.11_{-0.78}^{+0.14}$& $0.07_{-0.30}^{+0.72}$& $0.54_{-0.07}^{+0.06}$ \\ 
SDSS3 & $13.14_{-0.86}^{+0.58}$& $-2.76_{-0.78}^{+0.45}$& $-1.37_{-0.36}^{+0.54}$& $0.51_{-0.11}^{+0.15}$ \\ 
SDSS4 & $13.38_{-0.09}^{+0.52}$& $-3.00_{-0.78}^{+0.09}$& $-1.59_{-0.29}^{+0.09}$& $0.48_{-0.01}^{+0.12}$ \\ 
SDSS5 & $13.38_{-0.09}^{+0.62}$& $-3.00_{-0.90}^{+0.07}$& $-1.57_{-0.31}^{+0.12}$& $0.47_{-0.00}^{+0.14}$ \\ 
SDSS6 & $14.57_{-0.47}^{+0.14}$& $-4.98_{-0.32}^{+0.90}$& $-2.13_{-0.09}^{+0.27}$& $0.80_{-0.19}^{+0.07}$ \\ 
SDSS7 & $14.29_{-0.78}^{+0.13}$& $-4.37_{-0.26}^{+1.04}$& $-1.78_{-0.10}^{+0.48}$& $0.73_{-0.22}^{+0.08}$ \\ 
SDSS8 & $13.82_{-0.06}^{+0.70}$& $-3.70_{-1.14}^{+0.02}$& $-1.45_{-0.38}^{+0.13}$& $0.54_{--0.01}^{+0.33}$ \\ 
REFLEX\,II (R) & $14.33_{-0.38}^{+0.26}$& $-4.30_{-0.53}^{+0.52}$& $-1.62_{-0.27}^{+0.35}$& $0.79_{-0.17}^{+0.21}$ \\ 
GAMA5+SDSS5 (GS) & $14.35_{-0.60}^{+0.36}$& $-4.38_{-0.73}^{+0.93}$& $-1.96_{-0.18}^{+0.37}$& $0.60_{-0.10}^{+0.15}$ \\ 
GAMA5+REFLEX\,II (GR) & $13.72_{-0.43}^{+0.64}$& $-3.44_{-0.93}^{+0.26}$& $-1.29_{-0.41}^{+0.32}$& $0.55_{-0.08}^{+0.26}$ \\ 
SDSS5+REFLEX\,II (SR) & $14.44_{-0.38}^{+0.25}$& $-4.52_{-0.58}^{+0.66}$& $-1.85_{-0.18}^{+0.21}$& $0.79_{-0.18}^{+0.23}$ \\ 
GAMA5+SDSS5+REFLEX\,II (GSR) & $14.13_{-0.40}^{+0.43}$& $-3.96_{-0.82}^{+0.55}$& $-1.68_{-0.24}^{+0.21}$& $0.63_{-0.11}^{+0.25}$ \\ 
FIXED-$\alpha$ (FIX) & $14.47_{-0.15}^{+0.13}$& $-4.57_{-0.26}^{+0.27}$& $-1.86_{-0.00}^{+0.00}$& $0.80_{-0.12}^{+0.14}$ \\ 
Omega & $14.43_{-0.15}^{+0.11}$& $-4.49_{-0.24}^{+0.29}$& $-1.85_{-0.03}^{+0.03}$& $0.77_{-0.11}^{+0.11}$ \\ \hline
\end{tabular}
\end{table*}

From Fig.\,\ref{fig:gamahmf} (main panel) we see that the recovered HMF peaks in the mass bin centred at $M \sim 10^{12.9}$\,M$_{\odot}$, and adopt $M=10^{12.8}$\,M$_{\odot}$ as our mass completeness limit but see also the discussion below in Section\,\ref{sec:detect}.
Note that while $1/V_{\rm max}$ corrects for the diminishing volume, the method only works up to the point at which all sub-classes within the mass range are sampled, i.e. one cannot correct for what one does not detect. Eventually as groups with large flux gaps between their brightest and $n^{\rm{th}}$ brightest member become undetectable, the mass function gradually becomes incomplete. Here we will take the empirical approach of only fitting to our mass function up to the point at which this turn down becomes apparent, and this is indicated by the solid (complete bins) and open (incomplete bins) symbols in Fig.~\ref{fig:gamahmf} (main panel). In due course this incompleteness needs to be modelled, but once again ties the analysis tightly to simulations which comes with its own issues (see Section\,\ref{sec:detect}).

Overlaid on the main panel of Fig.\,\ref{fig:gamahmf} is the predicted HMF from \cite{murray2021} (dashed black line), also using the Planck 2018 Cosmology. The data agree within the errors with the HMF prediction, with a similar but slightly shallower low mass slope and comparable abundance at the characteristic mass. This is in part due to the size of the errorbars, which are very much dominated by the Monte-Carlo error (see Columns 5, 6 \& 7 in Table\,\ref{tab:gamahmf}). Typically the average halo mass error in the $N_{\rm FoF} \geq 5$ sample, is $\sigma_{\log_{10}M} \sim 0.25$, i.e. a factor of 2.1, and essentially reflects the significant mass uncertainties acting through the Eddington bias.

The solid blue line shows the optimal MRP functional fit to the GAMA data, while the fainter blue lines show a fraction of the Monte-Carlo refits. Note that we Monte-Carlo by jostling each individual data point by its allowed random error ($\sigma_{\rm Combined}$), and additionally the entire dataset systematically by its cosmic variance error ($\sigma_{\rm CosVar}$). We then refit again using Eqn.\,\ref{eqn:fit}. The Monte-Carlo MRP fits exhibit a range of curves that encompass the MRP expectation from $\Lambda$CDM. The range of the Monte-Carlo MRP fitted values are represented in Fig.\,\ref{fig:gamacov}, which shows the covariance of the fitted parameters. Clearly apparent is the tight correlation between $M_*$ and $\phi_*$, but we also note the trade-off between $M_{*}$ and the $\beta$ parameter. At face value the range of the individual errors is quite broad, suggesting that the fit is fairly poor; but this belies the significant covariance of the fitted parameters, which generally form fairly narrow distributions. The black circle shows the $\Lambda$CDM expectation from \cite{murray2021}, while the blue crosses our best fit value. The red cross shows the mean of the individual parameter distributions, which do not necessarily combine to provide a sensible fit. The dotted red lines show the standard deviation of each dataset which do not necessarily align with the 1, 2 and 3 $\sigma$ error contours (grey lins and colour shading).

Hence, we can conclude, that while the GAMA HMF data show apparent consistency with expectation, formally the errors do suggest some very mild tension. This is most easily seen in the apparent excess of haloes in the intermediate mass range in Fig.\,\ref{fig:gamahmf}, generally demanding a higher normalisation. This tension could potentially be released by modifying (i.e. fitting for) the mass calibration parameter $A$. However, our preference is to keep this parameter fixed for now, and in due course replace it with an empirically derived value in future analysis. In the meantime we look to improve our constraint by incorporating literature datasets, in which different fundamental mass calibrations are adopted, and hence averaging over a range of mass assumptions/calibrations. However, before folding in these external datasets, we first discuss some of the systematics at play.

\begin{figure}
	\centering     
	\includegraphics[width=\columnwidth]{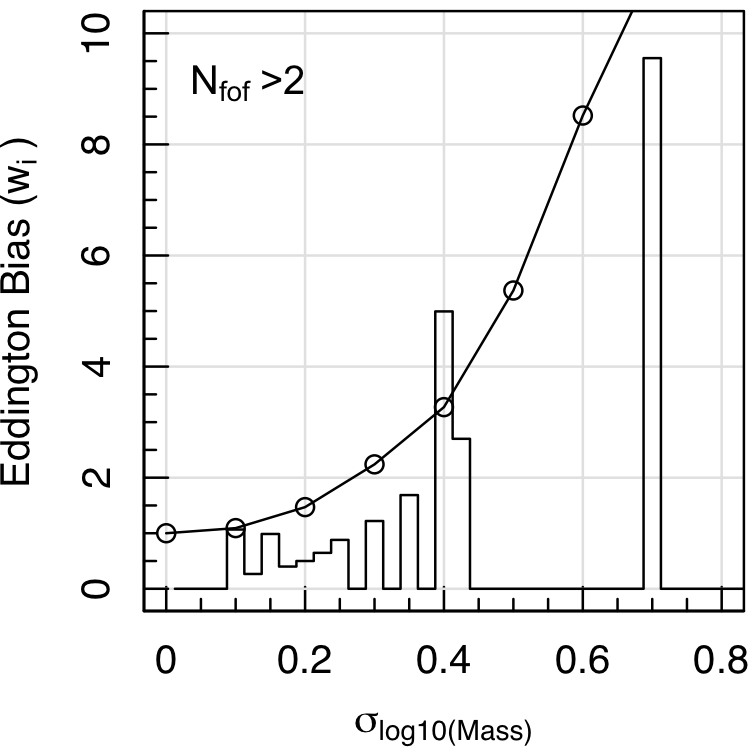}
	\caption{A demonstration of the severe impact of the mass uncertainty ($\sigma_{\log_{10}M}$) on the order of magnitude of the mean Eddington bias correction (circles and connecting lines).  Also shown, against arbitrary units, is the histogram of the mass error distribution for the GAMA groups used to derive our HMF.}
	\label{fig:edb}
\end{figure}

\begin{figure}
	\includegraphics[width=\columnwidth]{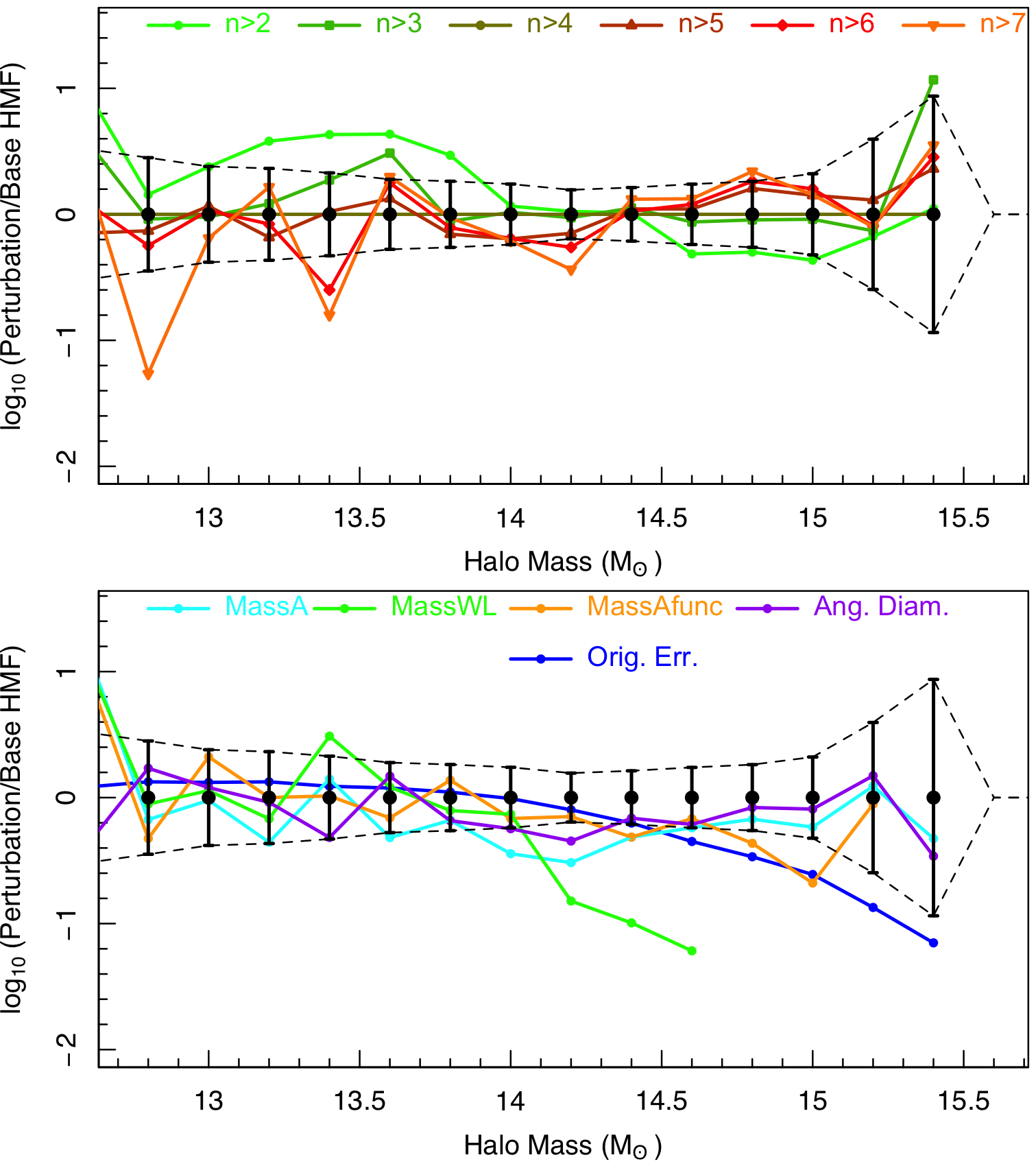}
	\caption{(both panels) The variation of the GAMA HMF based on different underlying choices as indicated in the labels. The black data points show the base GAMA HMF with associated errors and which is taken as the $N_{\rm FoF} \geq 5$ case. The upper panel shows variation based on multiplicity cut while the lower panel different choices in the analysis process. Generally in both panels the variation is enclosed within the errors with the exception of the weak lensing masses (lower panel, green).
	\label{fig:perturb}}
\end{figure}

\begin{figure}
	\includegraphics[width=\columnwidth]{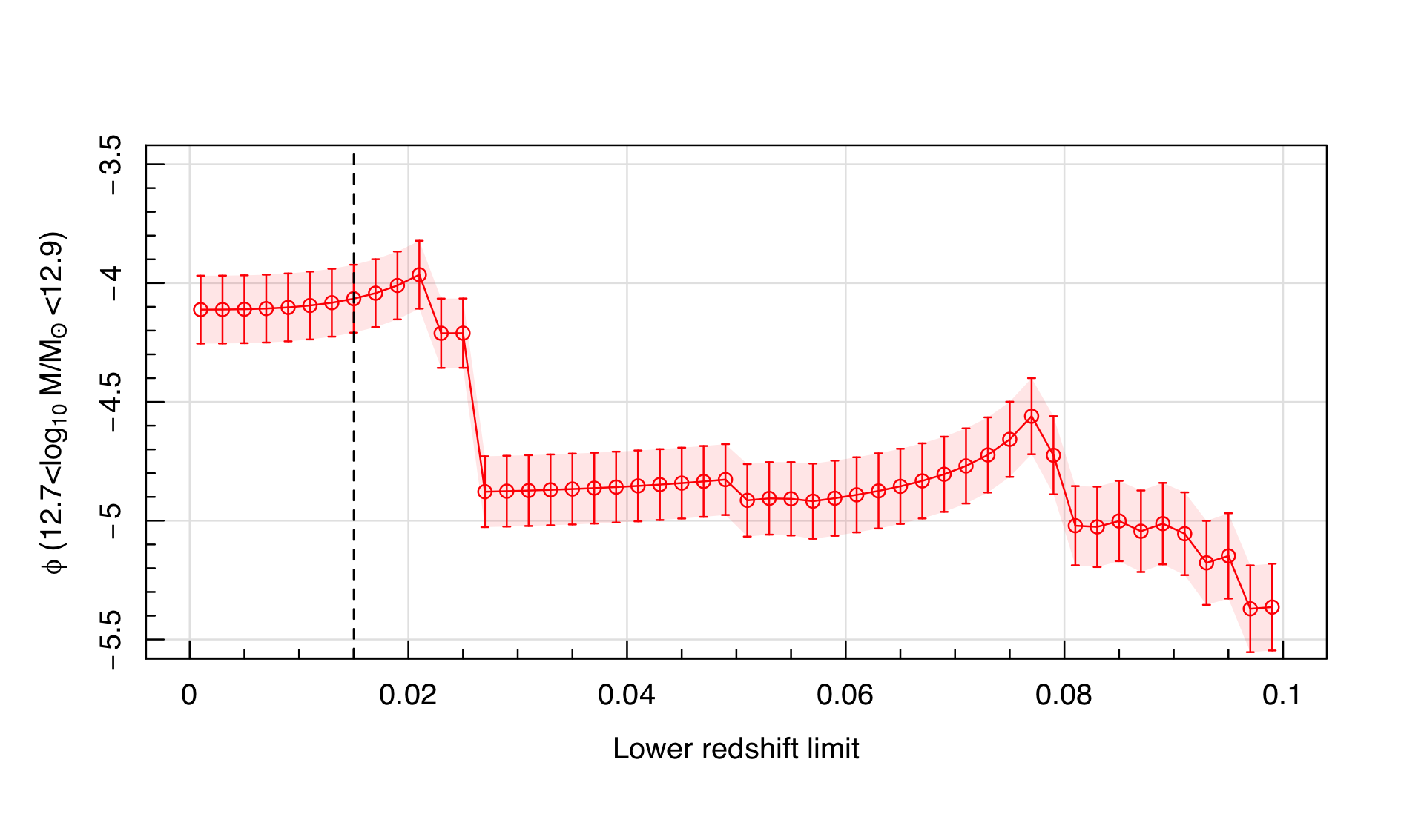}
	\caption{The dependence of the density of haloes in the range $10^{12.7}$M$_{\odot} -- 10^{12.9}$M$_{\odot}$ on the lower limiting redshift used for the deriviation of the GAMA HMF.
	\label{fig:detect}}
\end{figure}

\subsection{Some continued digressions}

\subsubsection{Severity of the Eddington bias}
Our measured HMF has a strong dependency on the Eddington bias correction. To emphasise this we show, as the black line with circles in Fig.\,\ref{fig:edb}, the global value for $w_i$ (vertical axis) if we were to adopt the constant mass error indicated on the horizontal axis. Hence for a {\it constant} mass error of $\sigma_{\log_{10}M}=0.5$ we find an amplification, or multiplication of the derived HMF $\phi$ values of a factor ($w_i$) of $\sim 5.5$. To some extent this is acceptable if known and corrected for appropriately, but this approach fundamentally relies on the precision to which the mass errors are known. Ultimately the smaller the mass uncertainty, the smaller is the required Eddington bias correction, and the more robust the HMF. The histogram in Fig.\,\ref{fig:edb} shows the actual distribution of the mass errors for the full sample (i.e. $N_{\rm FoF} \geq 3$), with an arbitrary vertical scaling. 
Ultimately to keep the Eddington bias manageable (i.e. $w_i < 1.25 \times$) would require mass errors below $\sigma_{\log_{10}M} \sim 0.1$. The difficulty in achieving this may well prove to be an fundamental obstacle in advancing our empirical measurements of the HMF. However we once again it is not necessarily the size of the Edidngton bias that matters most but our ability to correct for it robustly.

\subsubsection{The impact of a larger mass error}
As noted in Fig.\,\ref{fig:masserr} our adopted errors, while consistent with weak lensing \citep{viola2015} and the studies of \cite{old2014}, are a good deal smaller than those originally advocated by \cite{robotham2011}. Here we rerun our HMF calculation again with a multiplicity limit of 5, but this time using the original errors. We show the perturbations of the recovered HMF relative to the original as a fraction in Fig.\,\ref{fig:perturb} (lower panel, dark blue line). In general the agreement is good (i.e. within the errorbars), except at high masses where the larger error gives rise to a systematic bias that exceeds the Monte-Carlo errors. We believe, for the reasons outlined earlier, that the original mass errors were overestimated. Note that Fig.\,\ref{fig:appendix2} (centre right) shows the derived HMF using the original mass errors.

\subsubsection{The impact of alternative mass estimates}
Fig.\,\ref{fig:perturb} (lower panel) also shows the perturbations from a recalculation of our HMF for $N_{\rm FoF}=5$ but now using {\sc MassA} with $A=10$ (cyan line), {\sc MassAfunc} (orange line), and the masses derived via weak lensing from eqn.\,38 of \cite{viola2015} (green). As can be seen the data move fairly minimally for {\sc MassA} (essentially the scaling one would expected from moving from A=10.0 to A=13.9), and {\sc MassAfunc} (barely noticeable), but significantly with comparison to the weak lensing masses and in particular the high mass end is dramatically impacted, with the green curve falling off the plot at $10^{14.5}$\,M$_{\odot}$. It is hence possible that weak lensing may be less sensitive to higher-mass systems, possibly because of their relatively low space density. In this case the impact of trying to use the weak lensing mass predictions becomes catastrophic at the high mass end.

\subsubsection{Multiplicity cuts}
In Fig.\,\ref{fig:perturb} (upper panel) we show the perturbations to our mass function if we adopt a range of multiplicity cuts from $>2$ to $7$ (as indicated). In each case the sample size is significantly reduced but so too are the mass errors. Above our completeness limits we see this impact as a systematic bias towards higher abundances at higher masses but generally the data lie within the quoted errors, which is reassuring.
We also note that for very low multiplicity cuts we start to see an excess at lower halo masses that might suggest either contamination or incompleteness starting to impact high multiplicity cuts (or both).

\subsubsection{The onset of incompleteness \label{sec:detect}}
At some point the HMF turns down, despite the $1/V_{\rm max}$ correction. This will typically occur when subsets within a bin are no longer represented at all. For example, groups where the $5^{th}$ brightest member is fainter than our detection threshold at our lower redshift limit. An obvious solution is to determine incompleteness via comparison to simulations. The key issue here is that the precise flux distribution of intrinsically faint galaxies within a halo is poorly constrained within simulations. Essentially, this pathway requires the simulations to be fundamentally correct at a level at or below the mass resolution of the underlying numerical simulation. This is perhaps best illustrated by considering a Milky Way halo, where the stellar mass of the $5^{\rm{th}}$ most massive member is about $1000\times$ below that of the Milky Way, or $\sim 10^{9}$M$_{\odot}$. At this level the exact prescription adopted for dynamical friction, which dictates merger timescales etc., becomes important. Hence by calibrating to a specific mock, one ties one's detection limit, or incompleteness correction, to a specific semi-analytic prescription that may or may not match reality. 

The alternative is to find an empirical pathway. Here we propose that incompleteness becomes more apparent as one imposes arbitrarily more stringent selection limits, e.g. by increasing our lower redshift limit cut, or by increasing our multiplicity constraint. Fig.\,\ref{fig:perturb} (upper panel) shows how each of the density bins varies with multiplicity cut. Here we see that in the lowest bin used in our fitting, i.e. $10^{12.7} < \log_{10})M/M_{\odot} < 10^{12.9}$ is relatively stable, within the quoted error, suggesting resilience to incompleteness. Similarly Fig.\,\ref{fig:detect} shows how the density in the same halo mass bin varies as we increase the minimum redshift from 0.0 to 0.1. Once again one can see that as we increase the minimum redshift limit, the density remains stable until a limit of 0.02 at which point the density drops dramatically, and we interpret as the onset of irrecoverable incompleteness.

While the empirical method has the advantage of not requiring any knowledge of the intrinsic properties of the groups {\it a priori\/}, its weakness is it cannot correct for entirely dark haloes in which no gas has been processed into stars. In this sense our HMF is therefore a measure of the `astrophysically active' haloes only (i.e. haloes that are purely plasma filled and where no star-formation or gas cooling has occured). As the fraction of `astrophysically inactive' haloes is predicted to increase with decreasing halo mass, one might expect that at some point our empirically measured HMF will under-predict the HMF advocated by simulations.

\subsubsection{The impact of switching to the angular diameter distance}
Fig.\,\ref{fig:perturb} (lower panel, purple line) shows the perturbations when defining group size using the angular diameter distance rather than the comoving distance (as discussed in Section\,2.1.2). At the redshifts we are considering this effect is fairly minimal and remains within the errors.

For all of the variant HMFs we report the fitted MRP HMF values in Table\,\ref{tab:mrpfits} and show the figures in Appendix~A. Overall these results are reassuring in that the errors from our favoured GAMA HMF fit generally encompass the range of variation we see through the different choices. The principal exception is at very high masses where the use of the original errors or weak lensing mass constraints would lead to a dramatically different and more sharply truncated HMF.  We take GAMA5 (i.e. the GAMA sample with $N_{\rm FoF} \geq 5$) as being our optimal parameter set, and note the tendency for slightly higher multiplicity cuts to favour higher-$M*$ and lower-$\phi*$ but also note the strong degeneracy of these two parameters. 

\section{Including external halo mass function constraints}
\label{sec:ExternalEstimates}
Despite the importance of the HMF, there have been relatively few attempts at direct empirical measurement, although we note the efforts in measuring the velocity dispersion distributions. This is primarily due to the difficulty and complexity in constructing group catalogues that typically require large scale spectroscopic programs or wide area X-ray observations. Most notable are the HMFs derived by the two-degree field galaxy redshift survey team (2PIGG; \citeauthor{eke2006} \citeyear{eke2006}), and that derived from X-ray measurements with ROSAT (e.g. REFLEX\,II; \citeauthor{bohringer2017} \citeyear{bohringer2017}). In addition, attempts have been made to study the stellar-mass halo relation using the SDSS group catalogue. Here we attempt to briefly review and combine these external datasets to improve our estimate of the MRP parameters.

\subsection{2PIGG}
The two-degree field galaxy redshift survey team (2dFGRS; \citeauthor{colless2001} \citeyear{colless2001}), measured the distances to $\sim 250\,000$ galaxies at greater than 80 per cent spectroscopic completeness. From these data a group catalogue was constructed by \cite{eke2004} using a Percolation Inferred algorithm. This is in essence very similar to the friends-of-friends algorithm employed by the GAMA team \citep{robotham2011}, with linking-lengths calibrated to numerical simulations. 

The 2PIGG catalogue resulted in $\sim$ 29\,000 pairs or groups, and over 7\,000 with 4 or more members, a median redshift of $\sim 0.11$, and a median velocity dispersion of $\sigma \sim 260$\,km\,s$^{-1}$. In \cite{eke2006}, the team published the 2dFGRS HMF and found good agreement with the $\Lambda$CDM expectation, given the associated errors, over the mass range of $10^{13.5}$\,M$_{\odot}$ to $10^{15.25}$\,M$_{\odot}$.

The 2PIGG HMF data were derived for a cosmology with $\Omega_M=0.3$ and $\Omega_\Lambda=0.7$ and with $H_0=100$\,km\,s$^{-1}$. Here we use the reported 2PIGG HMF values \citep{eke2006}, and modify the group masses and number densities by $h^{-1}$ and $h^3$ respectively, but do not attempt to correct for the small shift from the 2PIGG native cosmology to Planck 2018 cosmology. We note in our final analysis we will not use 2PIGG results, but do include them in our figures for completeness. In Figs.\,\ref{fig:allhmf}\,\&\,\ref{fig:omegacrit}, the 2PIGG data are shown as grey data points and look to be consistent with the other datasets albeit with a slightly flatter slope.

\begin{table*}
\caption{The values derived in this work for the SDSS group catalogue of \citefix{tempel2017} and plotted in Fig.\,\ref{fig:allhmf} (as purple data points). As for Table\,\ref{tab:gamahmf}, Column\,4 rows\,1:16 represent our derived SDSS HMF. The errors are given as linear fractions. \label{tab:sdsshmf}}
\begin{tabular}{crcccccc} \hline
$\log_{10}(M/M_{\odot})$\fixv & N & $\log_{10}\phi$ & $\log_{10}\phi_{\rm corr}$ & $\sigma_{\rm Poisson}$ & $\sigma_{\rm Monte-Carlo}$ & $\sigma_{\rm CosVar}$ & $\sigma_{\rm Combined}$\\ 
(bin centre) & (linear) & Mpc$^{-3}$dex$^{-1}$ & Mpc$^{-3}$dex$^{-1}$ & (Frac.) & (Frac.) & (Frac.) & (Frac.)\\ \hline  \hline
$15.05$\fixv&$4$&$-6.889$&$-5.989$&$0.50$&$0.41$&$0.07$&$0.64$ \\ 
$14.95$&$5$&$-6.792$&$-6.094$&$0.45$&$0.37$&$0.07$&$0.58$ \\ 
$14.85$&$11$&$-6.450$&$-5.676$&$0.30$&$0.28$&$0.07$&$0.41$ \\ 
$14.75$&$14$&$-6.345$&$-5.672$&$0.27$&$0.24$&$0.07$&$0.36$ \\ 
$14.65$&$26$&$-6.055$&$-5.282$&$0.20$&$0.24$&$0.07$&$0.31$ \\ 
$14.55$&$25$&$-6.093$&$-5.535$&$0.20$&$0.20$&$0.07$&$0.29$ \\ 
$14.45$&$60$&$-5.708$&$-4.952$&$0.13$&$0.18$&$0.07$&$0.22$ \\ 
$14.35$&$71$&$-5.638$&$-4.970$&$0.12$&$0.18$&$0.07$&$0.21$ \\ 
$14.25$&$102$&$-5.460$&$-4.768$&$0.10$&$0.17$&$0.07$&$0.20$ \\ 
$14.15$&$152$&$-5.246$&$-4.471$&$0.08$&$0.16$&$0.07$&$0.18$ \\ 
$14.05$&$183$&$-5.057$&$-4.210$&$0.07$&$0.16$&$0.07$&$0.18$ \\ 
$13.95$&$240$&$-4.985$&$-4.173$&$0.06$&$0.16$&$0.07$&$0.17$ \\ 
$13.85$&$302$&$-4.755$&$-3.798$&$0.06$&$0.15$&$0.07$&$0.16$ \\ 
$13.75$&$332$&$-4.650$&$-3.663$&$0.05$&$0.16$&$0.07$&$0.17$ \\ 
$13.65$&$362$&$-4.586$&$-3.597$&$0.05$&$0.16$&$0.07$&$0.17$ \\ 
$13.55$&$367$&$-4.529$&$-3.532$&$0.05$&$0.16$&$0.07$&$0.17$ \\ 
$13.45$&$391$&$-4.456$&$-3.428$&$0.05$&$0.16$&$0.07$&$0.16$ \\ 
$13.35$&$351$&$-4.393$&$-3.327$&$0.05$&$0.17$&$0.07$&$0.18$ \\ 
$13.25$&$338$&$-4.359$&$-3.270$&$0.05$&$0.17$&$0.07$&$0.18$ \\ 
$13.15$&$302$&$-4.272$&$-3.103$&$0.06$&$0.19$&$0.07$&$0.20$ \\ 
$13.05$&$258$&$-4.422$&$-3.398$&$0.06$&$0.20$&$0.07$&$0.21$ \\
$12.95$&$217$&$-4.254$&$-3.041$&$0.07$&$0.21$&$0.07$&$0.22$ \\ \hline
$12.85$\fixv&$182$&$-4.472$&$-3.447$&$0.07$&$0.22$&$0.07$&$0.23$ \\ 
$12.75$&$152$&$-4.633$&$-3.713$&$0.08$&$0.23$&$0.07$&$0.24$ \\ 
$12.65$&$111$&$-4.562$&$-3.518$&$0.09$&$0.25$&$0.07$&$0.27$ \\ \hline
\end{tabular}
\end{table*}

\begin{figure*}
	\centering     
	\includegraphics[width=\textwidth]{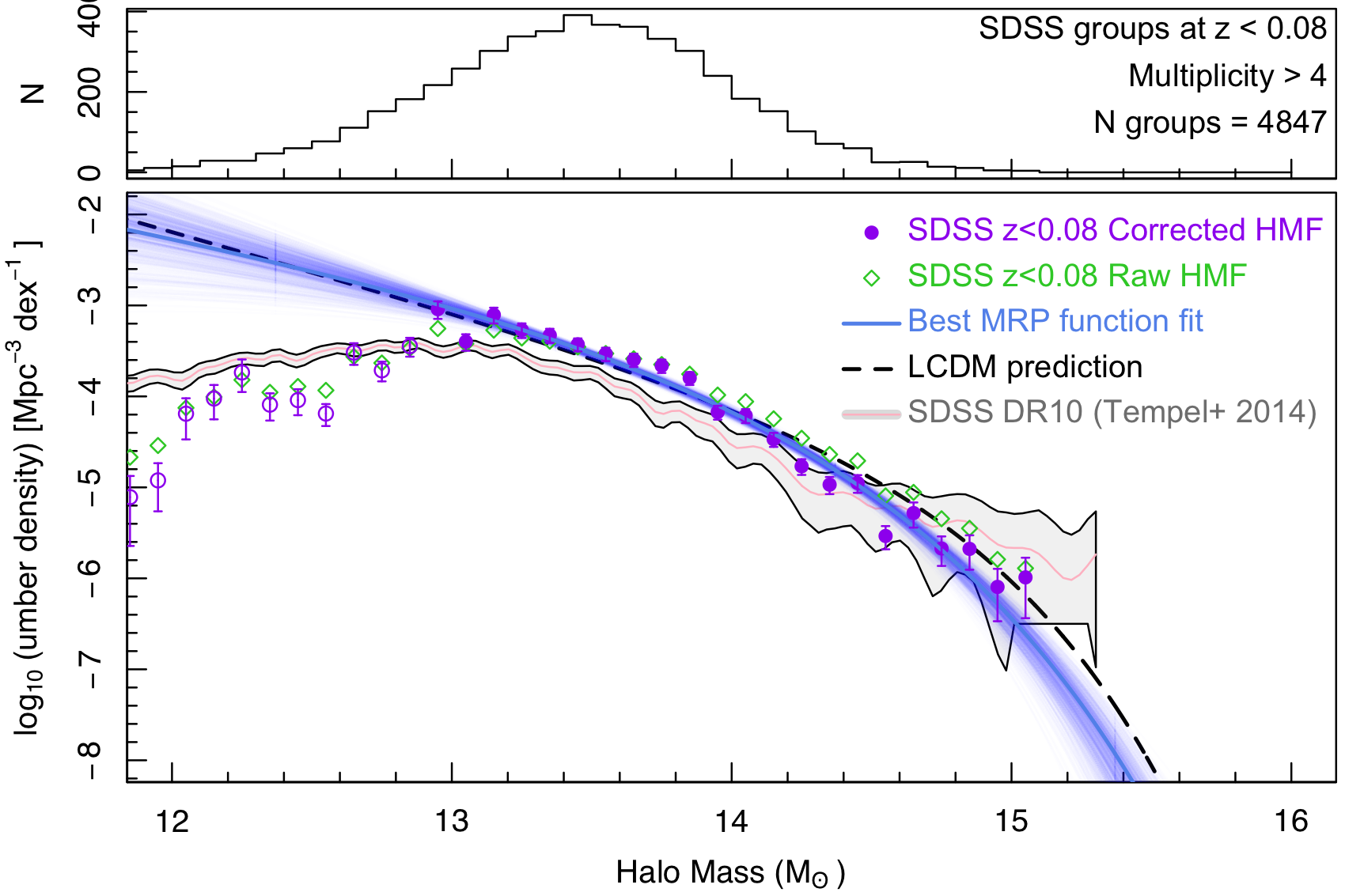}
	\caption{As for Fig.\,\ref{fig:gamahmf} except now for the SDSS data of \citefix{tempel2017}.
	\label{fig:sdsshmf}}
\end{figure*}

\begin{figure*}
	\centering     
	\includegraphics[width=\textwidth]{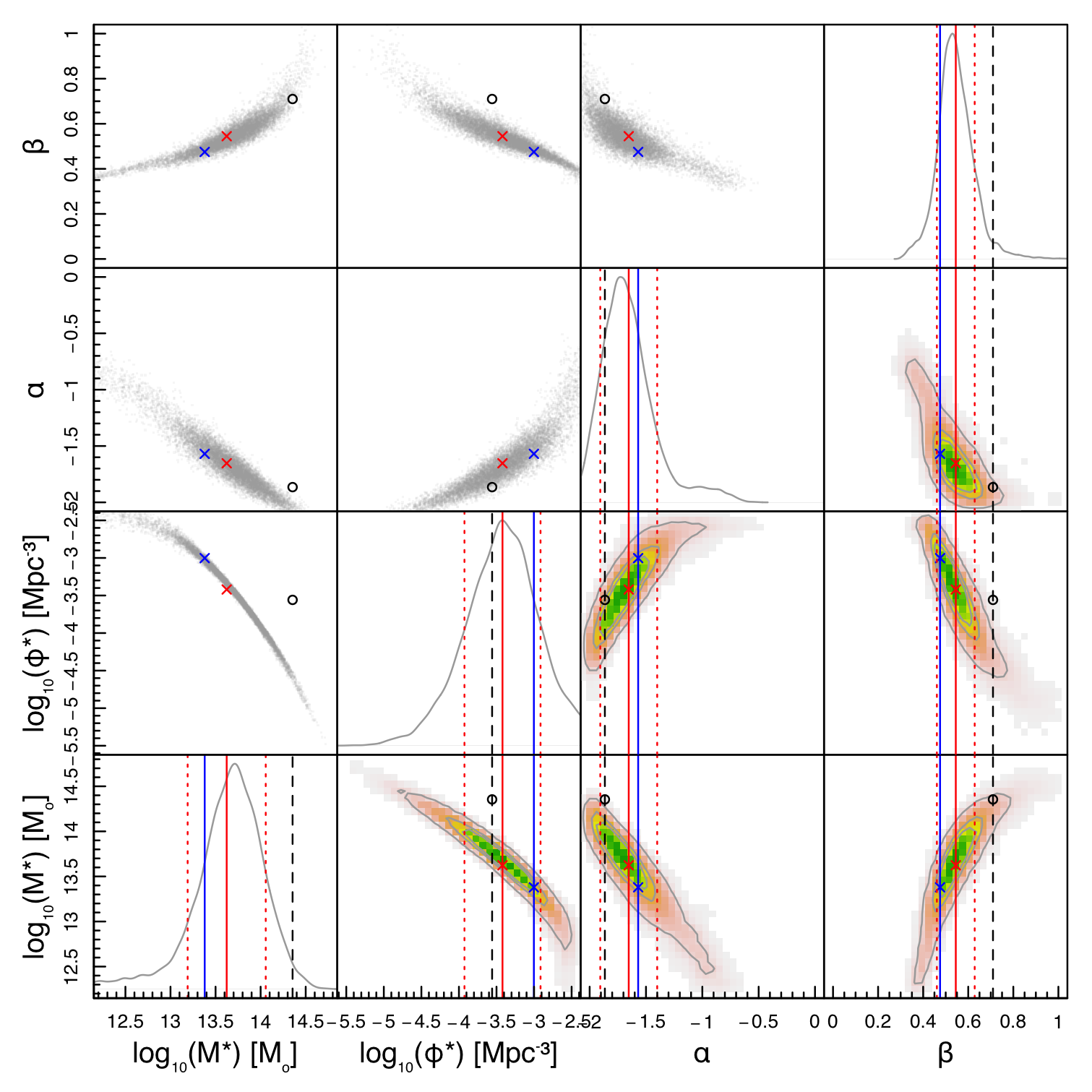}
	\caption{As for Fig.\,\ref{fig:gamacov} except now for the SDSS data of \citefix{tempel2017}.
	\label{fig:sdsscov}}
\end{figure*}

\subsection{SDSS}
Recently, \cite{tempel2017} applied their friends-of-friends group finding algorithm to the Sloan Digital Sky Survey Data Release 12 \citep{alam2015}. The final catalogue contains a total of 88\,662 galaxy pairs or groups, 37\,365 with 3 or more members and 10\,087 with 5 or more members. This is derived from the SDSS\,DR12 parent catalogue of 584\,449 galaxies within a 7\,221 square degree area of the SDSS Main Survey footprint (see also \citealt{tempel2014}). 

A particular focus of the work was the identification of merging groups, which can often be mistaken for a single high mass cluster, and care was taken to disentangle these cases. The online catalogues were downloaded from \url{http://cosmodb.to.ee}. The catalogue contains $M_{200}$ values that are derived from a combination of the velocity dispersions, which are also based on the {\sc gapper} method to convert the redshift distributions to velocity dispersions, and the group sizes given as $R_{\rm vir}$. \cite{old2014} conducted a fairly exhaustive study of mass estimates and derived errors on the masses for three mass intervals (see their table\,3). \cite{old2014} report approximate mass errors appropriate for the \citeauthor{tempel2014} group finder (i.e. $\sigma_{\log_{10}M} \approx 0.25$ at $10^{15}$\,M$_{\odot}$ rising to $\sigma_{\log_{10}M} \approx 0.35$ at $10^{13.0}$\,M$_{\odot}$ (our cutoff for the SDSS data). However these do not include the impact of multiplicity that we see as the dominant error. It is worth noting that \cite{tempel2014} attempt to circumvent the mass calibration issue by fixing $A$ to the value implied from a pure NFW mass profile.


We now implement an identical method as described for GAMA to determine the SDSS HMF. The results of our derived SDSS HMF are tabulated in Table\,\ref{tab:sdsshmf} to the nominal mass limit of $10^{12.9}$\,M$_{\odot}$, and plotted in Fig.\,\ref{fig:sdsshmf} as the purple data points. Also shown in Fig.\,\ref{fig:sdsshmf} are the volume-limited ($-18$) results from \cite{tempel2014}. Contrary to GAMA, the SDSS data at higher masses tend to fall marginally below the $\Lambda$CDM expectation, and in general have a slightly lower amplitude. We note that the data are also slightly inconsistent with the SDSS\,DR10 estimate of \cite{tempel2014}. This could be due to the distinct methods applied, $1/V_{\rm max}$ with Eddington bias correction versus a purely volume limited sample (i.e. constant $V_{\rm max}$ without an Eddington bias correction. It could also be due to the effort invested in the DR12 catalogue \citep{tempel2017}, in identifying and separating erroneously merged groups, leading to a modest reduction of high mass systems and a modest increase in intermediate mass systems.

Fig.\,\ref{fig:sdsscov} shows the covariance of the fitted parameters that are generally better behaved than for GAMA (cf. Fig.\,\ref{fig:gamacov}). Again we see a slight offset in the $M_*$ parameter but note the degeneracy between $\beta$ and $M_*$. We also see a slightly steeper low mass slope of between $\alpha=-1.57^{+0.12}_{-0.31}$ and closer to expectation ($\alpha=-1.89$).


\subsection{ROSAT and REFLEX\,II}
\begin{figure*}
	\centering     
	\includegraphics[width=\textwidth]{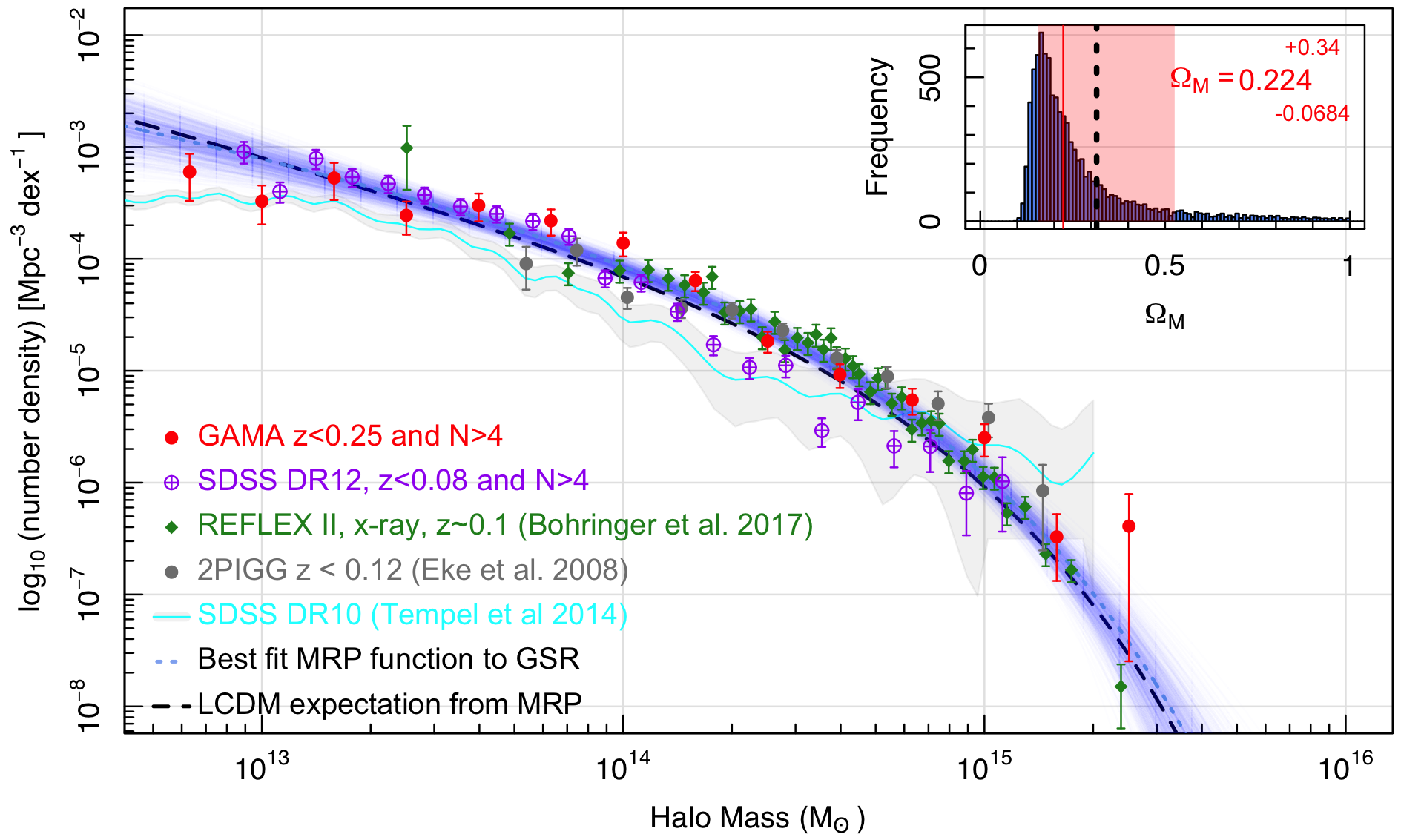}
	\caption{The combined empirical HMF data (as indicated). Shown as a black and blue dashed lines are the $\Lambda$CDM prediction and the best MRP function fit to the combined GAMA, SDSS and REFLEX\,II data along with the spread of MRP fits in blue that show the results from our Monte-Carlo refitting. The inset panel shows the integral of the Monte-Carlo MRP fits to zero mass (blue histogram) with the red band showing the 1\,$\sigma$ error range. The vertical black dashed line shows the Planck 2018 value for $\Omega_{\rm M}$.}
	\label{fig:allhmf}
\end{figure*}

Probably the most compelling HMF measurements to date come from X-ray wavelengths, and in particular the ROSAT All Sky Survey dataset. From this dataset the ROSAT-ESO Flux Limited X-ray Galaxy Cluster Survey (REFLEX\,II) was formed containing 910 clusters out to z=0.4 \citep{bohringer2013}. These data were combined into an X-ray luminosity function in \cite{bohringer2014} and used to derive constraints on $\Omega_M$ and $\sigma_8$ for a specified halo-model. In \cite{bohringer2017} a sample of 863 clusters from the same sample were transformed into a mass function using the Luminosity-Mass relation ($L_X-M$), see \cite{bohringer2017}. These data are shown in Fig.\,\ref{fig:allhmf} as the green diamonds, and contain 20 clusters per bin except for the highest mass bin, which contains 3 clusters. The errors shown are simply $\sqrt{n}$. Uncertainty in the masses are estimated to be $\sigma_{\log_{10}M} \sim 0.1$ and hence an expected Eddington bias of $\sim \times 1.25$ (see Fig.\,\ref{fig:edb}). We do not correct for this. We also make no allowance for the fact that the effective redshift for this sample will be slightly lower than for GAMA.

\begin{figure*}
	\centering     
    \includegraphics[width=\textwidth]{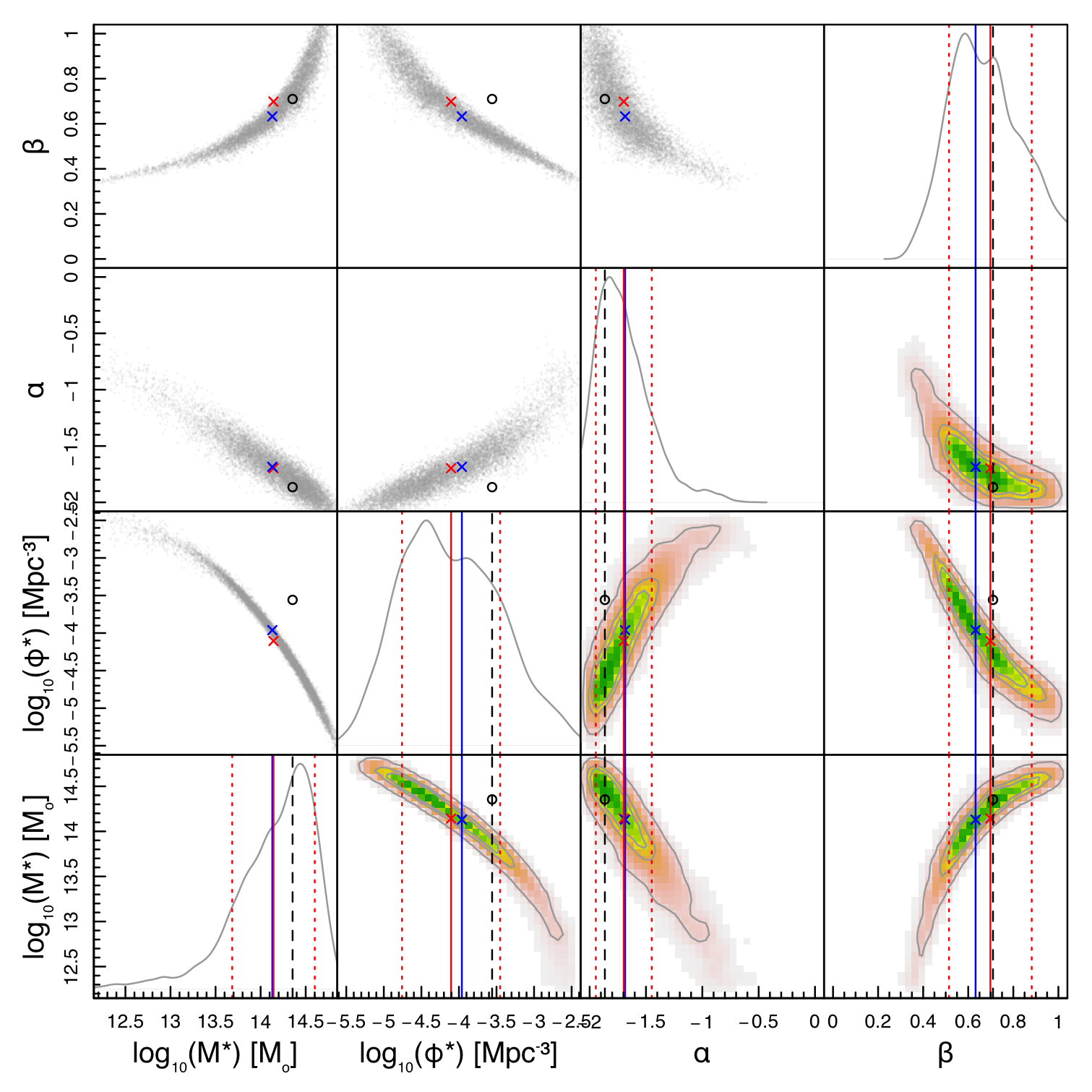}
	\caption{The covariances of the final fitted parameters to the combined GAMA, SDSS and REFLEX\,II data.}
	\label{fig:allcov}
\end{figure*}

\subsection{Joint constraints}
Fig.\,\ref{fig:allhmf} shows the combined dataset (as indicated). These form a dataset that scatters consistently around the expectation curve from $\Lambda$CDM (dashed black line). This agreement now extends over a mass range from $10^{15.5}$\,M$_{\odot}$ to $10^{12.7}$\,M$_{\odot}$, significantly expanding on previous studies, and providing strong corroboration of the $\Lambda$CDM prediction. In general the X-ray data (green data points) dominate the `knee' region, providing the tightest error constraints, while the GAMA and SDSS data (red and purple data points respectively) provide leverage on the low mass slope. The lack of detections in all three surveys at high mass, within their respective volumes, helps in constraining the high mass turn-down.

We fit the combined GAMA, SDSS, and REFLEX\,II data (and all combinations thereof), with the MRP function as before (see Eqn.\,\ref{eqn:mrp}). We find a good fit given by the blue dashed line with perhaps some greater scatter in normalisation. This reflects both the cosmic variance now acting independently on the three datasets and the independent methodologies for absolute mass calibration (i.e. the uncertainty in $A$). The fit has a convergent low halo mass slope of $-1.68^{+0.21}_{-0.24}$, just consistent with expectation \citep[$\alpha=-1.89$:][]{murray2021}. Integrating this curve to zero mass yields a total mass density of about half the expected value (see Fig.\,\ref{fig:allhmf} inset panel), although the range of values from our Monte-Carlo simulation (inset panel red shading) does enclose the Planck\,2018 value (vertical black dashed line). The uncertainties on the MRP parameters were again derived by independently perturbing each of the GAMA, SDSS and REFLEX\,II data points by their errors, and each dataset by their cosmic variance error, and refitting to form the spread of blue lines, showing the plausible range of fits. The uncertainty and co-variances of our final fitted parameters are represented in Fig.\,\ref{fig:allcov}, in general these are well behaved with most values of $\alpha$ convergent (i.e. $\alpha > -2$).  

\begin{figure*}
	\centering     
	\includegraphics[width=\textwidth]{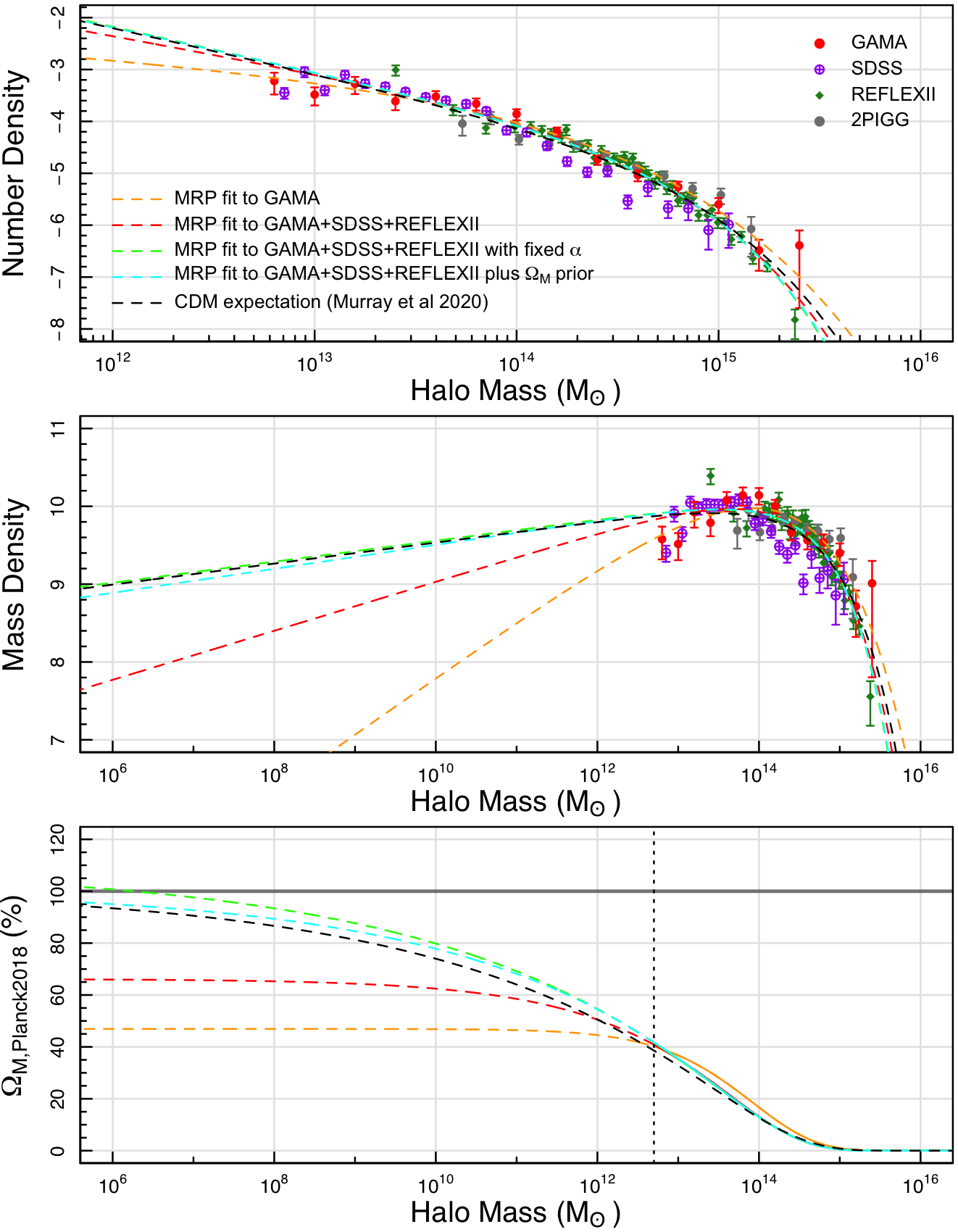}
	\caption{Various fits to the data (as indicated) and showing (top panel) the HMF from Fig.\,\ref{fig:allhmf} but now over a broader range, (central panel) the contribution of each mass interval to the total density, and (lower panel) the cumulative contribution of the various models to the Planck 2018 mass density. The vertical dashed line in the lower panel indicates the halo mass limit to which our data extends.}
	\label{fig:omegacrit}
\end{figure*}

\begin{figure*}
	\centering     
	\includegraphics[width=\textwidth]{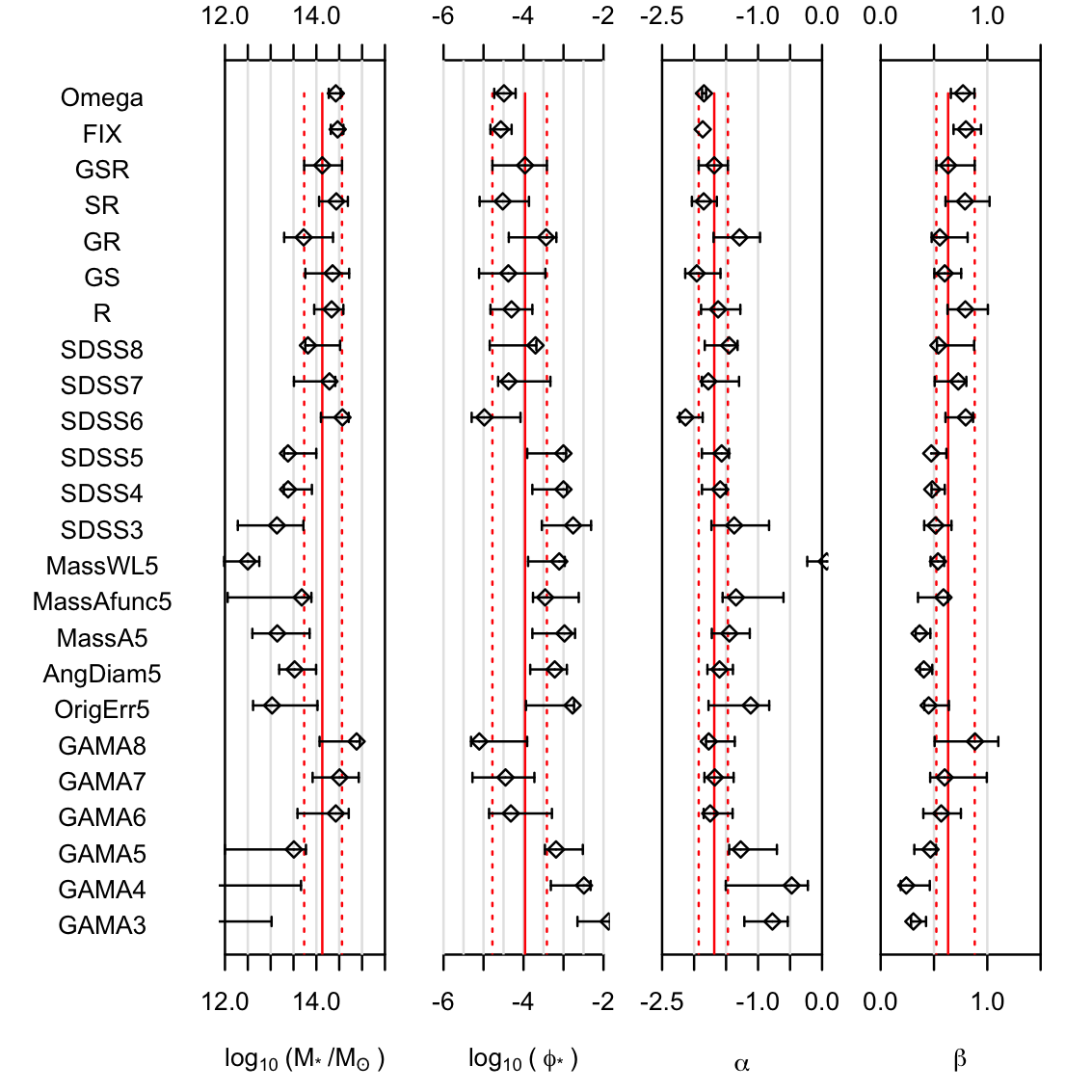}
	\caption{The full set of our fitted MRP results highlighting the variation through our various trials and digressions. The GSR dataset is our final favoured fit and the red solid and dashed lines show the mean and errors of our best fit parameters extrapolated across the plot for comparison purposes. In general most fits are consistent with our final fitted parameters. Note also that the errors plotted in this fashion do not incorporate the strong covariances between the four MRP parameters.}
	\label{fig:errors}
\end{figure*}

 Fig.\,\ref{fig:omegacrit} shows: (upper) our best fit MRP function over a broader halo mass range; (centre) the same distribution but now expressed in terms of the contribution to the mass density (rather than number density); and (lower) the cumulative contribution, by integrating from high mass to low mass, and expressed as a fraction of the Planck\,2018 matter density ($\Omega_M=0.3147 \pm 0.0074$). The MRP fit to the GAMA data only (GAMA5) is shown in red, and the fit to the final combined GAMA+SDSS+REFLEX\,II data is shown in orange. The joint fit has significantly tighter errors not represented here, but more apparent in Fig\,\ref{fig:errors}, which shows all the MRP fitted parameters for all data combinations explored in the main text and Appendix.

We now consider two extensions: firstly, fixing $\alpha$, and secondly, incorporating the Planck\,2018 $\Omega_{M}$ prior.
We show in Fig.\,\ref{fig:omegacrit} in green, the optimal MRP fit if we fix the $\alpha$ parameter to the value recommended by \cite{murray2021}. The red, orange and green all match the data reasonably well, but as can be seen in the lower panel, all three either under- or over-predict the total matter-density when integrated to very low halo masses. Hence we now fold in the Planck\,2018 $\Omega_{M}$ value as an additional constraint on the MRP fitting process. To do this we minimise the following expression:
\begin{equation}
\eqalign{
&\frac{1}{N_{\rm bins}}\sum{\frac{[\phi(M_{\rm emp})-\phi(M_{\rm mod})]^2}{\sigma_{\rm Combined}^2}} \cr
&+\; \frac{\left[\int^{\infty}_{0} (M_h.\phi_{(M_*,\phi_*,\alpha,\beta)}dM_h/\rho_{\rm crit})-\Omega_{M}\right]^2}{\sigma_{\Omega_M}^2}.
}
\end{equation}
Implementing this strategy results in the cyan curve in Fig.\,\ref{fig:omegacrit}, which now very closely matches the expectation curve (black line) and by construction integrates to the Planck\,2018 matter density value. 

\begin{figure*}
	\centering     
	\includegraphics[width=\columnwidth]{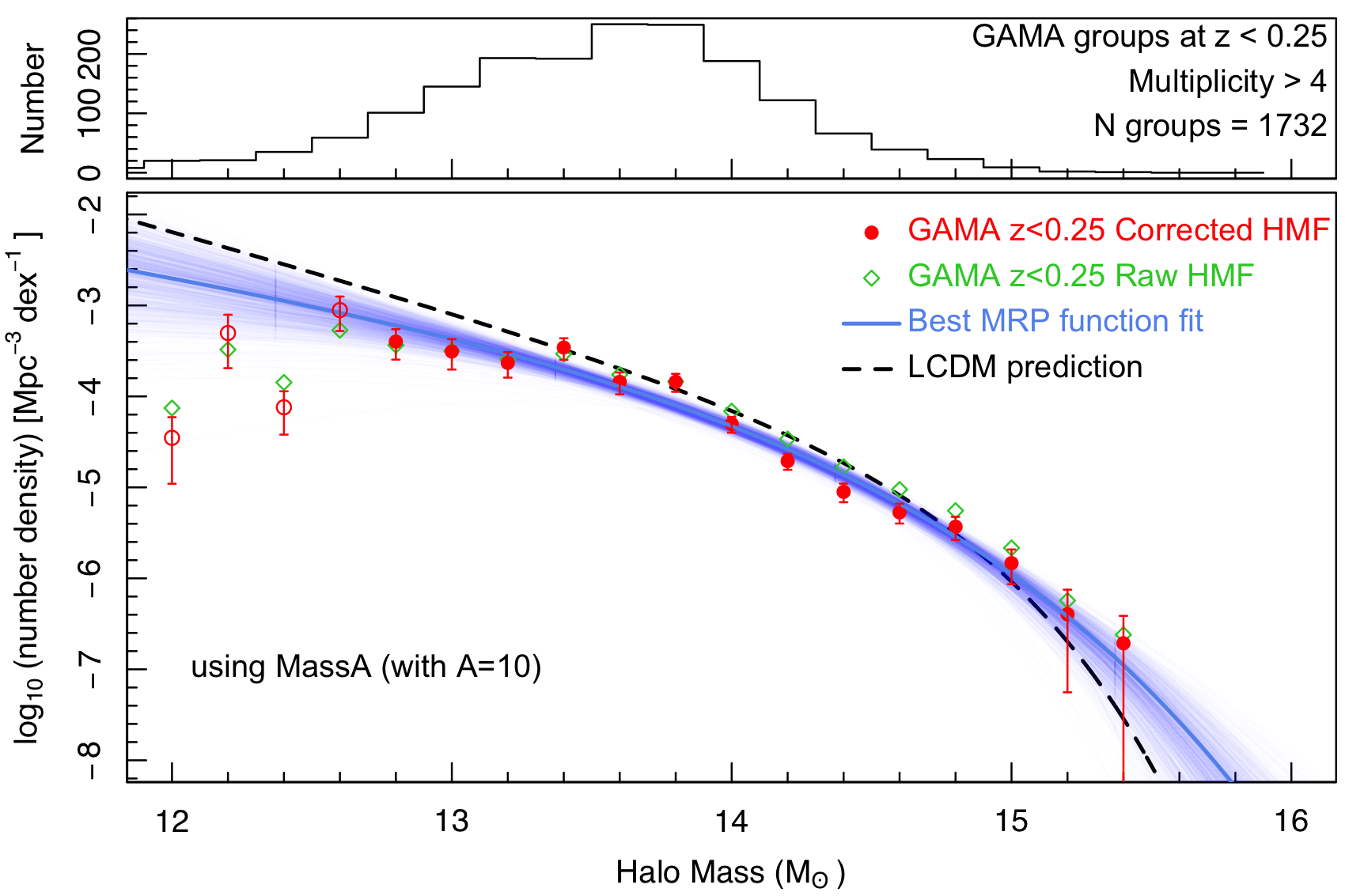}
	\includegraphics[width=\columnwidth]{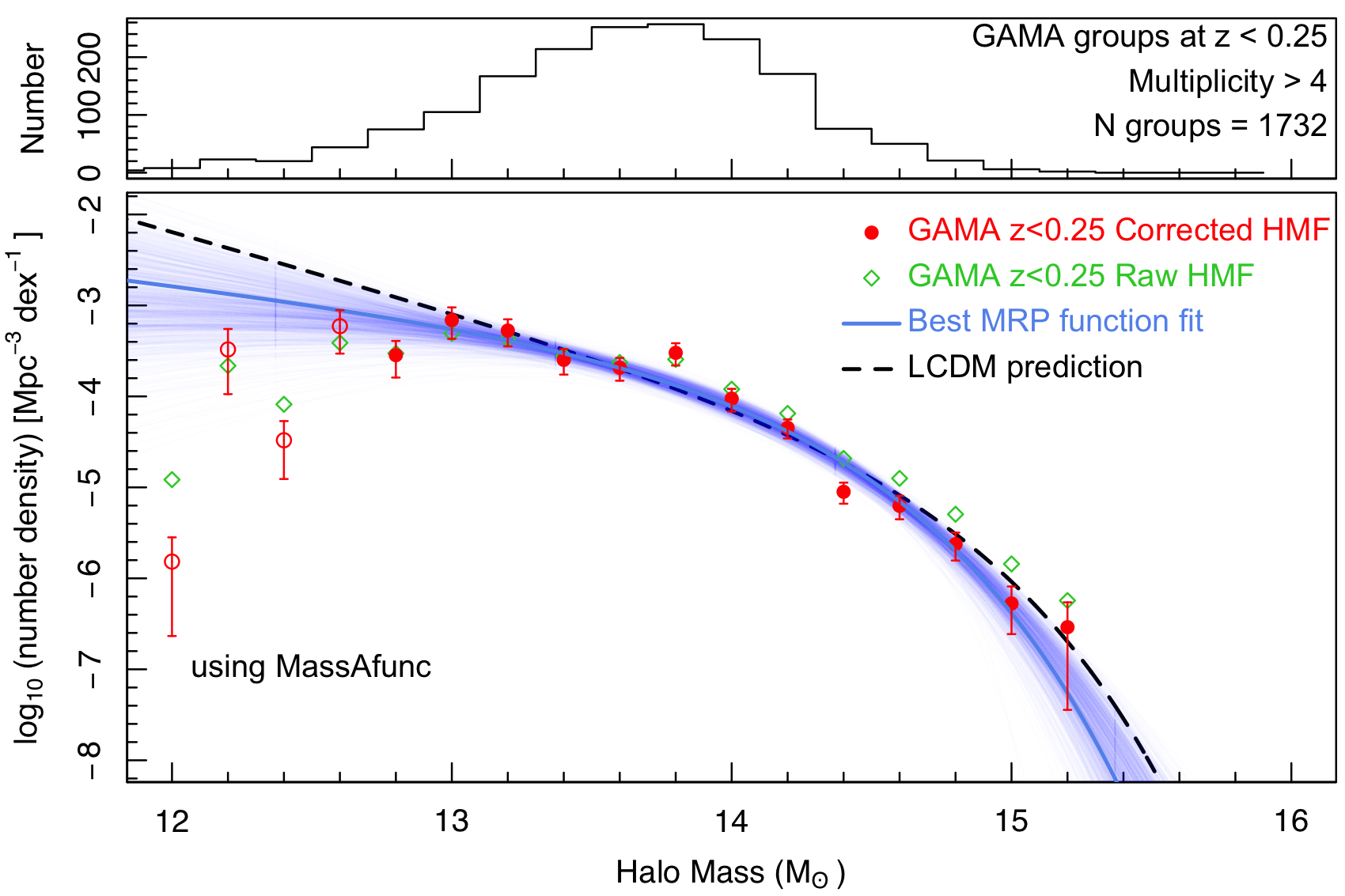}
	\includegraphics[width=\columnwidth]{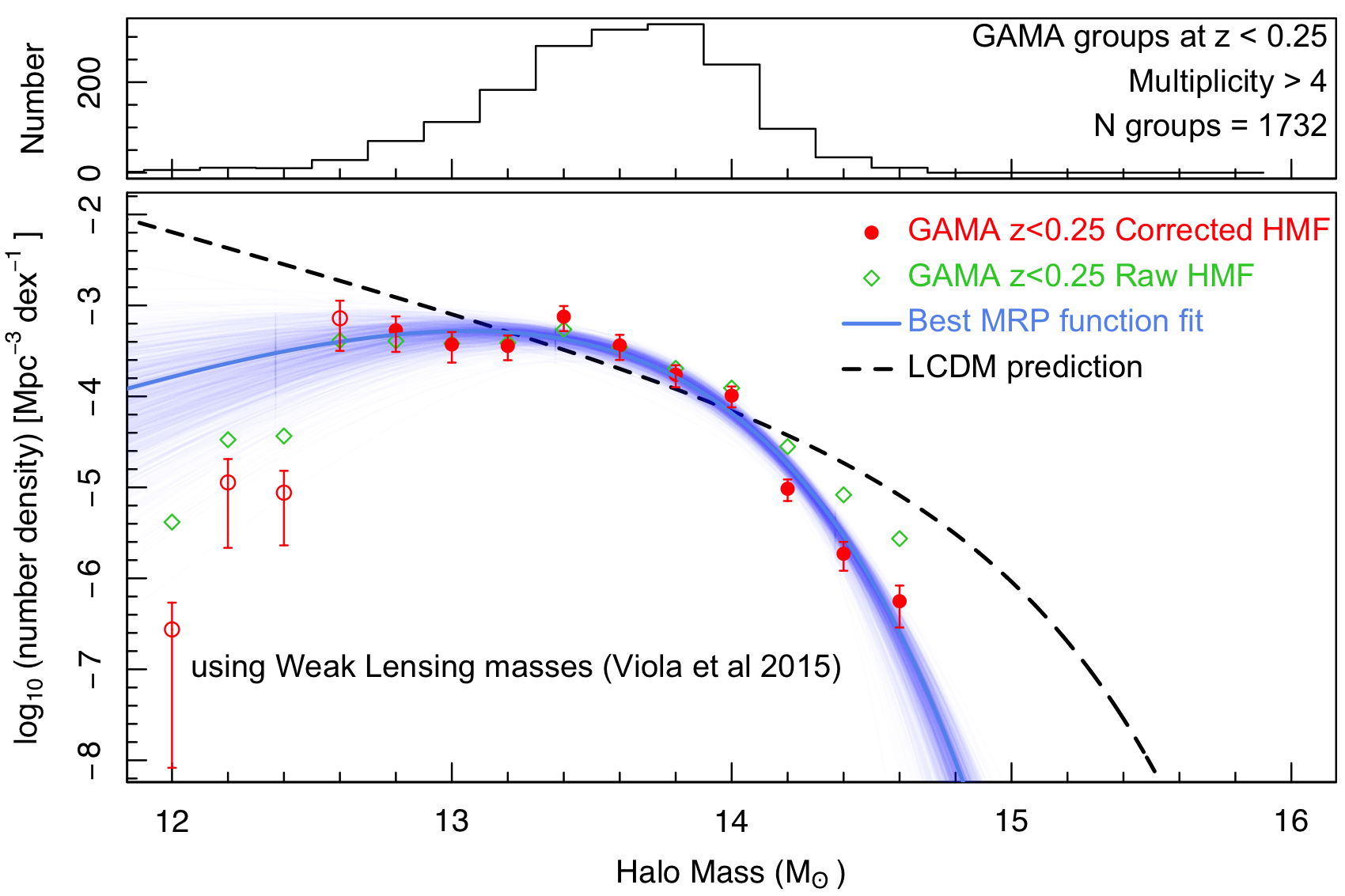}
	\includegraphics[width=\columnwidth]{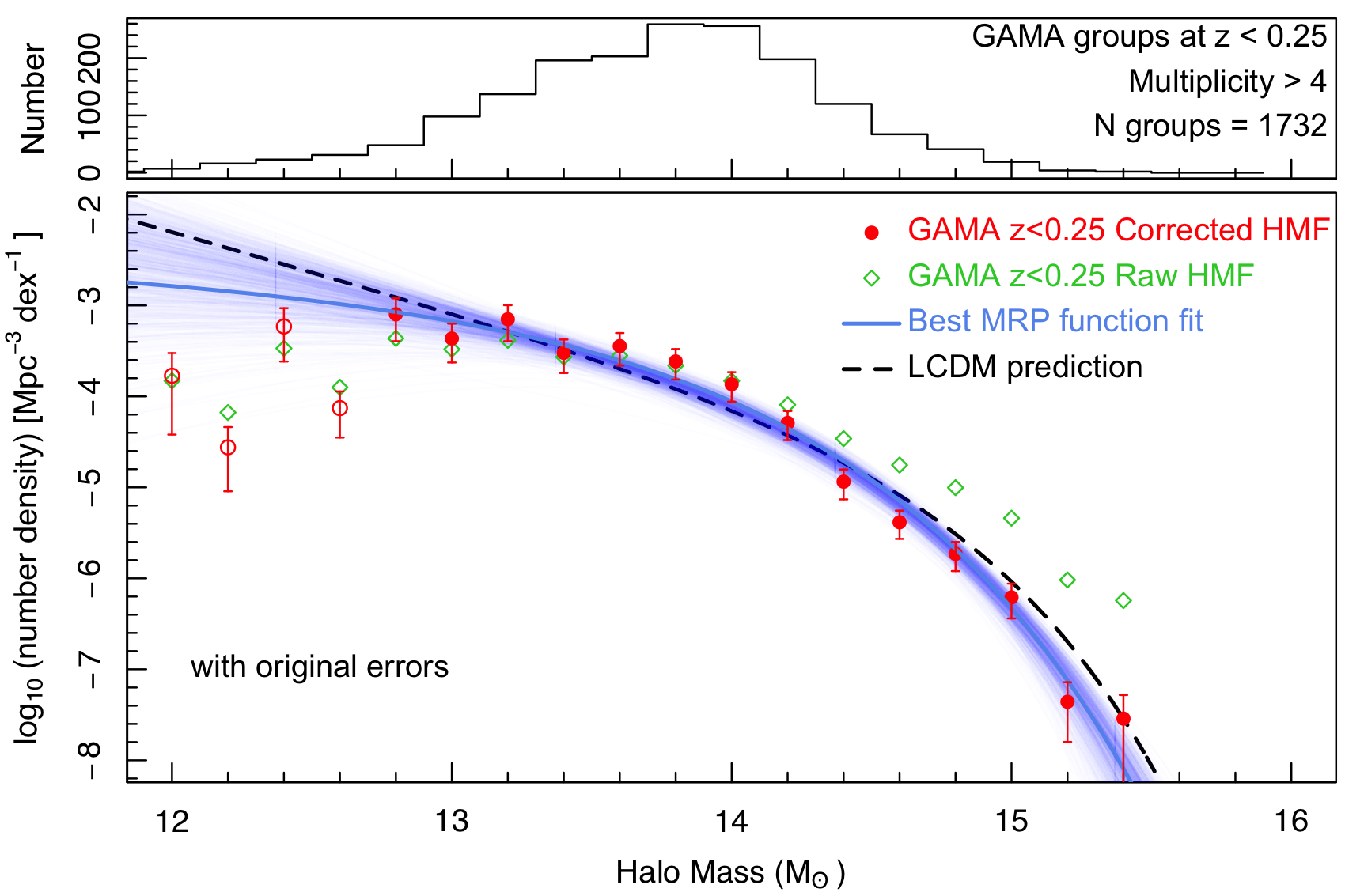}
	\includegraphics[width=\columnwidth]{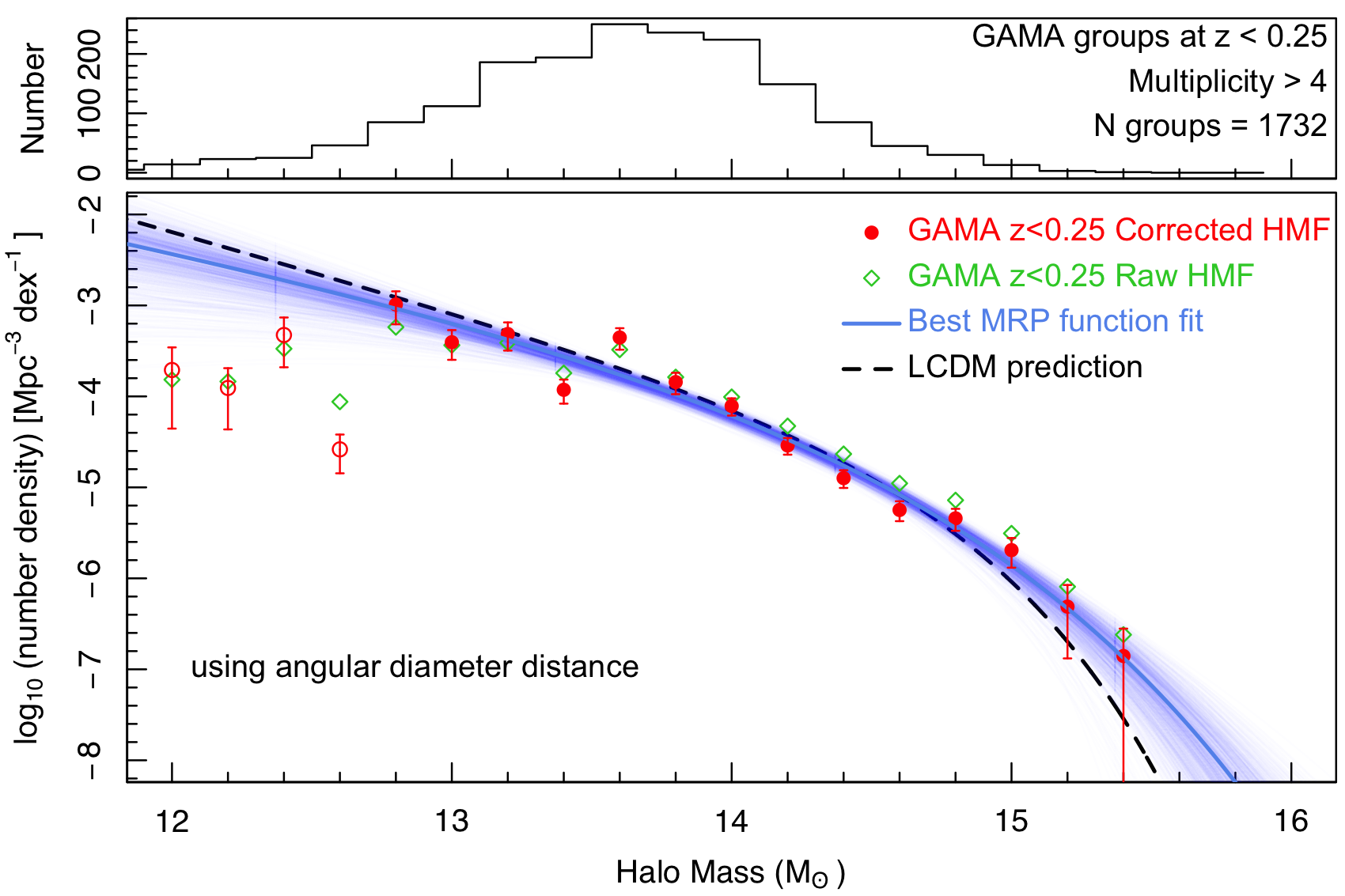}
	\caption{GAMA HMF derivations for different mass, error, and distance choices but all with a multiplicity cut of $N_{\rm FoF} > 4$. \label{fig:appendix2}}
\end{figure*}

Note that to some extent this is not entirely surprising, as the data now primarily constrain the shape of the high-mass end (i.e. $M_*,\phi_*,\beta$), while the $\Omega_M$ constraint is then forcing the low mass slope parameter. That most data within their errors overlap with this fit is reassuring, but does perhaps indicate some mild mass bias in the G3C mass estimates at the lower halo masses. This is perhaps also indicated in the weak lensing mass estimates, and it is noticeable that the HMF based on the weak lensing masses, while fitting poorly at the high mass end, fits better in the $10^{13}$\,M$_{\odot}$ to $10^{14}$\,M$_{\odot}$ range (see Fig.\,\ref{fig:appendix2}, centre left).

Hence it is plausible that at high halo masses, velocity dispersion mass estimates are more robust, while at lower halo masses (where interlopers may be more prevalent), weak lensing masses are more robust. This does open a pathway to constructing HMFs from the combination of both methods, which we leave for future study. Nevertheless, it has become clear through this work that refining our mass estimates and uncertainties is critical, even more so than larger or lower halo mass samples.

The parameters of all the fits shown in Fig.\,\ref{fig:omegacrit} and on the various plots throughout this paper, are listed in Table\,\ref{tab:mrpfits}. Our strong recommendation is to adopt the values using the $\Omega_M$ constraint when extrapolating to halo masses below $10^{12.7}$\,M$_{\odot}$, and to use the combined (i.e. GAMA, SDSS and REFLEX\,II data or GSR) MRP fit when seeking to represent the data, which should be deemed valid over the range $10^{12.7}$\,M$_{\odot}$ to $10^{15.8}$\,M$_{\odot}$.

New survey programmes such as DESI and WAVES will soon produce much larger group catalogues, but it is clear that two factors will be critical if we are to advance our empirical measurement of the HMF:

\noindent
(1) high spectroscopic completeness ($>90$ per cent) is critical in ensuring low mass haloes are detected and sparse haloes, or those with mass gaps, are not biased against. 

\noindent
(2) significant work needs to be done to improve our mass estimates of galaxy groups especially at low multiplicity and low masses, and this should include comparison of velocity dispersion, weak lensing and other mass estimates on a halo-by-halo basis as well as for halo populations.

Overall from our work, and from studies based on simulations by \cite{chauhan2021}, we see that it is important to sample at least 5 members of a galaxy group in order to construct a credible mass estimate, and hence a robust HMF. It is sobering to note that for the Local Group this implies probing down to NGC3109, or the SMC, which are about $1000\times$ fainter than the Milky Way or Andromeda.

\section{Summary} 
In this work we have attempted to reconstruct the Halo Mass Function (HMF) via a predominantly model-free empirical pathway, and to highlight the key areas for future consideration along the way. In general we conclude that over a broad range in halo mass we can confirm the form and amplitude of the HMF, as predicted from numerical simulations and analytic calculations. In particular we find the dominant error to be the Eddington bias, which arises from the significant mass uncertainties inherent in group determination. To second order, significantly larger samples are also required and over broader local volumes to probe to lower halo masses with meaningful statistics. In combination with X-ray data we see good consistency, and hence complementarity, between the GAMA, SDSS and REFLEX\,II data. In combination these datasets are almost enough to constrain the functional form of the HMF, and we find a MRP function fit with the following parameters,

\begin{align}
\log_{10}(\phi_*\,\rm{Mpc}^{3}) &= -3.96^{+0.55}_{-0.82}, \nonumber\\
\log_{10}(M_*\,\rm{M}_{\odot}^{-1}) &=14.13^{+0.43}_{-0.40}, \nonumber\\
\alpha &=-1.68^{+0.21}_{-0.24}, \nonumber\\
\beta &=0.63^{+0.25}_{-0.11}.\nonumber
\end{align}
Note that the MRP fit is for an effective redshift of $\tilde{z} \approx 0.1$ and to convert to $z = 0$ one should add 0.075\,dex to $\log_{10}(M_*/M_{\odot})$ and subtract 0.075\,dex from $\log_{10}(\phi*)$.

Fig.\,\ref{fig:errors} shows graphically the complete set of all the MRP fits shown in Table\,\ref{tab:mrpfits}, and it generally portrays a consistent picture (i.e. most errorbars encompass the favoured values).
Our final MRP parameters that best represents the data are the third from top value (GSR), and the red lines trace the final fitted values down the figure. It is clear that the REFLEX\,II data are the dominant dataset and currently the X-ray pathway looks like the more secure method to derive the HMF. Nevertheless the inclusion of the GAMA and SDSS data do significantly tighten the error bars of the fit while remaining consistent with the REFLEX\,II-only values. 

At this stage it is worth noting that almost {\it all} the fits are formally covergent, but the error ranges include divergent solutions. In general the implied $\Omega_M$  values are consistent with expectation but with relatively broad errors. For example our combined GSR fit returns the range of $\Omega_M$ values shown in Fig.~\ref{fig:allhmf} (inset panel) with the median and 16\%/84\% quantiles as indicated by the vertical red solid line and red band. Shown in black is the $\Omega_M$ expectation from Planck\,2018.  If we use the Planck\,2018 $\Omega_M$ value as a prior, then we can recover significantly more stringent MRP fitted values, with quite modest errors, which by design converges to the expected dark matter density. These values are:

\begin{align}
\log_{10}(\phi_*\,\rm{Mpc}^{3})&= -4.49^{+0.29}_{-0.24}, \nonumber\\
\log_{10}(M_*\,\rm{M}_{\odot}^{-1})&=14.43^{+0.11}_{-0.15}, \nonumber\\
\alpha &=-1.85^{+0.03}_{-0.03},\nonumber\\ 
\beta &=0.77^{+0.11}_{-0.11}.\nonumber
\end{align}
Note that the MRP fit is for an effective redshift of $\tilde{z} \approx 0.1$ and to convert to $z = 0$ one should add 0.075\,dex to $\log_{10}(M_*/M_{\odot})$ and subtract 0.075\,dex from $\log_{10}(\phi*)$.

\noindent
We note there is no real rationale for expecting the HMF to observe a strictly power-law distribution to zero mass, but the test does demonstrate plausible consistency between the preferred MRP values, the Planck\,2018 matter density value, and our combined GAMA, SDSS and REFLEX\,II data.

While a stringent independent constraint on $\Omega_M$ from an HMF analysis alone is not yet viable, we can ask how much of the total matter density predicted by Planck 2018 is resolved into haloes down to our mass limit. Integrating down to $M_h = 10^{12.7}$\,M$_{\odot}$, we find approximately $41 \pm 5$ per cent of the expected matter content (regardless of which fit is adopted). This is significant, as it supports the notion that the majority of mass is contained within group haloes, and suggests that it should be possible to build direct dark matter maps from galaxy group catalogues. Such maps would have significantly higher resolution than that achievable via either weak lensing, redshift space distortions, or peculiar velocity studies. 

Finally, it is worth reiterating that our mostly empirically derived HMF shows broad general agreement with the expectation from $\Lambda$CDM, and in doing so provides some confirmation of $\Lambda$CDM. However, it is also worth highlight that this also represents a coming of age of galaxy group catalogues. This is important, as most simulations and observations highlight how the galaxy formation process is strongly dependent on the underlying dark-matter halo mass. This is perhaps most obvious in the  numerical studies of star-formation efficiency as a function of halo mass \citep{behroozi2013,weschler2018}, but also in the transition from high-mass haloes dominated by hot plasma to low mass haloes dominated by cold gas and the `goldilocks zone' in which star-formation is most efficient.

Paramount to capitalising on a group-centric approach to extra-galactic astronomy is the need for robust masses. In this work, as with most studies, we rely on the viral thereom embodied in the very simple Eqn.\,\ref{eqn:mass} and the calibration of a simple constant $A$ via simulations. This approach does raise concerns over non-orthogonality between the measurements and the models (circularity?), but also the simple virial mass estimate clearly does not use all the information to hand, and does not acknowledge or incorporate factors such as partial-virialization, onging/lingering halo mergers, halo morphologies, and the additional information encoded in the full spatial distribution of the detected galaxy population, its velocity dispersion distribution, and the combination thereof. 

There are two obvious and orthogonal pathways for addressing the above issues. The first is to tie the construction of group catalogues more closely to simulations and to incorporate this calibration process into the mass error estimation for each individual group. The second is to pursue a vigorous programme for deriving improved empirical constraints on the group masses. One obvious starting point is to use the 120 $G^3C$ groups for which 20+ members are known and to undertake further study of these systems via X-ray, radio, strong-lensing, and Sunyaev-Zeldovich techniques. Such a project will be critical if we aim to use galaxy groups as truly direct empirical probes of the population of dark matter haloes.

\section*{Acknowledgements}
ASGR and DO are recipients of Australian Research Council Future Fellowships
(FT200100375 and FT190100083) funded by the Australian Government.

GAMA is a joint European-Australasian project based around a spectroscopic campaign using the Anglo-Australian Telescope. The GAMA input catalogue is based on data taken from the Sloan Digital Sky Survey and the UKIRT Infrared Deep Sky Survey. Complementary imaging of the GAMA regions is being obtained by a number of independent survey programmes including GALEX MIS, VST KiDS, VISTA VIKING, WISE, Herschel-ATLAS, GMRT and ASKAP providing UV to radio coverage. GAMA is funded by the STFC (UK), the ARC (Australia), the AAO, and the participating institutions. The GAMA website is \url{http://www.gama-survey.org/} .

Based on observations made with ESO Telescopes at the La Silla Paranal Observatory under programme ID 179.A-2004.

Based on observations made with ESO Telescopes at the La Silla Paranal Observatory under programme ID 177.A-3016.

This work was supported by resources provided by the Pawsey Supercomputing Centre with funding from the Australian Government and the Government of Western Australia.

The analysis and figures were made using the R programming language \url{https://www.r-project.org} with R packages designed for astronomical calculations and plotting, these include: {\sc celestial} \citep{celestial} and {\sc magicaxis} \citep{magicaxis}.

\section*{Data Availability}
The data underlying this article are available from the GAMA Date Release 4 website, at \url{http://www.gama-survey.org/dr4}, or for external datasets via the links provided in the text.

\bibliography{paperlib.bib}

\begin{thebibliography}{}
\makeatletter
\relax
\def\mn@urlcharsother{\let\do\@makeother \do\$\do\&\do\#\do\^\do\_\do\%\do\~}
\def\mn@doi{\begingroup\mn@urlcharsother \@ifnextchar [ {\mn@doi@}
  {\mn@doi@[]}}
\def\mn@doi@[#1]#2{\def\@tempa{#1}\ifx\@tempa\@empty \href
  {http://dx.doi.org/#2} {doi:#2}\else \href {http://dx.doi.org/#2} {#1}\fi
  \endgroup}
\def\mn@eprint#1#2{\mn@eprint@#1:#2::\@nil}
\def\mn@eprint@arXiv#1{\href {http://arxiv.org/abs/#1} {{\tt arXiv:#1}}}
\def\mn@eprint@dblp#1{\href {http://dblp.uni-trier.de/rec/bibtex/#1.xml}
  {dblp:#1}}
\def\mn@eprint@#1:#2:#3:#4\@nil{\def\@tempa {#1}\def\@tempb {#2}\def\@tempc
  {#3}\ifx \@tempc \@empty \let \@tempc \@tempb \let \@tempb \@tempa \fi \ifx
  \@tempb \@empty \def\@tempb {arXiv}\fi \@ifundefined
  {mn@eprint@\@tempb}{\@tempb:\@tempc}{\expandafter \expandafter \csname
  mn@eprint@\@tempb\endcsname \expandafter{\@tempc}}}

\bibitem[\protect\citeauthoryear{{Alam} et~al.,}{{Alam}
  et~al.}{2015}]{alam2015}
{Alam} S.,  et~al., 2015, \mn@doi [\apjs] {10.1088/0067-0049/219/1/12}, \href
  {https://ui.adsabs.harvard.edu/abs/2015ApJS..219...12A} {219, 12}

\bibitem[\protect\citeauthoryear{{Allen}, {Evrard}  \& {Mantz}}{{Allen}
  et~al.}{2011}]{allen2011}
{Allen} S.~W.,  {Evrard} A.~E.,   {Mantz} A.~B.,  2011, \mn@doi [\araa]
  {10.1146/annurev-astro-081710-102514}, \href
  {https://ui.adsabs.harvard.edu/abs/2011ARA&A..49..409A} {49, 409}

\bibitem[\protect\citeauthoryear{{Anderson}, {Gaspari}, {White}, {Wang}  \&
  {Dai}}{{Anderson} et~al.}{2015}]{anderson2015}
{Anderson} M.~E.,  {Gaspari} M.,  {White} S. D.~M.,  {Wang} W.,   {Dai} X.,
  2015, \mn@doi [\mnras] {10.1093/mnras/stv437}, \href
  {https://ui.adsabs.harvard.edu/abs/2015MNRAS.449.3806A} {449, 3806}

\bibitem[\protect\citeauthoryear{{Asencio}, {Banik}  \& {Kroupa}}{{Asencio}
  et~al.}{2021}]{elgordo2}
{Asencio} E.,  {Banik} I.,   {Kroupa} P.,  2021, \mn@doi [\mnras]
  {10.1093/mnras/staa3441}, \href
  {https://ui.adsabs.harvard.edu/abs/2021MNRAS.500.5249A} {500, 5249}

\bibitem[\protect\citeauthoryear{{Bahcall} \& {Cen}}{{Bahcall} \&
  {Cen}}{1993}]{bahcall1993}
{Bahcall} N.~A.,  {Cen} R.,  1993, \mn@doi [\apjl] {10.1086/186803}, \href
  {https://ui.adsabs.harvard.edu/abs/1993ApJ...407L..49B} {407, L49}

\bibitem[\protect\citeauthoryear{{Baldry} et~al.,}{{Baldry}
  et~al.}{2014}]{baldry2014}
{Baldry} I.~K.,  et~al., 2014, \mn@doi [\mnras] {10.1093/mnras/stu727}, \href
  {https://ui.adsabs.harvard.edu/abs/2014MNRAS.441.2440B} {441, 2440}

\bibitem[\protect\citeauthoryear{{Beers}, {Flynn}  \& {Gebhardt}}{{Beers}
  et~al.}{1990}]{beers1990}
{Beers} T.~C.,  {Flynn} K.,   {Gebhardt} K.,  1990, \mn@doi [\aj]
  {10.1086/115487}, \href
  {https://ui.adsabs.harvard.edu/abs/1990AJ....100...32B} {100, 32}

\bibitem[\protect\citeauthoryear{{Behroozi}, {Wechsler}  \&
  {Conroy}}{{Behroozi} et~al.}{2013}]{behroozi2013}
{Behroozi} P.~S.,  {Wechsler} R.~H.,   {Conroy} C.,  2013, \mn@doi [\apj]
  {10.1088/0004-637X/770/1/57}, \href
  {https://ui.adsabs.harvard.edu/abs/2013ApJ...770...57B} {770, 57}

\bibitem[\protect\citeauthoryear{{Binney} \& {Tremaine}}{{Binney} \&
  {Tremaine}}{2008}]{BT}
{Binney} J.,  {Tremaine} S.,  2008, {Galactic Dynamics: Second Edition}

\bibitem[\protect\citeauthoryear{{B{\"o}hringer}, {Chon}, {Collins}, {Guzzo},
  {Nowak}  \& {Bobrovskyi}}{{B{\"o}hringer} et~al.}{2013}]{bohringer2013}
{B{\"o}hringer} H.,  {Chon} G.,  {Collins} C.~A.,  {Guzzo} L.,  {Nowak} N.,
  {Bobrovskyi} S.,  2013, \mn@doi [\aap] {10.1051/0004-6361/201220722}, \href
  {https://ui.adsabs.harvard.edu/abs/2013A&A...555A..30B} {555, A30}

\bibitem[\protect\citeauthoryear{{B{\"o}hringer}, {Chon}  \&
  {Collins}}{{B{\"o}hringer} et~al.}{2014}]{bohringer2014}
{B{\"o}hringer} H.,  {Chon} G.,   {Collins} C.~A.,  2014, \mn@doi [\aap]
  {10.1051/0004-6361/201323155}, \href
  {https://ui.adsabs.harvard.edu/abs/2014A&A...570A..31B} {570, A31}

\bibitem[\protect\citeauthoryear{{B{\"o}hringer}, {Chon}  \&
  {Fukugita}}{{B{\"o}hringer} et~al.}{2017}]{bohringer2017}
{B{\"o}hringer} H.,  {Chon} G.,   {Fukugita} M.,  2017, \mn@doi [\aap]
  {10.1051/0004-6361/201731205}, \href
  {https://ui.adsabs.harvard.edu/abs/2017A&A...608A..65B} {608, A65}

\bibitem[\protect\citeauthoryear{{Bond}, {Cole}, {Efstathiou}  \&
  {Kaiser}}{{Bond} et~al.}{1991}]{bond1991}
{Bond} J.~R.,  {Cole} S.,  {Efstathiou} G.,   {Kaiser} N.,  1991, \mn@doi
  [\apj] {10.1086/170520}, \href
  {https://ui.adsabs.harvard.edu/abs/1991ApJ...379..440B} {379, 440}

\bibitem[\protect\citeauthoryear{{Bower}, {Benson}, {Malbon}, {Helly}, {Frenk},
  {Baugh}, {Cole}  \& {Lacey}}{{Bower} et~al.}{2006}]{bower2006}
{Bower} R.~G.,  {Benson} A.~J.,  {Malbon} R.,  {Helly} J.~C.,  {Frenk} C.~S.,
  {Baugh} C.~M.,  {Cole} S.,   {Lacey} C.~G.,  2006, \mn@doi [\mnras]
  {10.1111/j.1365-2966.2006.10519.x}, \href
  {https://ui.adsabs.harvard.edu/abs/2006MNRAS.370..645B} {370, 645}

\bibitem[\protect\citeauthoryear{{Brainerd} \& {Villumsen}}{{Brainerd} \&
  {Villumsen}}{1992}]{brainerd1992}
{Brainerd} T.~G.,  {Villumsen} J.~V.,  1992, \mn@doi [\apj] {10.1086/171593},
  \href {https://ui.adsabs.harvard.edu/abs/1992ApJ...394..409B} {394, 409}

\bibitem[\protect\citeauthoryear{{Brainerd}, {Blandford}  \&
  {Smail}}{{Brainerd} et~al.}{1996}]{brainerd1996}
{Brainerd} T.~G.,  {Blandford} R.~D.,   {Smail} I.,  1996, \mn@doi [\apj]
  {10.1086/177537}, \href
  {https://ui.adsabs.harvard.edu/abs/1996ApJ...466..623B} {466, 623}

\bibitem[\protect\citeauthoryear{{Brimioulle}, {Seitz}, {Lerchster}, {Bender}
  \& {Snigula}}{{Brimioulle} et~al.}{2013}]{brimioulle2013}
{Brimioulle} F.,  {Seitz} S.,  {Lerchster} M.,  {Bender} R.,   {Snigula} J.,
  2013, \mn@doi [\mnras] {10.1093/mnras/stt525}, \href
  {https://ui.adsabs.harvard.edu/abs/2013MNRAS.432.1046B} {432, 1046}

\bibitem[\protect\citeauthoryear{{Caldwell}, {McCarthy}, {Baldry}, {Collins},
  {Schaye}  \& {Bird}}{{Caldwell} et~al.}{2016}]{caldwell2016}
{Caldwell} C.~E.,  {McCarthy} I.~G.,  {Baldry} I.~K.,  {Collins} C.~A.,
  {Schaye} J.,   {Bird} S.,  2016, \mn@doi [\mnras] {10.1093/mnras/stw1892},
  \href {https://ui.adsabs.harvard.edu/abs/2016MNRAS.462.4117C} {462, 4117}

\bibitem[\protect\citeauthoryear{{Chauhan}, {Lagos}, {Stevens}, {Bravo},
  {Rhee}, {Power}, {Obreschkow}  \& {Meyer}}{{Chauhan}
  et~al.}{2021}]{chauhan2021}
{Chauhan} G.,  {Lagos} C. d.~P.,  {Stevens} A. R.~H.,  {Bravo} M.,  {Rhee} J.,
  {Power} C.,  {Obreschkow} D.,   {Meyer} M.,  2021, \mn@doi [\mnras]
  {10.1093/mnras/stab1925}, \href
  {https://ui.adsabs.harvard.edu/abs/2021MNRAS.506.4893C} {506, 4893}

\bibitem[\protect\citeauthoryear{{Colless} et~al.,}{{Colless}
  et~al.}{2001}]{colless2001}
{Colless} M.,  et~al., 2001, \mn@doi [\mnras]
  {10.1046/j.1365-8711.2001.04902.x}, \href
  {https://ui.adsabs.harvard.edu/abs/2001MNRAS.328.1039C} {328, 1039}

\bibitem[\protect\citeauthoryear{{Davies} et~al.,}{{Davies}
  et~al.}{2018}]{davies2018}
{Davies} L.~J.~M.,  et~al., 2018, \mn@doi [\mnras] {10.1093/mnras/sty1553},
  \href {https://ui.adsabs.harvard.edu/abs/2018MNRAS.480..768D} {480, 768}

\bibitem[\protect\citeauthoryear{{Dey} et~al.,}{{Dey} et~al.}{2019}]{desi}
{Dey} A.,  et~al., 2019, \mn@doi [\aj] {10.3847/1538-3881/ab089d}, \href
  {https://ui.adsabs.harvard.edu/abs/2019AJ....157..168D} {157, 168}

\bibitem[\protect\citeauthoryear{{Doi}, {Fukugita}, {Okamura}  \&
  {Turner}}{{Doi} et~al.}{1995}]{doi1995}
{Doi} M.,  {Fukugita} M.,  {Okamura} S.,   {Turner} E.~L.,  1995, \mn@doi [\aj]
  {10.1086/117379}, \href
  {https://ui.adsabs.harvard.edu/abs/1995AJ....109.1490D} {109, 1490}

\bibitem[\protect\citeauthoryear{{Dong}, {Zhang}, {Yang}, {Zhang}  \&
  {Luo}}{{Dong} et~al.}{2019}]{dong2019}
{Dong} F.,  {Zhang} J.,  {Yang} X.,  {Zhang} J.,   {Luo} W.,  2019, \mn@doi
  [\apj] {10.3847/1538-4357/ab3a9d}, \href
  {https://ui.adsabs.harvard.edu/abs/2019ApJ...883..155D} {883, 155}

\bibitem[\protect\citeauthoryear{{Driver}}{{Driver}}{2021}]{driver2021}
{Driver} S.,  2021, \mn@doi [Nature Astronomy] {10.1038/s41550-021-01441-w},
  \href {https://ui.adsabs.harvard.edu/abs/2021NatAs.tmp..144D} {}

\bibitem[\protect\citeauthoryear{{Driver} et~al.,}{{Driver}
  et~al.}{2009}]{driver2009}
{Driver} S.~P.,  et~al., 2009, \mn@doi [Astronomy and Geophysics]
  {10.1111/j.1468-4004.2009.50512.x}, \href
  {https://ui.adsabs.harvard.edu/abs/2009A&G....50e..12D} {50, 5.12}

\bibitem[\protect\citeauthoryear{{Driver} et~al.,}{{Driver}
  et~al.}{2011}]{driver2011}
{Driver} S.~P.,  et~al., 2011, \mn@doi [\mnras]
  {10.1111/j.1365-2966.2010.18188.x}, \href
  {https://ui.adsabs.harvard.edu/abs/2011MNRAS.413..971D} {413, 971}

\bibitem[\protect\citeauthoryear{{Driver}, {Davies}, {Meyer}, {Power},
  {Robotham}, {Baldry}, {Liske}  \& {Norberg}}{{Driver}
  et~al.}{2016}]{driver2016}
{Driver} S.~P.,  {Davies} L.~J.,  {Meyer} M.,  {Power} C.,  {Robotham}
  A.~S.~G.,  {Baldry} I.~K.,  {Liske} J.,   {Norberg} P.,  2016, in
  {Napolitano} N.~R.,  {Longo} G.,  {Marconi} M.,  {Paolillo} M.,   {Iodice}
  E.,  eds, The Universe of Digital Sky Surveys.

\bibitem[\protect\citeauthoryear{{Driver} et~al.,}{{Driver}
  et~al.}{2019}]{driver2019}
{Driver} S.~P.,  et~al., 2019, \mn@doi [The Messenger]
  {10.18727/0722-6691/5126}, \href
  {https://ui.adsabs.harvard.edu/abs/2019Msngr.175...46D} {175, 46}

\bibitem[\protect\citeauthoryear{{Duarte} \& {Mamon}}{{Duarte} \&
  {Mamon}}{2014}]{duarte2014}
{Duarte} M.,  {Mamon} G.~A.,  2014, \mn@doi [\mnras] {10.1093/mnras/stu378},
  \href {https://ui.adsabs.harvard.edu/abs/2014MNRAS.440.1763D} {440, 1763}

\bibitem[\protect\citeauthoryear{{Eke} et~al.,}{{Eke} et~al.}{2004}]{eke2004}
{Eke} V.~R.,  et~al., 2004, \mn@doi [\mnras]
  {10.1111/j.1365-2966.2004.07408.x}, \href
  {https://ui.adsabs.harvard.edu/abs/2004MNRAS.348..866E} {348, 866}

\bibitem[\protect\citeauthoryear{{Eke}, {Baugh}, {Cole}, {Frenk}  \&
  {Navarro}}{{Eke} et~al.}{2006}]{eke2006}
{Eke} V.~R.,  {Baugh} C.~M.,  {Cole} S.,  {Frenk} C.~S.,   {Navarro} J.~F.,
  2006, \mn@doi [\mnras] {10.1111/j.1365-2966.2006.10568.x}, \href
  {https://ui.adsabs.harvard.edu/abs/2006MNRAS.370.1147E} {370, 1147}

\bibitem[\protect\citeauthoryear{{Elahi}, {Power}, {Lagos}, {Poulton}  \&
  {Robotham}}{{Elahi} et~al.}{2018}]{elahi2018}
{Elahi} P.~J.,  {Power} C.,  {Lagos} C. d.~P.,  {Poulton} R.,   {Robotham} A.
  S.~G.,  2018, \mn@doi [\mnras] {10.1093/mnras/sty590}, \href
  {https://ui.adsabs.harvard.edu/abs/2018MNRAS.477..616E} {477, 616}

\bibitem[\protect\citeauthoryear{{Frenk}, {White}, {Davis}  \&
  {Efstathiou}}{{Frenk} et~al.}{1988}]{frenk1988}
{Frenk} C.~S.,  {White} S. D.~M.,  {Davis} M.,   {Efstathiou} G.,  1988,
  \mn@doi [\apj] {10.1086/166213}, \href
  {https://ui.adsabs.harvard.edu/abs/1988ApJ...327..507F} {327, 507}

\bibitem[\protect\citeauthoryear{{Han} et~al.,}{{Han} et~al.}{2015}]{han2015}
{Han} J.,  et~al., 2015, \mn@doi [\mnras] {10.1093/mnras/stu2178}, \href
  {https://ui.adsabs.harvard.edu/abs/2015MNRAS.446.1356H} {446, 1356}

\bibitem[\protect\citeauthoryear{{Hoekstra}, {Donahue}, {Conselice}, {McNamara}
   \& {Voit}}{{Hoekstra} et~al.}{2011}]{hoekstra2011}
{Hoekstra} H.,  {Donahue} M.,  {Conselice} C.~J.,  {McNamara} B.~R.,   {Voit}
  G.~M.,  2011, \mn@doi [\apj] {10.1088/0004-637X/726/1/48}, \href
  {https://ui.adsabs.harvard.edu/abs/2011ApJ...726...48H} {726, 48}

\bibitem[\protect\citeauthoryear{{Jee}, {Hughes}, {Menanteau}, {Sif{\'o}n},
  {Mandelbaum}, {Barrientos}, {Infante}  \& {Ng}}{{Jee} et~al.}{2014}]{elgordo}
{Jee} M.~J.,  {Hughes} J.~P.,  {Menanteau} F.,  {Sif{\'o}n} C.,  {Mandelbaum}
  R.,  {Barrientos} L.~F.,  {Infante} L.,   {Ng} K.~Y.,  2014, \mn@doi [\apj]
  {10.1088/0004-637X/785/1/20}, \href
  {https://ui.adsabs.harvard.edu/abs/2014ApJ...785...20J} {785, 20}

\bibitem[\protect\citeauthoryear{{Jenkins}, {Frenk}, {White}, {Colberg},
  {Cole}, {Evrard}, {Couchman}  \& {Yoshida}}{{Jenkins}
  et~al.}{2001}]{jenkins2001}
{Jenkins} A.,  {Frenk} C.~S.,  {White} S.~D.~M.,  {Colberg} J.~M.,  {Cole} S.,
  {Evrard} A.~E.,  {Couchman} H.~M.~P.,   {Yoshida} N.,  2001, \mn@doi [\mnras]
  {10.1046/j.1365-8711.2001.04029.x}, \href
  {https://ui.adsabs.harvard.edu/abs/2001MNRAS.321..372J} {321, 372}

\bibitem[\protect\citeauthoryear{{Knebe} et~al.,}{{Knebe}
  et~al.}{2013}]{knebe2013}
{Knebe} A.,  et~al., 2013, \mn@doi [\mnras] {10.1093/mnras/stt1403}, \href
  {https://ui.adsabs.harvard.edu/abs/2013MNRAS.435.1618K} {435, 1618}

\bibitem[\protect\citeauthoryear{{Kubo}, {Stebbins}, {Annis}, {Dell'Antonio},
  {Lin}, {Khiabanian}  \& {Frieman}}{{Kubo} et~al.}{2007}]{kubo2007}
{Kubo} J.~M.,  {Stebbins} A.,  {Annis} J.,  {Dell'Antonio} I.~P.,  {Lin} H.,
  {Khiabanian} H.,   {Frieman} J.~A.,  2007, \mn@doi [\apj] {10.1086/523101},
  \href {https://ui.adsabs.harvard.edu/abs/2007ApJ...671.1466K} {671, 1466}

\bibitem[\protect\citeauthoryear{{Lagos}, {Tobar}, {Robotham}, {Obreschkow},
  {Mitchell}, {Power}  \& {Elahi}}{{Lagos} et~al.}{2018}]{lagos2018}
{Lagos} C. d.~P.,  {Tobar} R.~J.,  {Robotham} A. S.~G.,  {Obreschkow} D.,
  {Mitchell} P.~D.,  {Power} C.,   {Elahi} P.~J.,  2018, \mn@doi [\mnras]
  {10.1093/mnras/sty2440}, \href
  {https://ui.adsabs.harvard.edu/abs/2018MNRAS.481.3573L} {481, 3573}

\bibitem[\protect\citeauthoryear{{Leauthaud} et~al.,}{{Leauthaud}
  et~al.}{2010}]{leauthaud2010}
{Leauthaud} A.,  et~al., 2010, \mn@doi [\apj] {10.1088/0004-637X/709/1/97},
  \href {https://ui.adsabs.harvard.edu/abs/2010ApJ...709...97L} {709, 97}

\bibitem[\protect\citeauthoryear{{Liske} et~al.,}{{Liske}
  et~al.}{2015}]{liske2015}
{Liske} J.,  et~al., 2015, \mn@doi [\mnras] {10.1093/mnras/stv1436}, \href
  {https://ui.adsabs.harvard.edu/abs/2015MNRAS.452.2087L} {452, 2087}

\bibitem[\protect\citeauthoryear{{Luki{\'c}}, {Heitmann}, {Habib}, {Bashinsky}
  \& {Ricker}}{{Luki{\'c}} et~al.}{2007}]{lukic2007}
{Luki{\'c}} Z.,  {Heitmann} K.,  {Habib} S.,  {Bashinsky} S.,   {Ricker} P.~M.,
   2007, \mn@doi [\apj] {10.1086/523083}, \href
  {https://ui.adsabs.harvard.edu/abs/2007ApJ...671.1160L} {671, 1160}

\bibitem[\protect\citeauthoryear{{Merson} et~al.,}{{Merson}
  et~al.}{2013}]{merson2013}
{Merson} A.~I.,  et~al., 2013, \mn@doi [\mnras] {10.1093/mnras/sts355}, \href
  {https://ui.adsabs.harvard.edu/abs/2013MNRAS.429..556M} {429, 556}

\bibitem[\protect\citeauthoryear{{Murray}, {Power}  \& {Robotham}}{{Murray}
  et~al.}{2013}]{murray2013}
{Murray} S.~G.,  {Power} C.,   {Robotham} A.~S.~G.,  2013, \mn@doi [\mnras]
  {10.1093/mnrasl/slt079}, \href
  {https://ui.adsabs.harvard.edu/abs/2013MNRAS.434L..61M} {434, L61}

\bibitem[\protect\citeauthoryear{{Murray}, {Robotham}  \& {Power}}{{Murray}
  et~al.}{2018}]{murray2018}
{Murray} S.~G.,  {Robotham} A.~S.~G.,   {Power} C.,  2018, \mn@doi [\apj]
  {10.3847/1538-4357/aaa552}, \href
  {https://ui.adsabs.harvard.edu/abs/2018ApJ...855....5M} {855, 5}

\bibitem[\protect\citeauthoryear{{Murray}, {Diemer}, {Chen}, {Neuhold},
  {Schnapp}, {Peruzzi}, {Blevins}  \& {Engelman}}{{Murray}
  et~al.}{2021}]{murray2021}
{Murray} S.~G.,  {Diemer} B.,  {Chen} Z.,  {Neuhold} A.~G.,  {Schnapp} M.~A.,
  {Peruzzi} T.,  {Blevins} D.,   {Engelman} T.,  2021, \mn@doi [Astronomy and
  Computing] {10.1016/j.ascom.2021.100487}, \href
  {https://ui.adsabs.harvard.edu/abs/2021A&C....3600487M} {36, 100487}

\bibitem[\protect\citeauthoryear{{Navarro}, {Frenk}  \& {White}}{{Navarro}
  et~al.}{1997}]{NFW1997}
{Navarro} J.~F.,  {Frenk} C.~S.,   {White} S. D.~M.,  1997, \mn@doi [\apj]
  {10.1086/304888}, \href
  {https://ui.adsabs.harvard.edu/abs/1997ApJ...490..493N} {490, 493}

\bibitem[\protect\citeauthoryear{{Old} et~al.,}{{Old} et~al.}{2014}]{old2014}
{Old} L.,  et~al., 2014, \mn@doi [\mnras] {10.1093/mnras/stu545}, \href
  {https://ui.adsabs.harvard.edu/abs/2014MNRAS.441.1513O} {441, 1513}

\bibitem[\protect\citeauthoryear{{Old} et~al.,}{{Old} et~al.}{2018}]{old2018}
{Old} L.,  et~al., 2018, \mn@doi [\mnras] {10.1093/mnras/stx3241}, \href
  {https://ui.adsabs.harvard.edu/abs/2018MNRAS.475..853O} {475, 853}

\bibitem[\protect\citeauthoryear{{Ondaro-Mallea}, {Angulo}, {Zennaro},
  {Contreras}  \& {Aric{\`o}}}{{Ondaro-Mallea}
  et~al.}{2021}]{ondaro-meallea2021}
{Ondaro-Mallea} L.,  {Angulo} R.~E.,  {Zennaro} M.,  {Contreras} S.,
  {Aric{\`o}} G.,  2021, arXiv e-prints, \href
  {https://ui.adsabs.harvard.edu/abs/2021arXiv210208958O} {p. arXiv:2102.08958}

\bibitem[\protect\citeauthoryear{{Peacock} \& {Heavens}}{{Peacock} \&
  {Heavens}}{1990}]{peacock1990}
{Peacock} J.~A.,  {Heavens} A.~F.,  1990, \mn@doi [\mnras]
  {10.1093/mnras/243.1.133}, \href
  {https://ui.adsabs.harvard.edu/abs/1990MNRAS.243..133P} {243, 133}

\bibitem[\protect\citeauthoryear{{Planck Collaboration} et~al.,}{{Planck
  Collaboration} et~al.}{2020}]{planck2018}
{Planck Collaboration} et~al., 2020, \mn@doi [\aap]
  {10.1051/0004-6361/201833910}, \href
  {https://ui.adsabs.harvard.edu/abs/2020A&A...641A...6P} {641, A6}

\bibitem[\protect\citeauthoryear{{Press} \& {Schechter}}{{Press} \&
  {Schechter}}{1974}]{press1974}
{Press} W.~H.,  {Schechter} P.,  1974, \mn@doi [\apj] {10.1086/152650}, \href
  {https://ui.adsabs.harvard.edu/abs/1974ApJ...187..425P} {187, 425}

\bibitem[\protect\citeauthoryear{{Rana}, {More}, {Miyatake}, {Nishimichi},
  {Takada}, {Robotham}, {Hopkins}  \& {Holwerda}}{{Rana}
  et~al.}{2021}]{rana2021}
{Rana} D.,  {More} S.,  {Miyatake} H.,  {Nishimichi} T.,  {Takada} M.,
  {Robotham} A. S.~G.,  {Hopkins} A.~M.,   {Holwerda} B.~W.,  2021, arXiv
  e-prints, \href {https://ui.adsabs.harvard.edu/abs/2021arXiv210705641R} {p.
  arXiv:2107.05641}

\bibitem[\protect\citeauthoryear{{Reed}, {Gardner}, {Quinn}, {Stadel},
  {Fardal}, {Lake}  \& {Governato}}{{Reed} et~al.}{2003}]{reed2003}
{Reed} D.,  {Gardner} J.,  {Quinn} T.,  {Stadel} J.,  {Fardal} M.,  {Lake} G.,
   {Governato} F.,  2003, \mn@doi [\mnras] {10.1046/j.1365-2966.2003.07113.x},
  \href {https://ui.adsabs.harvard.edu/abs/2003MNRAS.346..565R} {346, 565}

\bibitem[\protect\citeauthoryear{{Reed}, {Bower}, {Frenk}, {Jenkins}  \&
  {Theuns}}{{Reed} et~al.}{2007}]{reed2007}
{Reed} D.~S.,  {Bower} R.,  {Frenk} C.~S.,  {Jenkins} A.,   {Theuns} T.,  2007,
  \mn@doi [\mnras] {10.1111/j.1365-2966.2006.11204.x}, \href
  {https://ui.adsabs.harvard.edu/abs/2007MNRAS.374....2R} {374, 2}

\bibitem[\protect\citeauthoryear{{Robotham}}{{Robotham}}{2016a}]{celestial}
{Robotham} A. S.~G.,  2016a, {Celestial: Common astronomical conversion
  routines and functions} (\mn@eprint {ascl} {1602.011})

\bibitem[\protect\citeauthoryear{{Robotham}}{{Robotham}}{2016b}]{magicaxis}
{Robotham} A. S.~G.,  2016b, {magicaxis: Pretty scientific plotting with
  minor-tick and log minor-tick support} (\mn@eprint {ascl} {1604.004})

\bibitem[\protect\citeauthoryear{{Robotham} et~al.,}{{Robotham}
  et~al.}{2011}]{robotham2011}
{Robotham} A.~S.~G.,  et~al., 2011, \mn@doi [\mnras]
  {10.1111/j.1365-2966.2011.19217.x}, \href
  {https://ui.adsabs.harvard.edu/abs/2011MNRAS.416.2640R} {416, 2640}

\bibitem[\protect\citeauthoryear{{Schechter}}{{Schechter}}{1976}]{schechter1976}
{Schechter} P.,  1976, \mn@doi [\apj] {10.1086/154079}, \href
  {https://ui.adsabs.harvard.edu/abs/1976ApJ...203..297S} {203, 297}

\bibitem[\protect\citeauthoryear{{Singari}, {Ghosh}  \& {Khatri}}{{Singari}
  et~al.}{2020}]{singari2020}
{Singari} B.,  {Ghosh} T.,   {Khatri} R.,  2020, \mn@doi [\jcap]
  {10.1088/1475-7516/2020/08/028}, \href
  {https://ui.adsabs.harvard.edu/abs/2020JCAP...08..028S} {2020, 028}

\bibitem[\protect\citeauthoryear{{Sohn}, {Geller}, {Zahid}, {Fabricant},
  {Diaferio}  \& {Rines}}{{Sohn} et~al.}{2017}]{sohn2017}
{Sohn} J.,  {Geller} M.~J.,  {Zahid} H.~J.,  {Fabricant} D.~G.,  {Diaferio} A.,
    {Rines} K.~J.,  2017, \mn@doi [\apjs] {10.3847/1538-4365/aa653e}, \href
  {https://ui.adsabs.harvard.edu/abs/2017ApJS..229...20S} {229, 20}

\bibitem[\protect\citeauthoryear{{Springel} et~al.,}{{Springel}
  et~al.}{2005}]{springel2005}
{Springel} V.,  et~al., 2005, \mn@doi [\nat] {10.1038/nature03597}, \href
  {https://ui.adsabs.harvard.edu/abs/2005Natur.435..629S} {435, 629}

\bibitem[\protect\citeauthoryear{{Stanek}, {Evrard}, {B{\"o}hringer},
  {Schuecker}  \& {Nord}}{{Stanek} et~al.}{2006}]{stanek2006}
{Stanek} R.,  {Evrard} A.~E.,  {B{\"o}hringer} H.,  {Schuecker} P.,   {Nord}
  B.,  2006, \mn@doi [\apj] {10.1086/506248}, \href
  {https://ui.adsabs.harvard.edu/abs/2006ApJ...648..956S} {648, 956}

\bibitem[\protect\citeauthoryear{{Strigari}, {Bullock}, {Kaplinghat},
  {Diemand}, {Kuhlen}  \& {Madau}}{{Strigari} et~al.}{2007}]{strigari2007}
{Strigari} L.~E.,  {Bullock} J.~S.,  {Kaplinghat} M.,  {Diemand} J.,  {Kuhlen}
  M.,   {Madau} P.,  2007, \mn@doi [\apj] {10.1086/521914}, \href
  {https://ui.adsabs.harvard.edu/abs/2007ApJ...669..676S} {669, 676}

\bibitem[\protect\citeauthoryear{{Tempel} et~al.,}{{Tempel}
  et~al.}{2014}]{tempel2014}
{Tempel} E.,  et~al., 2014, \mn@doi [\aap] {10.1051/0004-6361/201423585}, \href
  {https://ui.adsabs.harvard.edu/abs/2014A&A...566A...1T} {566, A1}

\bibitem[\protect\citeauthoryear{{Tempel}, {Tuvikene}, {Kipper}  \&
  {Libeskind}}{{Tempel} et~al.}{2017}]{tempel2017}
{Tempel} E.,  {Tuvikene} T.,  {Kipper} R.,   {Libeskind} N.~I.,  2017, \mn@doi
  [\aap] {10.1051/0004-6361/201730499}, \href
  {https://ui.adsabs.harvard.edu/abs/2017A&A...602A.100T} {602, A100}

\bibitem[\protect\citeauthoryear{{Tempel}, {Kruuse}, {Kipper}, {Tuvikene},
  {Sorce}  \& {Stoica}}{{Tempel} et~al.}{2018}]{tempel2018}
{Tempel} E.,  {Kruuse} M.,  {Kipper} R.,  {Tuvikene} T.,  {Sorce} J.~G.,
  {Stoica} R.~S.,  2018, \mn@doi [\aap] {10.1051/0004-6361/201833217}, \href
  {https://ui.adsabs.harvard.edu/abs/2018A&A...618A..81T} {618, A81}

\bibitem[\protect\citeauthoryear{{Trevisan} \& {Mamon}}{{Trevisan} \&
  {Mamon}}{2017}]{trevisan2017}
{Trevisan} M.,  {Mamon} G.~A.,  2017, \mn@doi [\mnras] {10.1093/mnras/stx1656},
  \href {https://ui.adsabs.harvard.edu/abs/2017MNRAS.471.2022T} {471, 2022}

\bibitem[\protect\citeauthoryear{{Tucker}, {Walker}, {Mateo}, {Olszewski},
  {Geringer-Sameth}  \& {Miller}}{{Tucker} et~al.}{2020}]{tucker2020}
{Tucker} E.,  {Walker} M.~G.,  {Mateo} M.,  {Olszewski} E.~W.,
  {Geringer-Sameth} A.,   {Miller} C.~J.,  2020, \mn@doi [\apj]
  {10.3847/1538-4357/ab609d}, \href
  {https://ui.adsabs.harvard.edu/abs/2020ApJ...888..106T} {888, 106}

\bibitem[\protect\citeauthoryear{{Viola} et~al.,}{{Viola}
  et~al.}{2015}]{viola2015}
{Viola} M.,  et~al., 2015, \mn@doi [\mnras] {10.1093/mnras/stv1447}, \href
  {https://ui.adsabs.harvard.edu/abs/2015MNRAS.452.3529V} {452, 3529}

\bibitem[\protect\citeauthoryear{{Watson}, {Iliev}, {D'Aloisio}, {Knebe},
  {Shapiro}  \& {Yepes}}{{Watson} et~al.}{2013}]{watson2013}
{Watson} W.~A.,  {Iliev} I.~T.,  {D'Aloisio} A.,  {Knebe} A.,  {Shapiro} P.~R.,
    {Yepes} G.,  2013, \mn@doi [\mnras] {10.1093/mnras/stt791}, \href
  {https://ui.adsabs.harvard.edu/abs/2013MNRAS.433.1230W} {433, 1230}

\bibitem[\protect\citeauthoryear{{Wechsler} \& {Tinker}}{{Wechsler} \&
  {Tinker}}{2018}]{weschler2018}
{Wechsler} R.~H.,  {Tinker} J.~L.,  2018, \mn@doi [\araa]
  {10.1146/annurev-astro-081817-051756}, \href
  {https://ui.adsabs.harvard.edu/abs/2018ARA&A..56..435W} {56, 435}

\bibitem[\protect\citeauthoryear{{York} et~al.,}{{York} et~al.}{2000}]{sdss}
{York} D.~G.,  et~al., 2000, \mn@doi [\aj] {10.1086/301513}, \href
  {https://ui.adsabs.harvard.edu/abs/2000AJ....120.1579Y} {120, 1579}

\bibitem[\protect\citeauthoryear{{Zwicky}}{{Zwicky}}{1933}]{zwicky1933}
{Zwicky} F.,  1933, Helvetica Physica Acta, \href
  {https://ui.adsabs.harvard.edu/abs/1933AcHPh...6..110Z} {6, 110}

\makeatother
\end{thebibliography}

\section*{Appendix A}
All of the HMFs generated in this work are summarised in Fig.\,\ref{fig:perturb}.

\begin{figure*}
	\centering     
	\includegraphics[width=\columnwidth]{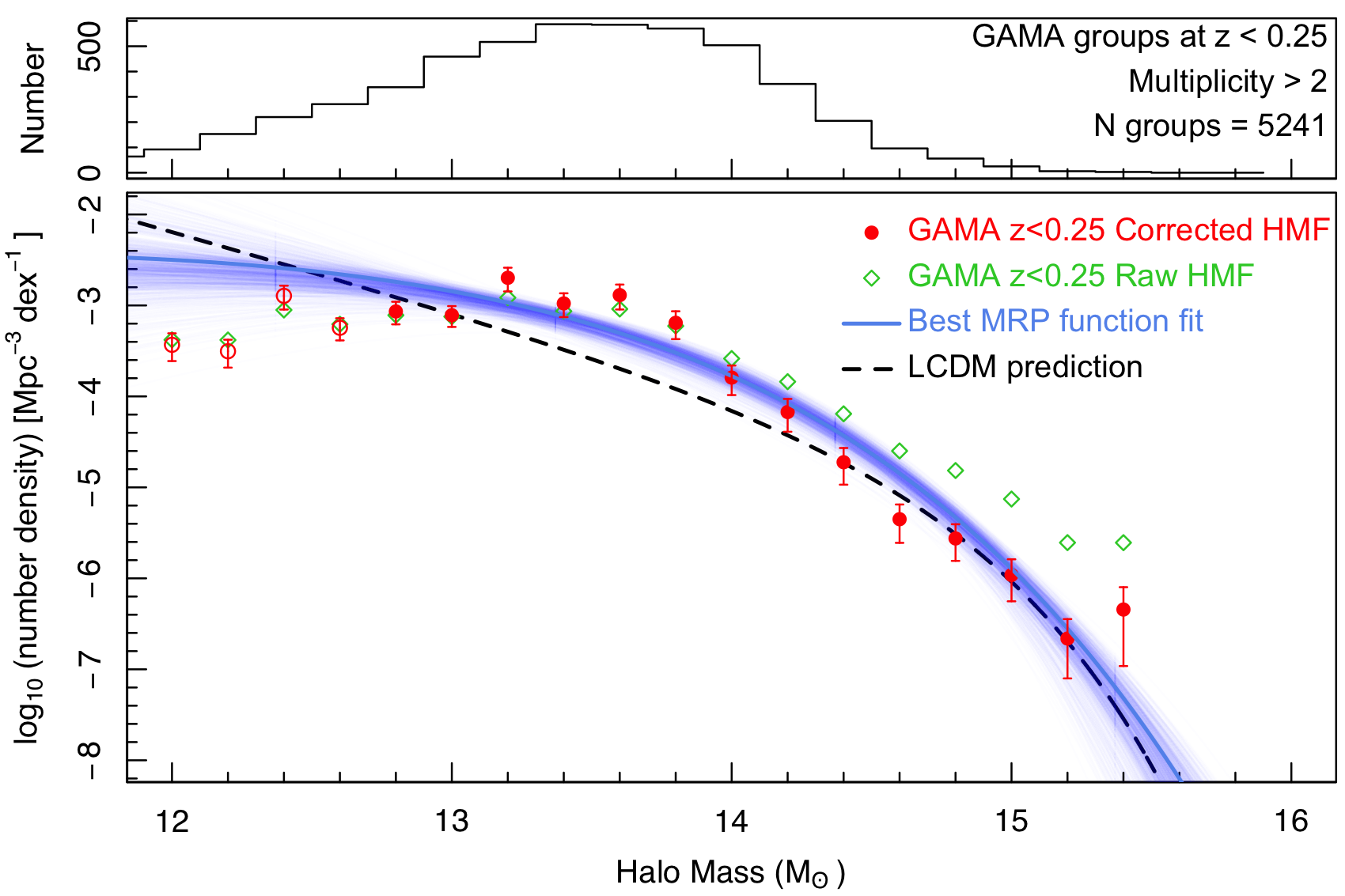}
	\includegraphics[width=\columnwidth]{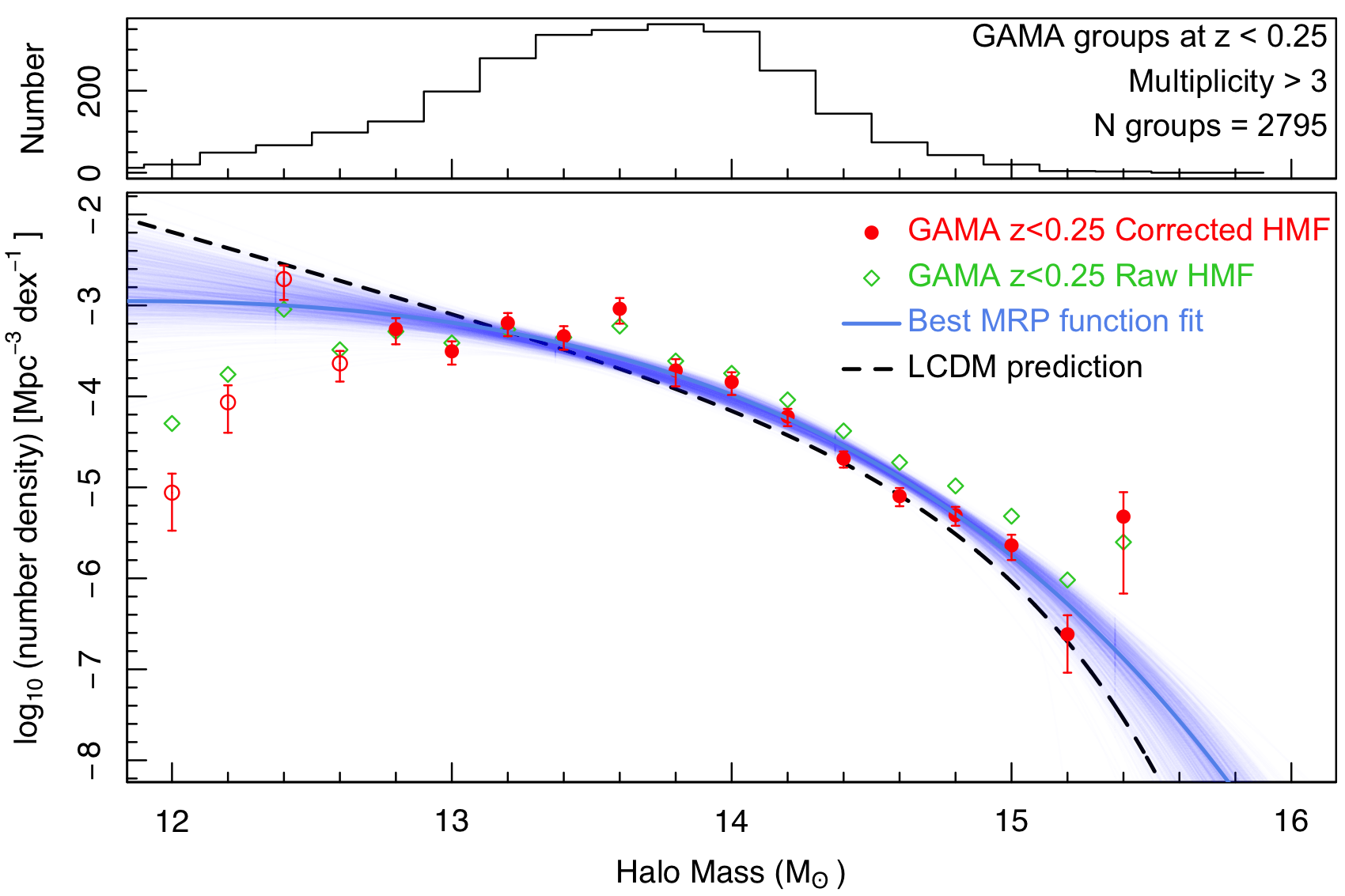}
	\includegraphics[width=\columnwidth]{figures/gamahmfGAMA5.png}
	\includegraphics[width=\columnwidth]{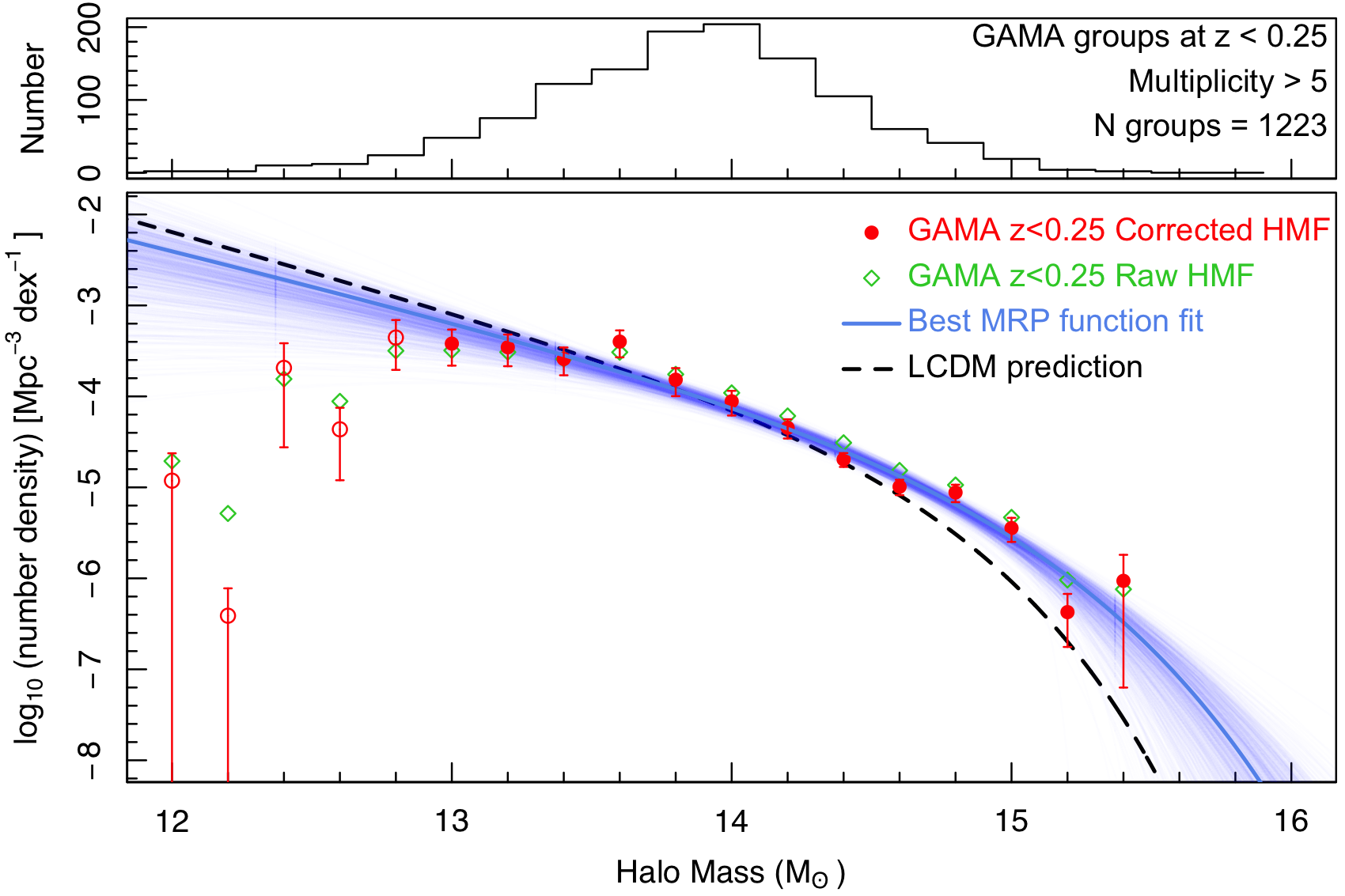}
	\includegraphics[width=\columnwidth]{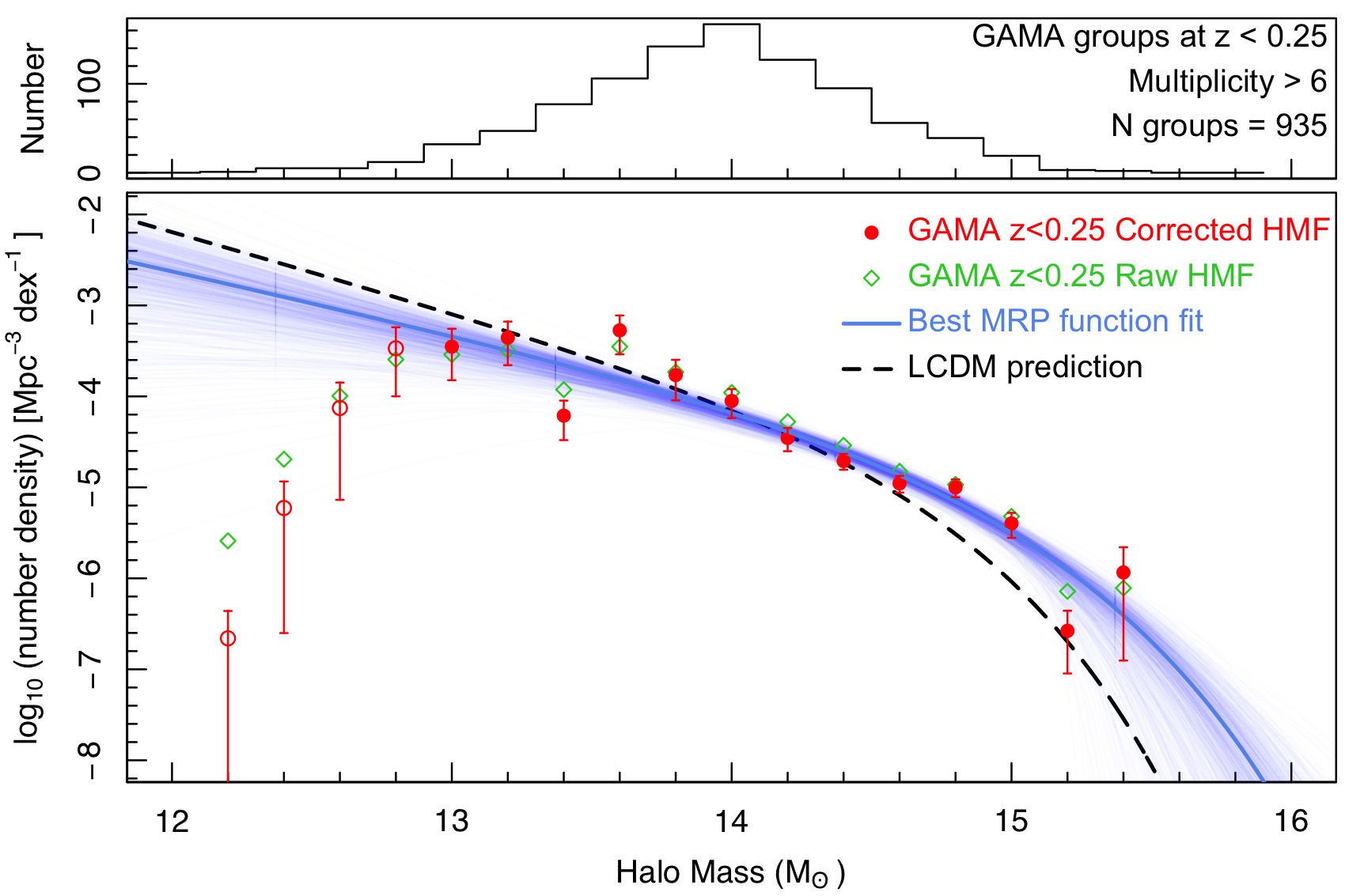}
	\includegraphics[width=\columnwidth]{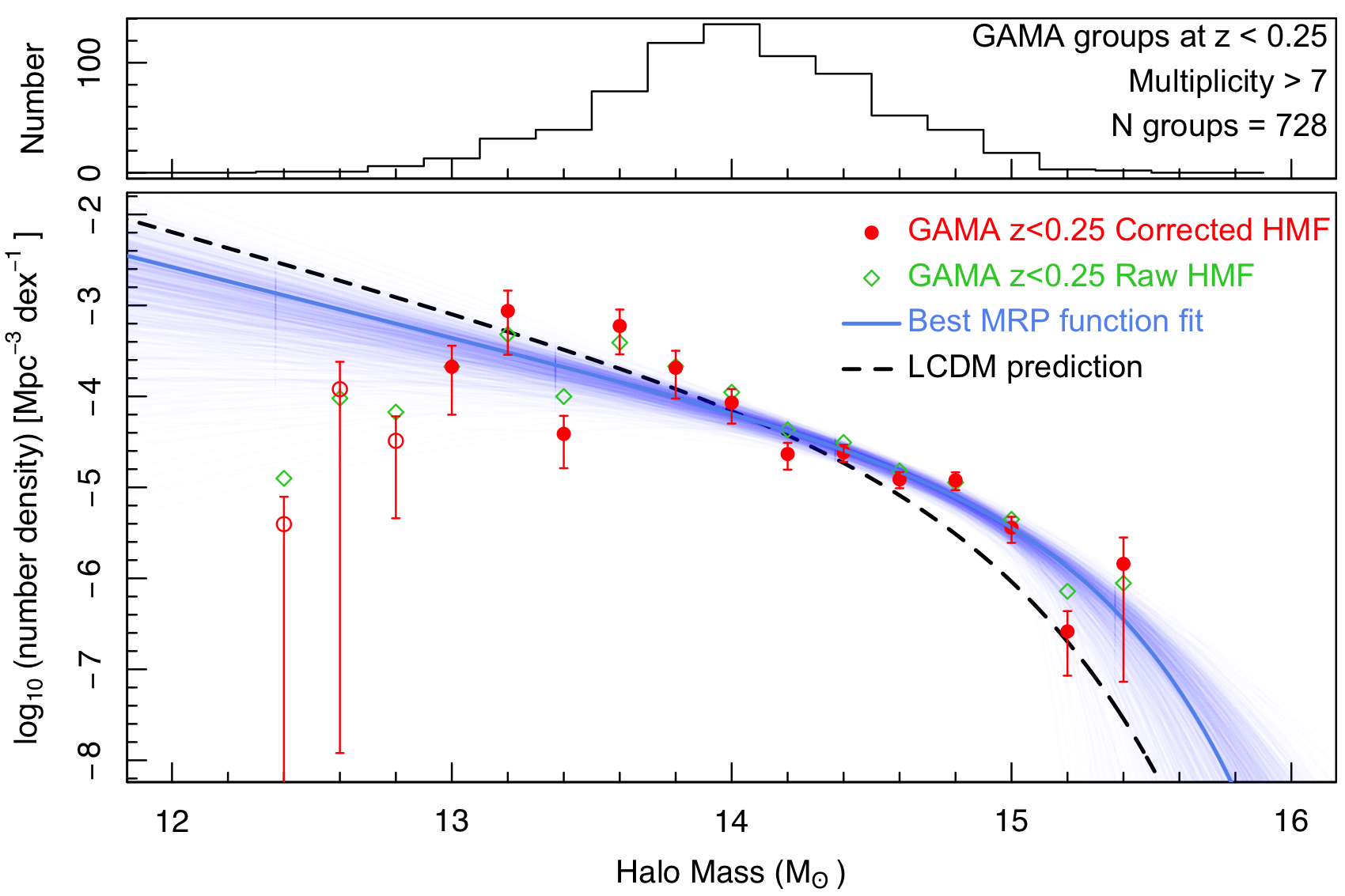}
	\caption{GAMA HMF derivations for different multiplicity cuts. \label{fig:appendix1}}
\end{figure*}

\begin{figure*}
	\centering     
	\includegraphics[width=\columnwidth]{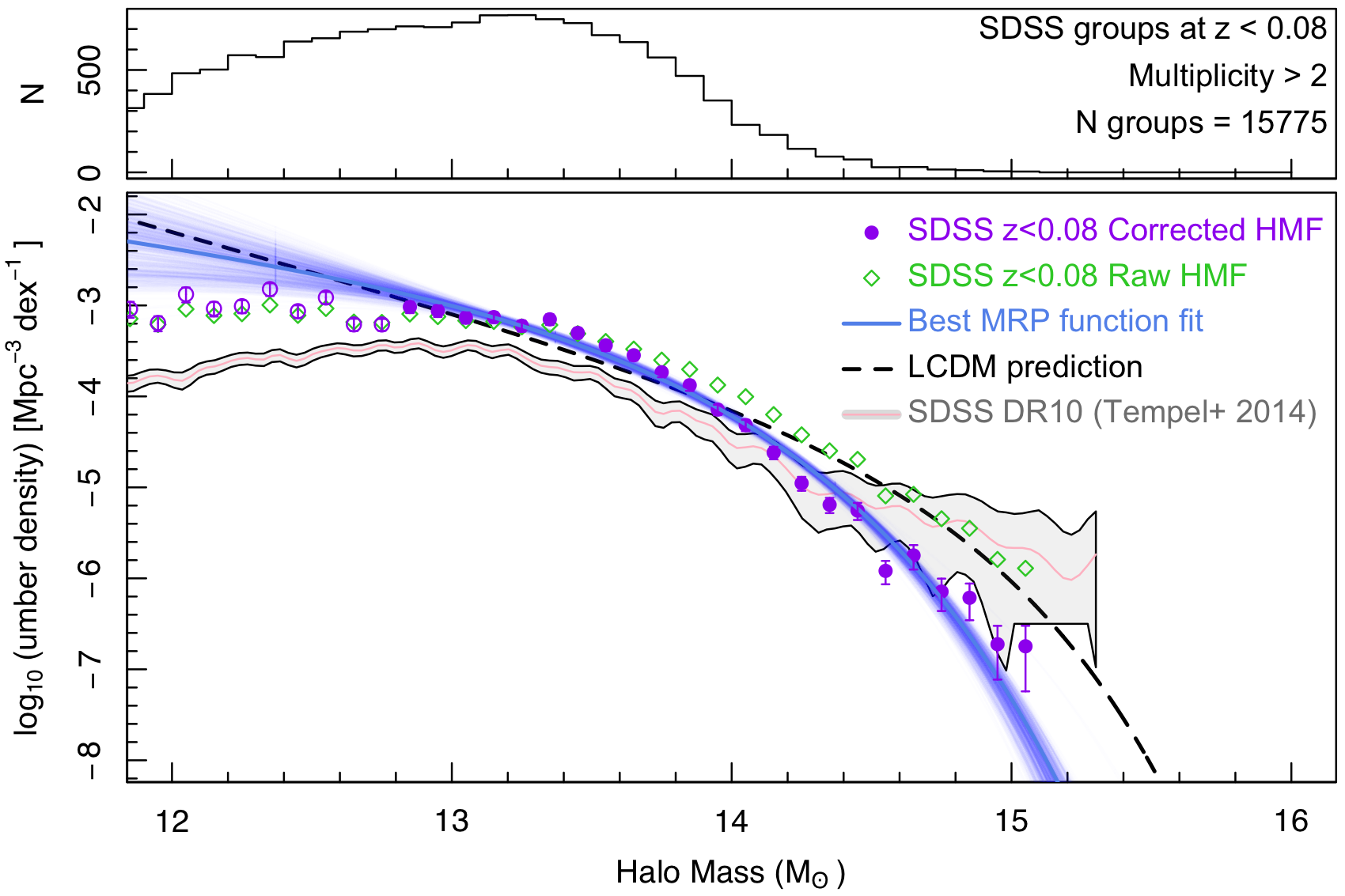}
	\includegraphics[width=\columnwidth]{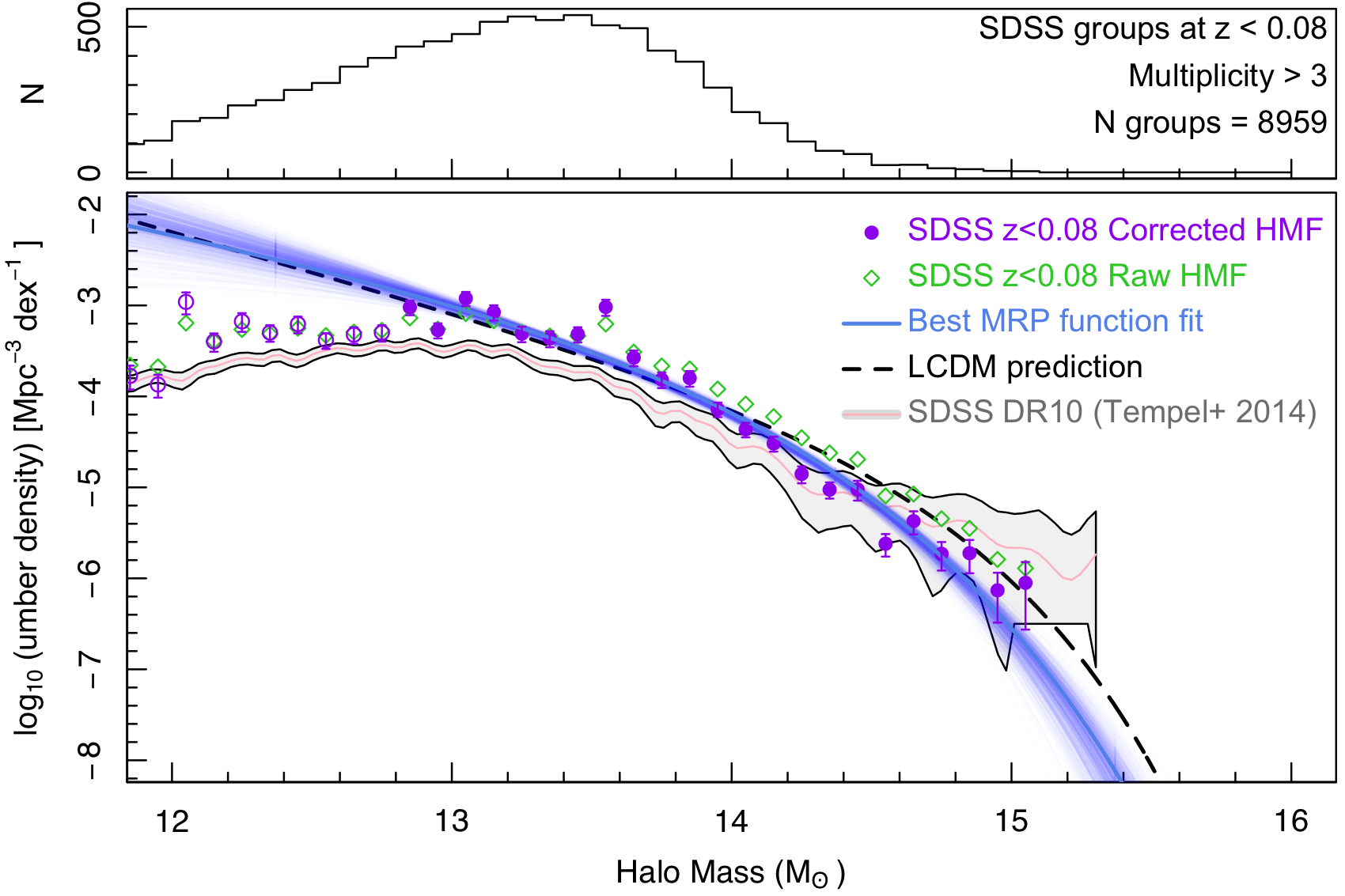}
	\includegraphics[width=\columnwidth]{figures/sdsshmf4.png}
	\includegraphics[width=\columnwidth]{figures/sdsshmf5.png}
	\includegraphics[width=\columnwidth]{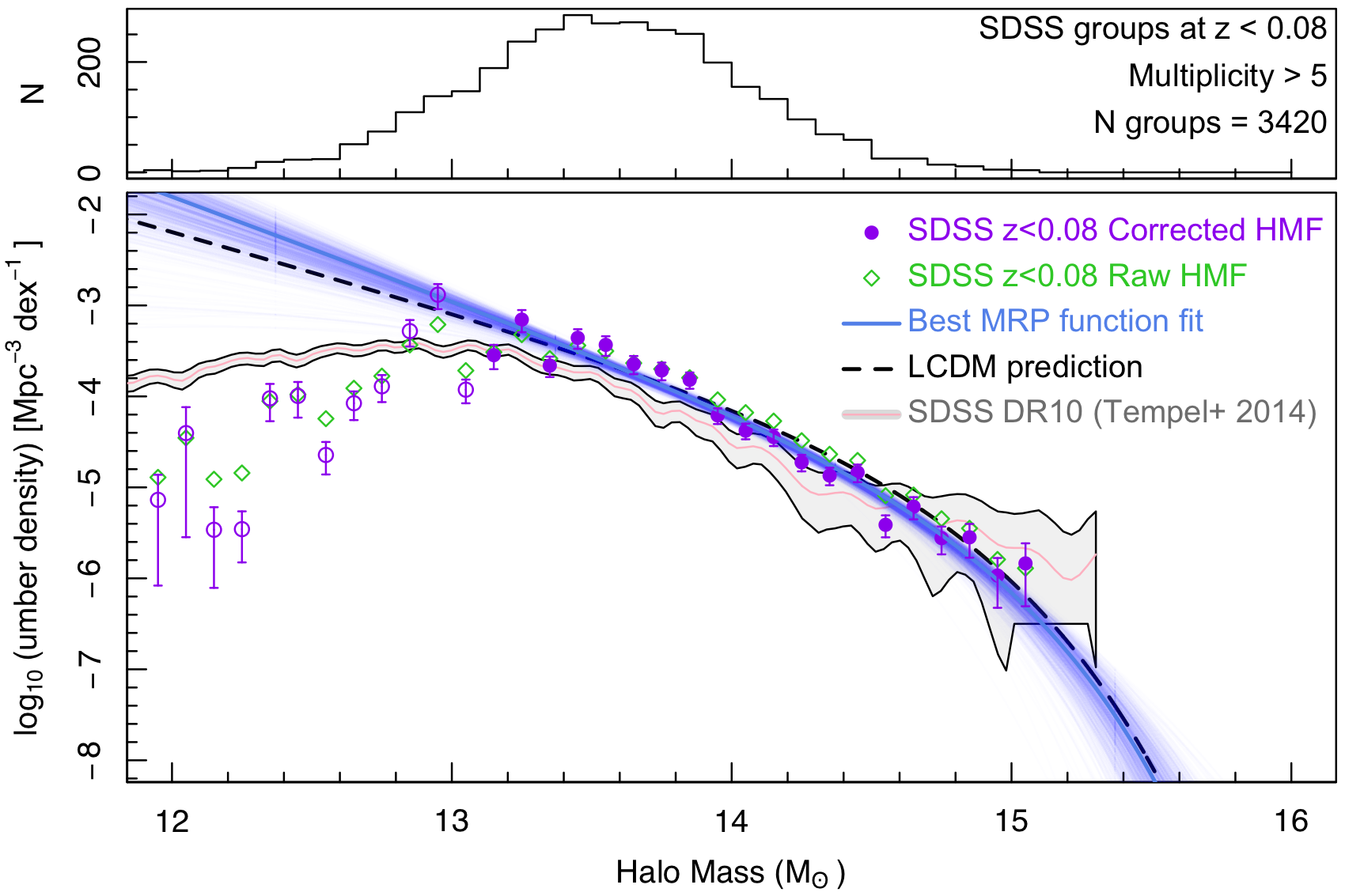}
	\includegraphics[width=\columnwidth]{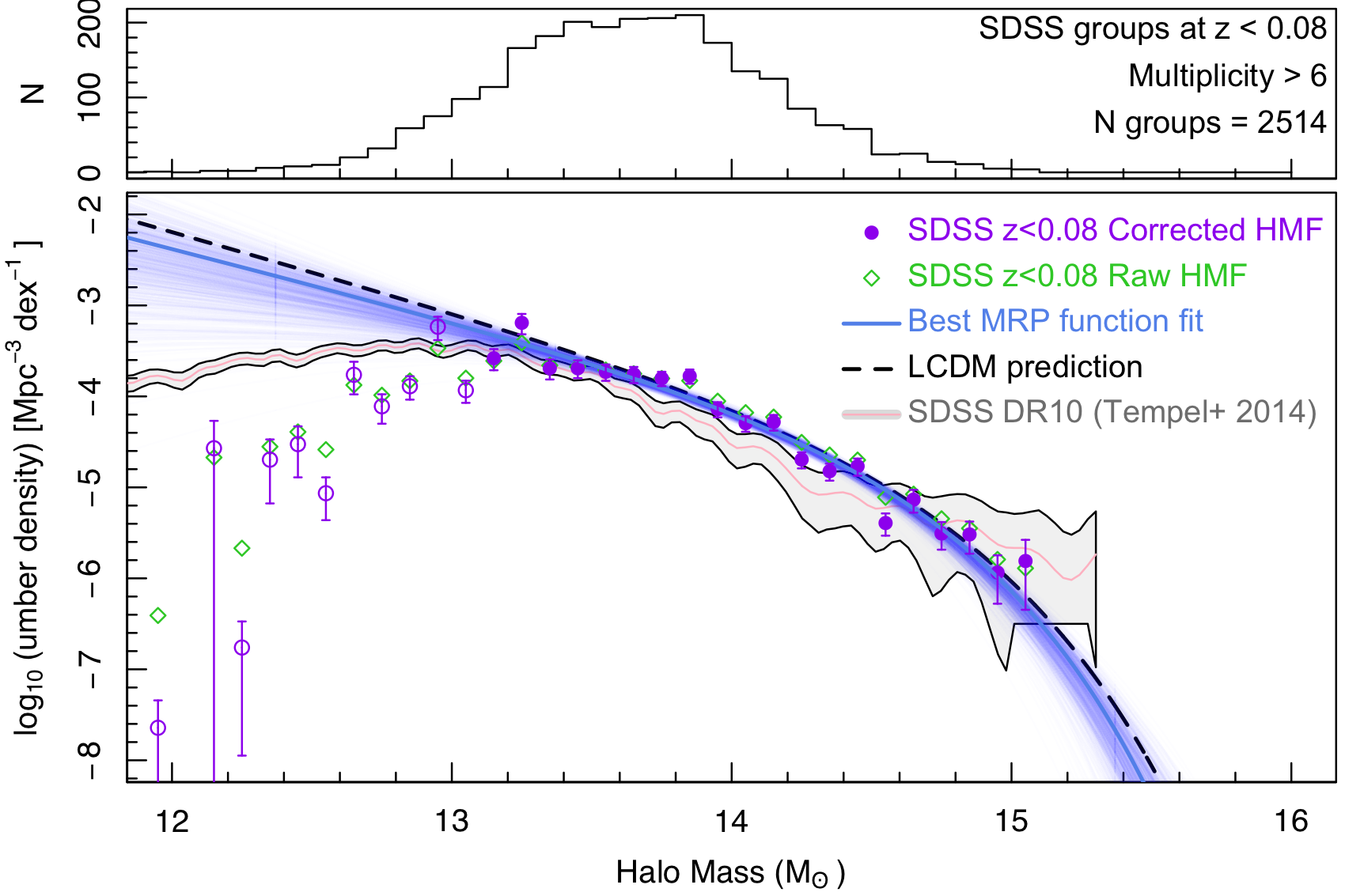}
	\includegraphics[width=\columnwidth]{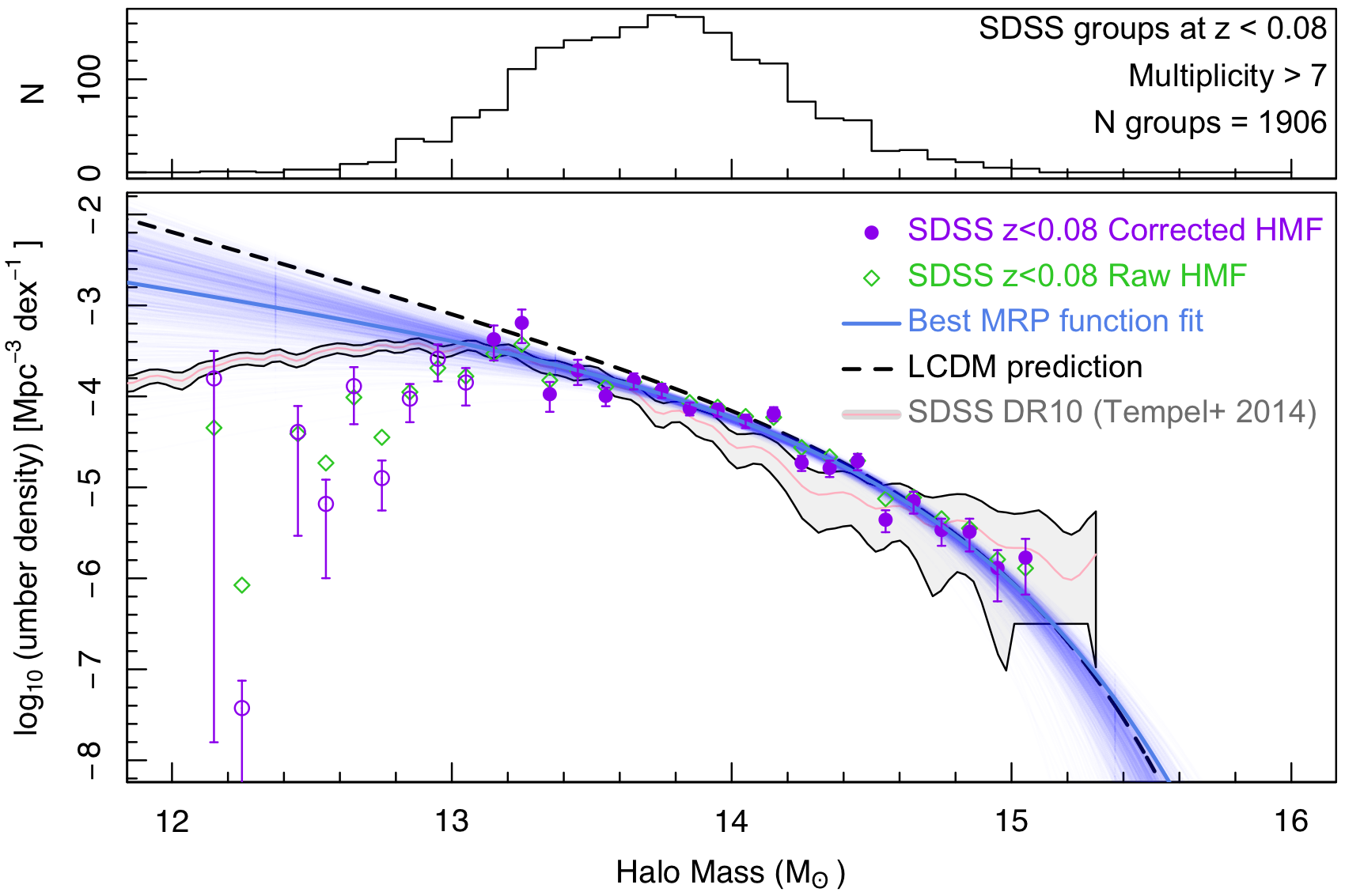}
	\caption{SDSS HMF derivations for different multiplicity cuts. \label{fig:appendix3}}
\end{figure*}

\begin{figure*}
	\centering     
	\includegraphics[width=\columnwidth]{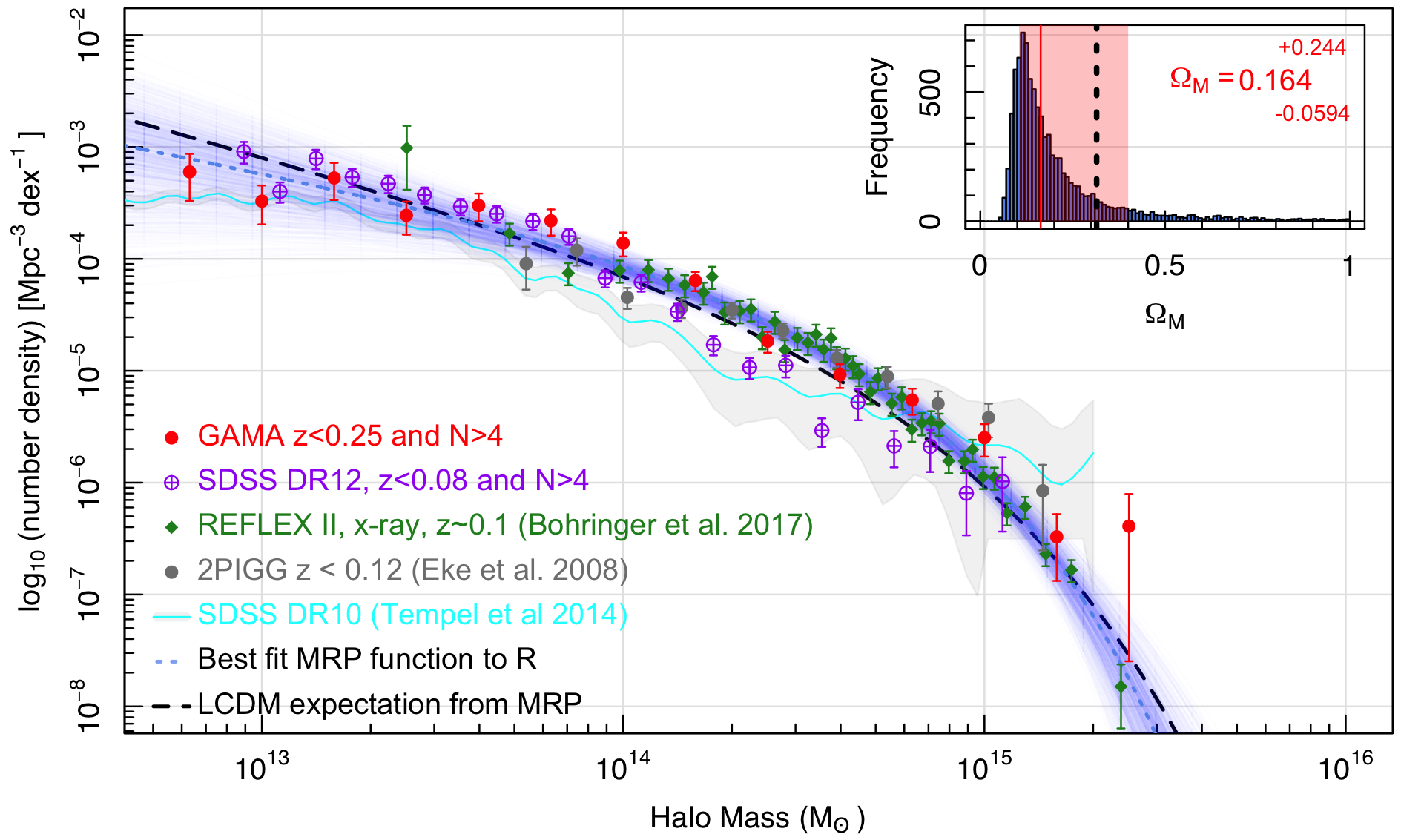}
	\includegraphics[width=\columnwidth]{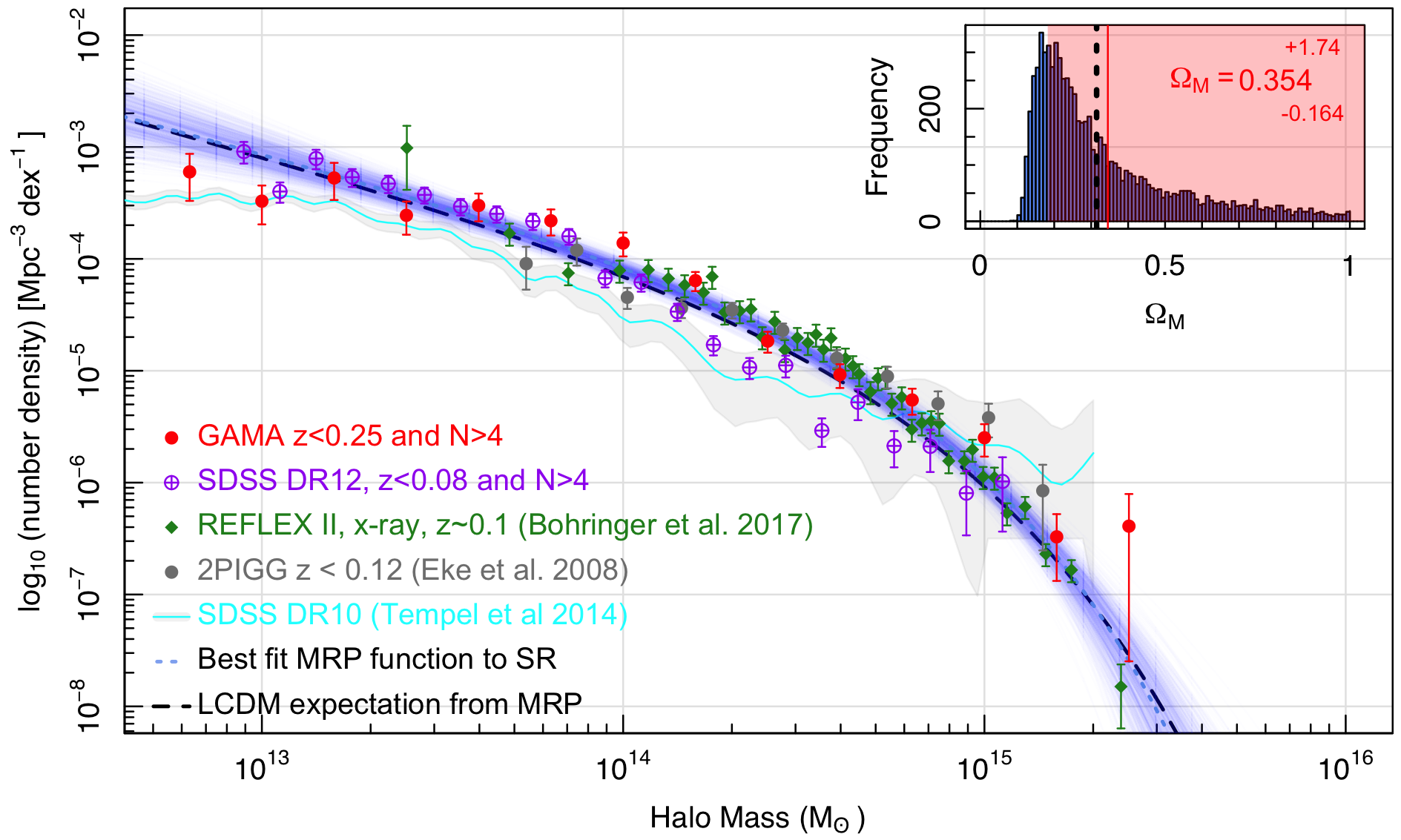}
	\includegraphics[width=\columnwidth]{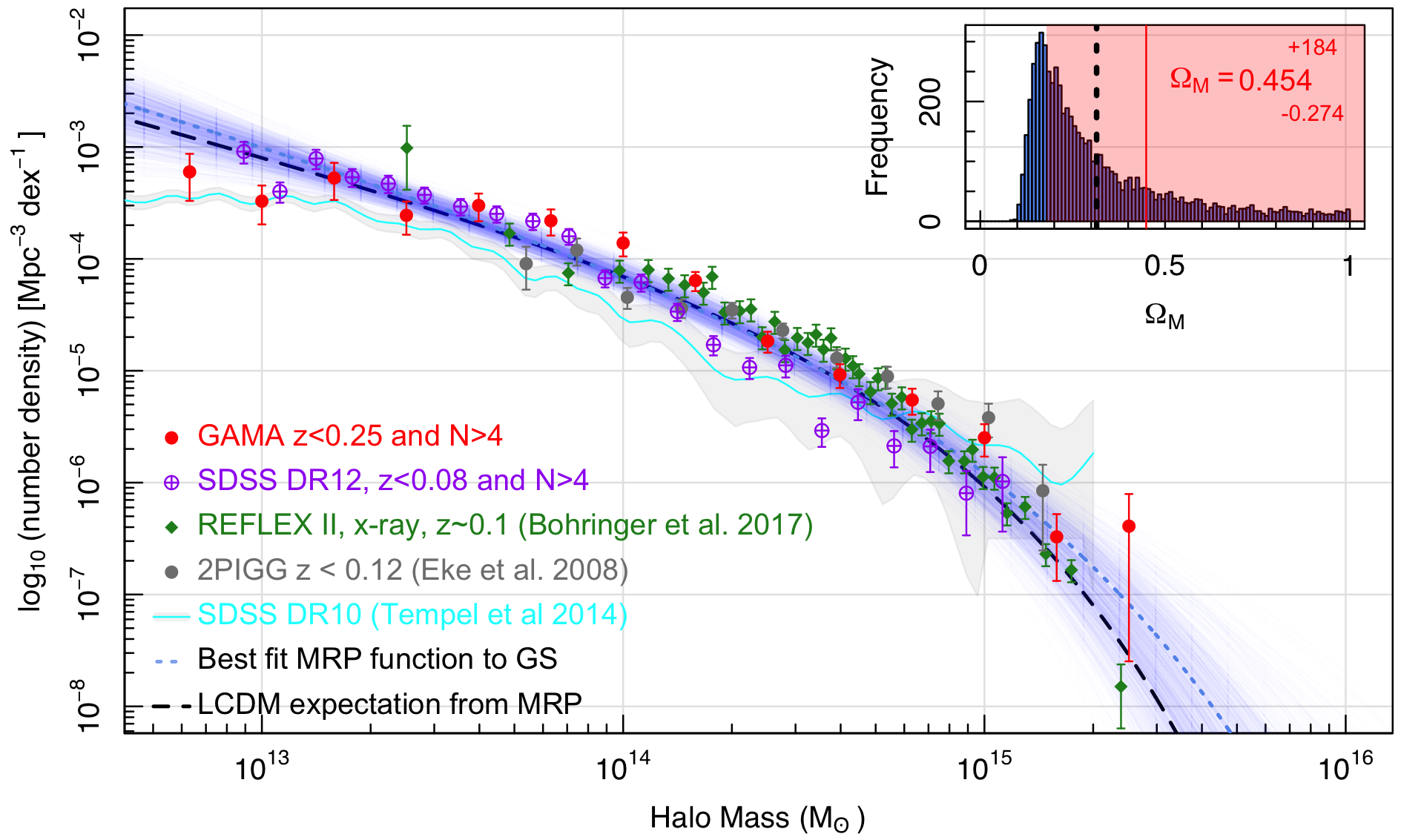}
	\includegraphics[width=\columnwidth]{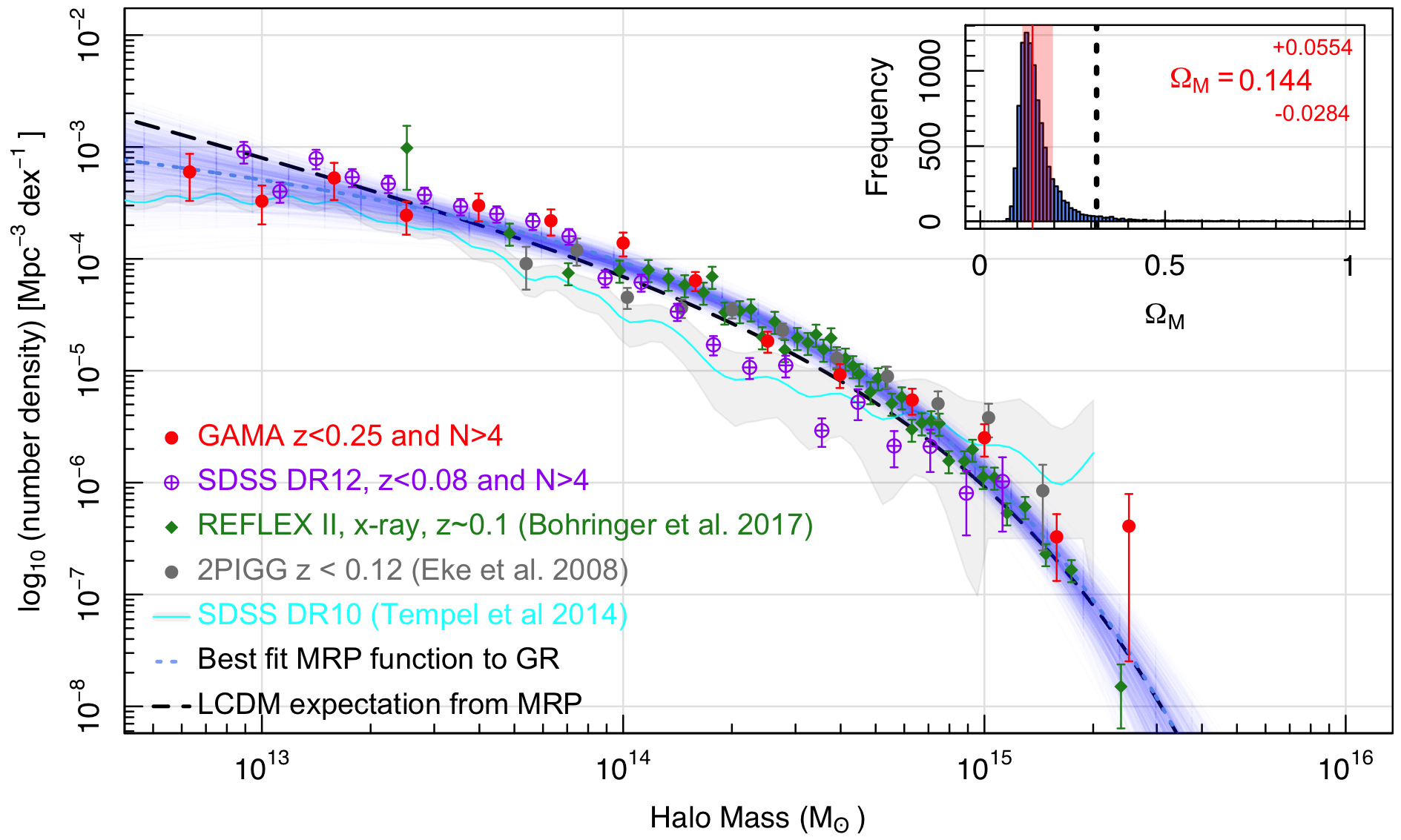}
	\includegraphics[width=\columnwidth]{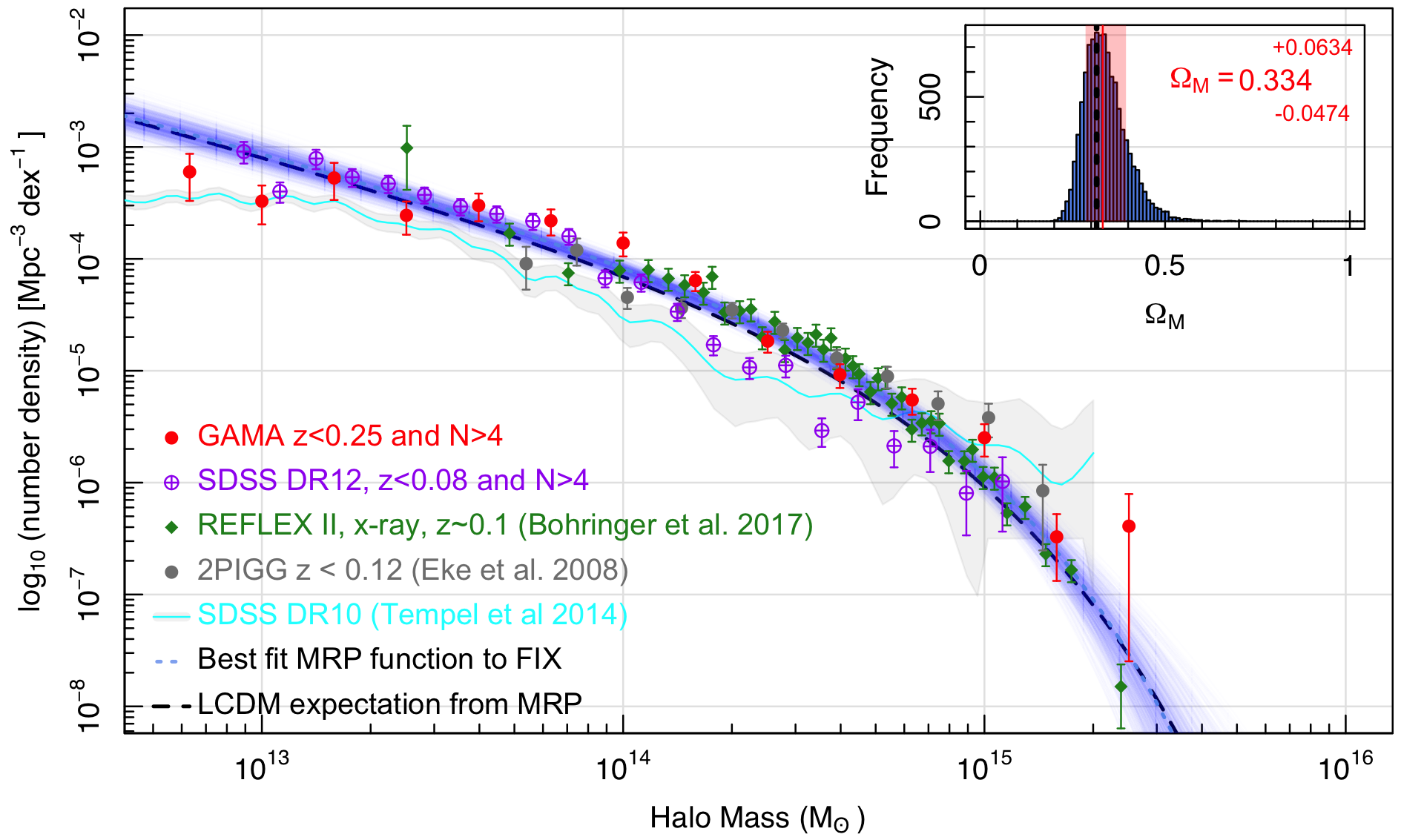}
	\includegraphics[width=\columnwidth]{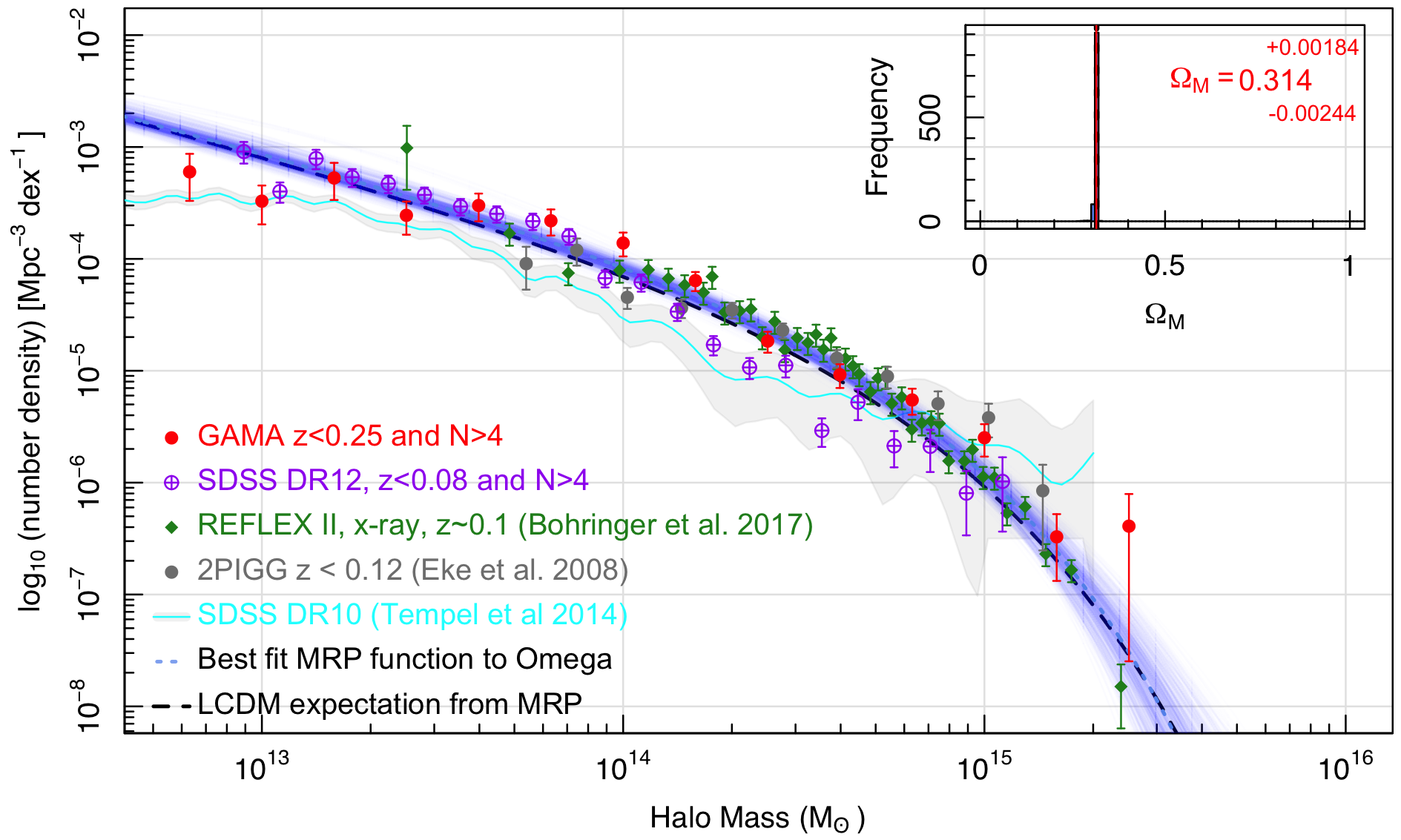}
	\caption{Different MRP fits to the GAMA, SDSS and REFLEX\,II data. \label{fig:appendix4}}
\end{figure*}

\end{document}